\documentclass[preprint, tighten, twocolumn]{aastex62}
\usepackage{apjfonts}
\usepackage{amsmath}
\usepackage{multirow}
\usepackage{textcomp}

\shorttitle{Black Hole Mass in NGC 3258}
\shortauthors{Boizelle et al.}

\usepackage{natbib}
\defcitealias{boi17}{Paper I}

\newcommand{\hst}{\ensuremath{HST}}
\newcommand{\kms}{km s\ensuremath{^{-1}}}
\newcommand{\vlos}{\ensuremath{v_\mathrm{LOS}}}
\newcommand{\vrad}{\ensuremath{v_\mathrm{rad}}}
\newcommand{\vext}{\ensuremath{v_\mathrm{ext}}}
\newcommand{\sigmalos}{\ensuremath{\sigma_\mathrm{LOS}}}
\newcommand{\vcst}{\ensuremath{v_\mathrm{c}^\star}}
\newcommand{\vsys}{\ensuremath{v_\mathrm{sys}}}
\newcommand{\rg}{\ensuremath{r_\mathrm{g}}}
\newcommand{\mbh}{\ensuremath{M_\mathrm{BH}}}
\newcommand{\msun}{\ensuremath{M_\odot}}
\newcommand{\lsun}{\ensuremath{L_\odot}}
\newcommand{\sigmaturb}{\ensuremath{\sigma_\mathrm{turb}}}
\newcommand{\vcstar}{\ensuremath{v_\mathrm{c}^\star}}
\newcommand{\per}{\ensuremath{^{-1}}}
\newcommand{\pertwo}{\ensuremath{^{-2}}}
\newcommand{\rfit}{\ensuremath{r_\mathrm{fit}}}
\newcommand{\chisq}{\ensuremath{\chi^2}}
\newcommand{\chisqnu}{\ensuremath{\chi^2_\nu}}
\newcommand{\ndof}{\ensuremath{N_\mathrm{dof}}}
\newcommand{\upsh}{\ensuremath{\Upsilon_H}}

\begin{document}

\title{A Precision Measurement of the Mass of the Black Hole in NGC 3258 \\ from High-Resolution ALMA Observations of its Circumnuclear Disk\footnote{Based on observations made with the NASA/ESA Hubble Space Telescope, obtained at the Space Telescope Science Institute, which is operated by the Association of Universities for Research in Astronomy, Inc., under NASA contract NAS 5-26555. These observations are associated with program \#14920.}}

\author[0000-0001-6301-570X]{Benjamin D. Boizelle}
\affiliation{George P. and Cynthia Woods Mitchell Institute for Fundamental Physics and Astronomy, 4242 TAMU, Texas A\&M University, College Station, TX, 77843-4242, USA}
\affiliation{Department of Physics and Astronomy, 4129 Frederick Reines Hall, University of California, Irvine, CA, 92697-4575, USA}
\email{bboizelle@tamu.edu}

\author[0000-0002-3026-0562]{Aaron J. Barth}
\affiliation{Department of Physics and Astronomy, 4129 Frederick Reines Hall, University of California, Irvine, CA, 92697-4575, USA}

\author[0000-0002-1881-5908]{Jonelle L. Walsh}
\affiliation{George P. and Cynthia Woods Mitchell Institute for Fundamental Physics and Astronomy, 4242 TAMU, Texas A\&M University, College Station, TX, 77843-4242, USA}

\author[0000-0002-3202-9487]{David A. Buote}
\affiliation{Department of Physics and Astronomy, 4129 Frederick Reines Hall, University of California, Irvine, CA, 92697-4575, USA}

\author[0000-0002-7892-396X]{Andrew J. Baker}
\affiliation{Department of Physics and Astronomy, Rutgers, the State University of New Jersey, 136 Frelinghuysen Road Piscataway, NJ 08854-8019, USA}

\author[0000-0003-2511-2060]{Jeremy Darling}
\affiliation{Center for Astrophysics and Space Astronomy, Department of Astrophysical and Planetary Sciences, University of Colorado, 389 UCB, Boulder, CO 80309-0389, USA}

\author[0000-0001-6947-5846]{Luis C. Ho}
\affiliation{Kavli Institute for Astronomy and Astrophysics, Peking University, Beijing 100871, China; Department of Astronomy, School of Physics, Peking University, Beijing 100871, China}

\begin{abstract}

We present $\sim0\farcs10-$resolution Atacama Large Millimeter/submillimeter Array (ALMA) CO(2$-$1) imaging of the arcsecond-scale ($r \approx 150$ pc) dusty molecular disk in the giant elliptical galaxy NGC 3258. The data provide unprecedented resolution of cold gas disk kinematics within the dynamical sphere of influence of a supermassive black hole, revealing a quasi-Keplerian central increase in projected rotation speed rising from 280 \kms\ at the disk's outer edge to $>400$ \kms\ near the disk center. We construct dynamical models for the rotating disk and fit beam-smeared model CO line profiles directly to the ALMA data cube. Our models incorporate both flat disks and tilted-ring disks that provide a better fit of the mildly warped structure in NGC 3258. We show that the exceptional angular resolution of the ALMA data makes it possible to infer the host galaxy's mass profile within $r=150$ pc solely from the ALMA CO kinematics, without relying on optical or near-infrared imaging data to determine the stellar mass profile. Our model therefore circumvents any uncertainty in the black hole mass that would result from the substantial dust extinction in the galaxy's central region. The best model fit yields $\mbh = 2.249\times10^9$ \msun\ with a statistical model-fitting uncertainty of just 0.18\%, and systematic uncertainties of 0.62\% from various aspects of the model construction and 12\% from uncertainty in the distance to NGC 3258. This observation demonstrates the full potential of ALMA for carrying out highly precise measurements of \mbh\ in early-type galaxies containing circumnuclear gas disks.
\end{abstract}


\section{Introduction}
\label{sec:intro}

Supermassive black holes (BHs), spanning a mass range of $\sim10^6-10^{10}$ \msun, are key constituents of the centers of likely all massive galaxies \citep{mag98,kor13}. Although BHs gravitationally dominate only the innermost regions of galaxies, their masses (\mbh) strongly correlate with several large-scale host galaxy properties, such as the stellar velocity dispersion \citep[$\sigma_\star$;][]{geb00,fer00} and bulge luminosity \citep[$L$;][]{kor95}. These local relationships encapsulate a fossil record of BH and galaxy growth through accretion and merger events, and suggest a co-evolution of central BHs and host galaxies. The local $\mbh-\sigma_\star$ and $\mbh-L$ relationships \citep{mcc13a,kor13,sag16,van16} are also widely employed in estimating \mbh\ for both nearby and distant galaxies across the Hubble sequence.

The BH census remains incomplete, particularly for the most luminous early-type galaxies (ETGs), including brightest cluster galaxies (BCGs) and brightest group galaxies (BGGs). Furthermore, a growing sample of BH masses reveals that the correlations are more complicated than initially thought and may not consistently apply to all galaxy types. For example, predicted \mbh\ values for the most luminous ETGs using their measured stellar velocity dispersions are in tension with masses estimated from the $\mbh-L$ relationship, with the discrepancy reaching an order of magnitude at $\mbh\sim10^{10}$ \msun\ \citep{lau07a,ber07}. The few BCGs with measured BH masses \citep{dal09,mcc12,rus13a} suggest a steeper $\mbh-\sigma_\star$ relationship and may point to different evolutionary processes within cluster centers \citep[e.g.,][]{kra18}. However, large uncertainties in the masses of several of the most massive BHs prevent any secure interpretation.

Presently, $\sim$100 dynamical \mbh\ measurements have been made, primarily by modeling stellar or ionized gas kinematics \citep[e.g.,][]{kor13,sag16}. Reliably measuring \mbh\ requires modeling the kinematics of tracers that originate within the BH sphere of influence $\rg \approx G\mbh/\sigma^2_\star$, where the BH dominates the host galaxy's gravitational potential. The confidence of a BH mass measurement hinges on how well the kinematic observations resolve \rg. Obtaining more than a couple of resolution elements across \rg\ remains challenging for the current generation of optical/near-infrared (NIR) telescopes, even when using adaptive optics (AO). \citet{rus13a} model the stellar kinematics of several luminous ETGs and find that the \mbh\ uncertainties, and the potential biases introduced by model systematics, increase when the angular resolution of the observations exceeds \rg. For stellar-dynamical modeling, these systematics include assumptions about the intrinsic galaxy shape, inclusion of a dark matter halo, and adoption of a spatially constant stellar mass-to-light ($M/L$) ratio \citep{geb09,van10,mcc13b}. Non-circular motion and the treatment of gas turbulence can bias gas-dynamical BH masses \citep[e.g.,][]{van98,wal10}. In the few cases where both stellar and gas-dynamical modeling techniques have been applied to the same galaxy, the inferred BH masses frequently disagree, and discrepancies of a factor $2-4$ are common \citep{geb11,rus11,kor13,wal13,bar16a}.

\begin{figure*}[ht]
\begin{center}
\includegraphics[trim=0mm 3mm 0mm 0mm, clip, width=\textwidth]{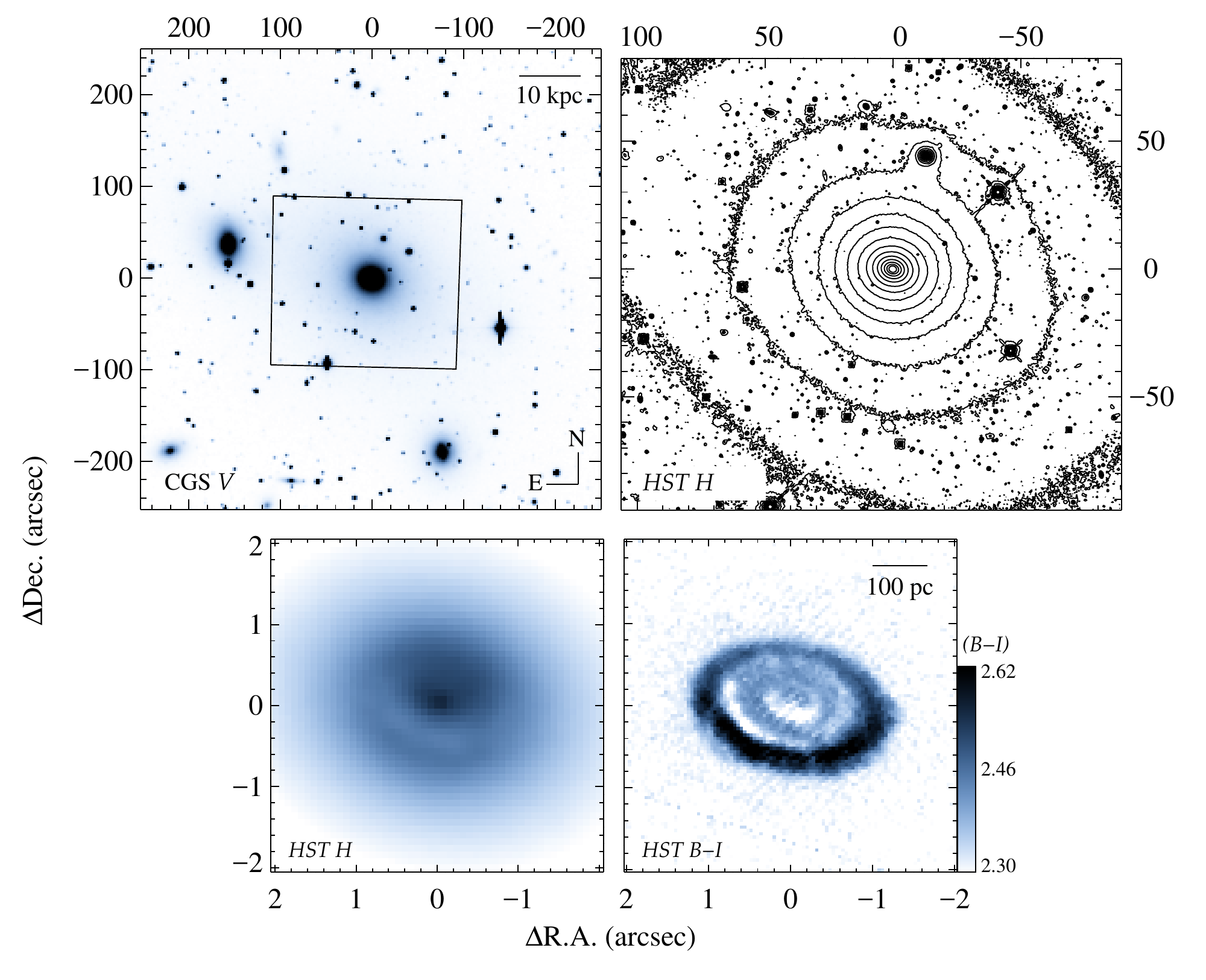}
\begin{singlespace}
  \caption{Optical and NIR imaging of NGC 3258. The wide-field CGS $V$-band image is shown with the footprint of the \hst\ $H$-band observation overlaid (\textit{upper left}). The \hst\ WFC3 $H$-band image is displayed as a contour map (\textit{upper right}). The $H$-band image shows dust obscuration from the circumnuclear disk within the central arcsecond of the galaxy (\textit{lower left}). A $2\farcs4-$wide dust disk is also evident in a $B-I$ color map constructed using \hst\ ACS observations (\textit{lower right}).}
\end{singlespace}
\label{fig:mosaic}
\end{center}
\end{figure*}

Data that \textit{highly} resolve \rg\ have the potential to avoid nearly all such serious systematics. Kinematic measurements of 22 GHz H$_2$O emission from radii $\ll \rg$ in megamaser disks enable \mbh\ determinations with percent-level precision \citep[e.g.,][]{miy95,kuo11}. Unfortunately, such disks are rare \citep[e.g.,][]{bra96} and tend to be found in late-type galaxies with black hole masses clustering in a narrow range about $\sim$10$^7$ \msun. Surveys to identify megamaser disks within ETGs have thus far been unsuccessful \citep[e.g.,][]{van16}, so a different method is needed to make precision BH mass measurements in the most massive galaxies.

Molecular gas tracers are a promising new avenue for reliably measuring BH masses, especially in ETGs. Recent $^{12}$CO surveys \citep{com07,you11,ala13,bol17,zab18} find central, regularly-rotating cold molecular gas in roughly 10\% of all nearby elliptical and S0 galaxies. Low turbulent velocity dispersions indicate that the molecular gas in these disks is a better tracer of the underlying gravitational potential than ionized gas in ETGs. Until recently, however, mm/sub-mm arrays were only able to resolve the nuclear gas kinematics at $r<\rg$ for a very small number of galaxies. For one such nearby ETG at $D\sim 16$ Mpc, \citet{dav13a} mapped rapid CO gas rotation at 0\farcs25 resolution with the Combined Array for Research in Millimeter-wave Astronomy (CARMA), and demonstrated that BH masses can be constrained using mm-wavelength molecular gas as tracers.

The Atacama Large Millimeter/submillimeter Array (ALMA) now offers the possibility of routinely carrying out molecular-line observations that resolve \rg, given its increased sensitivity and significantly higher angular resolution relative to previous facilities. ALMA observations are highly sensitive probes of molecular gas within the central $\sim$kpc region of luminous ETGs \citep[][hereinafter Paper I]{boi17} and have opened a new avenue for \mbh\ determination \citep{bar16a,bar16b,oni17,dav17b,dav18,smi19} via detection and modeling of the central high-velocity rotation around the BH. However, even when ALMA observations resolve \rg, a central nearly-Keplerian rise in rotation speed is not typically seen, indicating a dearth of gas at locations close to the BH. In many cases, the data reveal only a modest central rise in peak rotation speed originating from gas in the outer portion of the BH sphere of influence, suggesting the presence of a central hole in the CO distribution at $r<\rg$. Other cases are found to exhibit a resolved central hole in the CO distribution with radius larger than \rg\ \citepalias[e.g.,][]{boi17}. For high-precision measurement of \mbh, the ideal configuration is a disk with bright CO emission extending down to radii much smaller than \rg, from which the central rotation speed due to the BH's gravity would rise far above the rotation speed at larger radii due to the host galaxy's mass. ALMA observations published to date indicate that disks with these properties are fairly rare among the local ETG population, with only a very small fraction exhibiting signatures of very rapid central rotation from radii deep within \rg.

NGC 3258 was first observed by ALMA as part of the Cycle 2 program described in \citetalias{boi17}. That $\sim$0\farcs44 resolution imaging revealed bright CO(2$-$1) emission from a rapidly rotating nuclear gas disk, with a spatially unresolved central rise in line-of-sight velocity (\vlos) extending to $\sim$500 \kms\ relative to the systemic velocity (\vsys) and rising to $\gtrsim200$ \kms\ above the rotation speed of the outer disk. These attributes made NGC 3258 a promising target for high-resolution ALMA imaging in order to determine its BH mass to high precision. This E1 galaxy has a bulge stellar velocity dispersion of $\sigma_\star=260\pm10$ km s$^{-1}$ and $K$-band absolute magnitude of $M_K=-24.33\pm0.45$ mag \citep[from the HyperLeda database;][]{mak14}. We adopt a distance modulus $m-M=32.53\pm 0.27$ mag based on surface brightness fluctuation measurements \citep[SBF;][]{ton01}, which corresponds to a luminosity distance $D_L=31.9\pm 3.9$ Mpc. Using an observed redshift $z=0.0092091$ from our initial dynamical modeling results \citep[that is very close to other optical measurements;][]{dev91}, this $D_L$ corresponds to an angular size distance of 31.3 Mpc, for which $1\arcsec$ spans a physical scale of 151.8 pc. NGC 3258 is one of two BGGs that dominate the dynamically young Antlia cluster \citep{hes15}, a somewhat poor cluster with $\sim$400 member galaxies \citep{fer90}. Optical long-slit spectroscopy reveals only weak evidence for stellar rotation but a large central stellar velocity dispersion of $\sim$400 \kms\ \citep{kop00,bru04}. As no atomic gas reservoir is detected within this galaxy \citep{hes15}, the cold gas in NGC 3258 appears to be primarily molecular. Mid-infrared \textit{Spitzer} spectra show significant nuclear polycyclic aromatic hydrocarbon emission that, together with other nuclear diagnostic line diagrams, suggests a recent ($\sim$200 Myr) burst of star formation \citep{ram13}.

In this paper, we present $\sim$0\farcs10 resolution ALMA Cycle 4 CO(2$-$1) observations of NGC 3258. A factor of four improvement in angular resolution compared with the earlier Cycle 2 data fully resolves gas rotation within \rg\ and enables measurement of the BH mass to an unprecedented level of precision for a giant elliptical galaxy. The extraordinary resolution of the gas kinematics within \rg\ in NGC 3258 makes it possible to constrain the spatially extended host galaxy mass distribution within the galaxy's inner arcsecond solely from the ALMA kinematic data, in contrast to the traditional approach of using high-resolution optical/NIR imaging data to measure and deproject the host galaxy luminosity profile and assuming a spatially uniform stellar mass-to-light ratio. Measuring the host galaxy's mass profile from the kinematic data makes it possible to avoid an uncertainty of order several percent in \mbh\ that would result from the uncertain extinction of the host galaxy's central stellar luminosity profile. This method is particularly advantageous for systems such as NGC 3258 in which the central region of the galaxy is highly obscured by dust, as will be the case for nearly all CO-bright galaxies targeted for ALMA observations to measure \mbh.  Our measurement yields statistical model-fitting uncertainties that are significantly smaller than the systematic uncertainties resulting from issues such as localized irregularities in the gas disk kinematics. We carry out a variety of tests to estimate these model-fitting systematics and find that they are below the $\sim1\%$ level except for the uncertainty in the galaxy's distance, which contributes $>$10\% systematic uncertainty to the error budget, as is generally the case for nearly all dynamical BH mass measurements.

This paper is organized as follows. In Section~\ref{sec:oir}, we present \textit{Hubble Space Telescope} (\hst) optical and NIR broadband imaging of NGC 3258 and measurements of the galaxy's light profile. We describe models for extinction and reddening due to the inclined circumnuclear dust disk embedded within the galaxy, and demonstrate that the disk is very optically thick at visible wavelengths. We use the \hst\ data to derive dust-corrected models for the host galaxy's intrinsic luminosity distribution that we then deproject and employ as a component of the traditional approach for BH mass measurement. We introduce the new ALMA Cycle 4 observations in Section~\ref{sec:almadata}. In Section~\ref{sec:dynamical}, we describe our gas-dynamical modeling method and discuss results when fitting these models to the Cycle 2 and 4 ALMA data cubes. We present model-fitting results for the simple case of a geometrically flat disk, and for a tilted-ring model that more closely matches the disk's mildly warped structure. We compare results from models employing a dust-corrected stellar mass profile measured from \hst\ imaging and models based on a new method that determines the host galaxy's radial mass profile solely from the ALMA CO kinematics. In Section~\ref{sec:discussion}, we discuss the implications of high-precision ALMA BH mass measurements and place NGC 3258 in the context of \mbh$-$host galaxy relationships.

\section{Optical and Near-Infrared Observations}
\label{sec:oir}

A typically key input into gas-dynamical models is the stellar contribution to a galaxy's gravitational potential. We used \hst\ Wide Field Camera 3 (WFC3) NIR images to determine the luminous mass distribution in the galaxy's central region and Advanced Camera for Surveys (ACS) Wide Field Channel (WFC) observations to characterize the dust disk properties. In order to probe the galaxy's outskirts, we supplemented the \hst\ WFC3 data with ground-based, wide-field images from the Carnegie-Irvine Galaxy Survey \citep[CGS;][]{ho11}. Below, we summarize the observations, data reduction procedures, surface brightness measurements, and disk extinction models.

\subsection{HST Imaging}
\label{sec:hstimaging}

We observed NGC 3258 in one orbit on 5 June 2017 as part of program GO-14920 with \hst\ WFC3 through the IR channel using the F110W and F160W ($J$ and $H$) filters. We took four MULTIACCUM exposures in each filter with the SPARS25 (NSAMP$=$12$-$13) and STEP50 (NSAMP$=$13) modes, employing a large box dither pattern that always kept the galaxy nucleus in one corner of the detector. We processed the data through the \texttt{CALWF3} pipeline and used \texttt{AstroDrizzle} \citep{gon12} to produce cleaned, distortion-corrected images with a pixel scale of 0\farcs08 pixel\per\ in each filter. The final $J$ and $H$-band images cover a 3\farcm4$\times$3\farcm1 field of view, and have total integration times of 18 minutes and 23 minutes, respectively. The images are dominated by galaxy light over the full WFC3/IR field of view, and so we did not perform background subtraction at this stage. The final \hst\ images have an angular resolution of 0\farcs21 ($J$) and 0\farcs22 ($H$), determined by averaging the full widths at half maximum (FWHMs) of several foreground stars. In Figure~\ref{fig:mosaic}, we present the $H$-band mosaic of NGC 3258 and its inner 4\arcsec$\times$4\arcsec\ region, which illustrates the substantial extinction by the central dust disk.

\begin{figure}
\begin{center}
\includegraphics[trim=0mm 0mm 0mm 0mm, clip, width=\columnwidth]{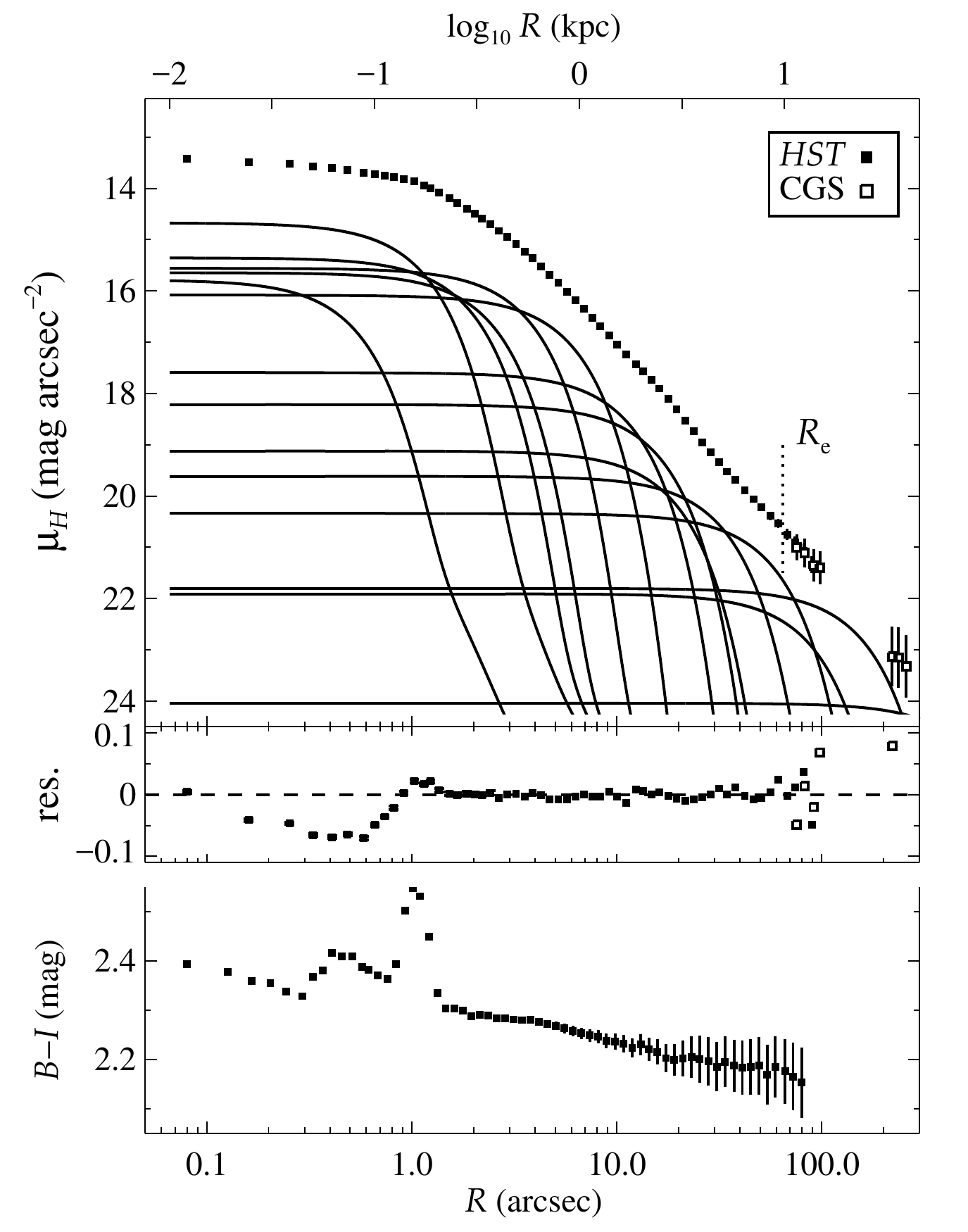}
\begin{singlespace}
  \caption{The surface brightness profile of NGC 3258 along the major axis and its 14-component, multi-Gaussian expansion (\textit{top}). The surface brightness was measured from the \hst\ $H$-band image, with CGS data spliced in at radii beyond $\sim$70\arcsec. The gap in CGS data points at $\sim$120\arcsec\ is the result of masking out a neighboring galaxy and a foreground star. The multi-Gaussian expansion is a good parameterization of the observed surface brightness, with model residuals (\textit{middle}) below the \textit{2}\% level at most locations except for the most dust-obscured points around $R\sim0\farcs5$. The arcsecond-scale dust disk is clearly identified in the $B-I$ color profile (\textit{bottom}).}
\end{singlespace}
\label{fig:mge}
\end{center}
\end{figure}

In addition, we retrieved ACS/WFC F435W and F814W ($B$ and $I$) images of NGC 3258 from the \hst\ archive. These $B$ and $I$ images were taken over three orbits as part of program GO-9427, and have integration times of 89 min and 38 min, respectively. We processed the raw ACS data using the \texttt{CALACS} pipeline, included corrections for charge transfer inefficiency, and then drizzled the geometrically rectified ACS exposures in each filter. The final images have an angular resolution of $\sim$0\farcs12 and cover the galaxy's central 3\farcm4$\times$3\farcm3 region. In Figure~\ref{fig:mosaic}, we show the $B-I$ map, in which a nearly azimuthally symmetric, $\sim$2\farcs4$-$wide dust disk is visible \citep[see also][]{bru04,cap05}.

\subsection{CGS Imaging}
\label{sec:cgs}

We used ground-based optical data from CGS to complement the \hst\ images. The CGS observations and data reduction are described by \citet{ho11}, and the processed images have a 9\arcmin$\times$9\arcmin\ field of view and a pixel scale of 0\farcs26. We selected the CGS $V$-band observation instead of a redder filter that would better trace old stellar populations in order to avoid the ``red halo'' effect. This instrumental effect, stronger in longer-wavelength filters, adds an extended feature to the point-spread function (PSF) wings, potentially affecting measurements of the galaxy's brightness profile \citep{hua13}. The sensitivity of the $V$-band CGS image reaches 26.9 mag arcsec\pertwo, and the image was taken in $\sim$1\arcsec\ seeing. The CGS $V$-band image of NGC 3258 is displayed in Figure~\ref{fig:mosaic}.

\subsection{Stellar Luminosity Profile}
\label{sec:stellarprofile}

After masking out foreground stars and galaxies, we measured NGC 3258's surface brightness from the $H$ and $V$-band images in regions spaced logarithmically in radius and equally in angle \citep{cap02}. The average position angle (PA) of 77\degr\ was determined using the central $R<10$\arcsec\ region of the NIR mosaic. (Throughout this paper, we use $R$ to denote projected radius on the plane of the sky, and $r$ to denote radial distance within the galaxy.) Using the surface brightness measurements at radii between $70-100$\arcsec\ along the galaxy's major axis, we determined the $H$-band background level and the $V-H$ color needed to align the two profiles. We found a best-fit background level of $H=$ 20.8 mag arcsec\pertwo\ and a color of $V-H=$ 2.40 mag \citep[consistent with optical-NIR colors of elliptical galaxies at large radii; e.g.,][]{sch93}, which we then applied to the $H$-band surface brightness measurements at radii 0\farcs07$-$100\arcsec\ and to the $V$-band surface brightness measurements at radii 70$-$300\arcsec, respectively. This produced $H$-band surface brightness profiles measured at nineteen angles between $0-90\degr$ from the major axis and extending out to $\sim$300\arcsec\ ($\sim$45.5 kpc), or 5$-$7 times the estimated half-light radius \citep[$R_\mathrm{e}$;][]{lau89,dir03}. Although we used the surface brightness profile along the major axis to establish the $H$-band background and $V-H$ color, there was good visual agreement between the $H$ and scaled $V$-band surface brightness measurements at all angles. We also note that $V-H$ gradients are negligible at radii $\gtrsim60$\arcsec, and therefore we assumed that $V-H$ remains constant with radius when adjusting the large-scale $V$-band measurements.

\begin{deluxetable}{cccc}
\tabletypesize{\small}
\tablecaption{$H$-band MGE Parameters\label{tbl:mge}}
\tablewidth{0pt}
\tablehead{
\colhead{$j$} & 
\colhead{$\log_{10}$ $I_{H,j}$ (\lsun\ pc\pertwo)} & 
\colhead{$\sigma'_{j}$ (arcsec)} & 
\colhead{$q\arcmin_{j}$} \\[-1.5ex]
\colhead{(1)} & 
\colhead{(2)} & 
\colhead{(3)} & 
\colhead{(4)}
}
\startdata
1 & 3.75 & 0.37 & 0.78 \\
2 & 4.14 & 0.82 & 0.83 \\
3 & 3.85 & 1.37 & 0.75 \\
4 & 3.72 & 1.77 & 0.77 \\
5 & 3.75 & 2.68 & 0.79 \\
6 & 3.54 & 4.44 & 0.85 \\
7 & 2.93 & 8.41 & 0.84 \\
8 & 2.67 & 11.7 & 1.00 \\
9 & 2.32 & 14.0 & 0.85 \\
10 & 2.12 & 23.9 & 0.94 \\
11 & 1.83 & 41.6 & 0.81 \\
12 & 1.20 & 64.8 & 0.96 \\
13 & 1.24 & 115 & 0.72 \\
14 & 0.35 & 400 &  0.72
\enddata
\begin{singlespace}
  \tablecomments{MGE decomposition of the \hst\ $+$ CGS surface brightness measurements after masking out the most dust-obscured regions to the south of the nucleus. Column (1) lists the component number $j$, column (2) is the central surface brightness, assuming an $H$-band absolute magnitude of 3.37 mag for the Sun \citep{wil18}, column (3) gives the Gaussian standard deviation along the major axis, and col.\ (4) provides the axis ratio. Primes indicate projected quantities, and all the components have a PA of 77\degr.}
\end{singlespace}
\end{deluxetable}

We corrected for foreground Galactic reddening of $A_H=0.041$ mag \citep{sch11} and modeled the galaxy's $H$-band surface brightness with a two-dimensional (2D) multi-Gaussian expansion \citep[MGE;][]{em94,cap02} after masking out the most dust-obscured regions to the south of the nucleus between $R\sim0\farcs15-0\farcs8$. Although individual Gaussian components do not have physical meaning, MGEs are commonly used to represent a wide variety of surface brightness profiles and allow for the luminosity density to be determined through an analytical deprojection. We required that each Gaussian component have the same center and PA, and constrained the observed flattening (the ratio between the projected major and minor axis of the 2D Gaussian, $q^\prime$) to be $>0.72$. Such a restriction avoids highly flattened components that would limit the range of inclination angles ($i$) for which an MGE model can be deprojected. Prior to parameterizing the $H$-band surface brightness with an MGE model, we generated a Tiny Tim \citep{kri04} PSF, which was dithered and drizzled identically to the $H$-band mosaic of the galaxy. We applied the MGE formalism to the \hst\ PSF, modeling it as the sum of six concentric circular Gaussians. This six-component PSF was taken into account while fitting the galaxy's surface brightness using the MGE \citep{cap02}.

The final MGE model of the galaxy consists of 14 concentric elliptical Gaussians, for which the best-fit parameters are provided in Table~\ref{tbl:mge}. This MGE is a good description of the $H$-band surface brightness measurements, and we show a comparison between the MGE model and the data along the galaxy major axis in Figure~\ref{fig:mge}. The galaxy's total $H$-band luminosity is $L_H=1.1\times10^{11}$ \lsun, measured within the central 300\arcsec\ (45.5 kpc), and we find that $R_\mathrm{e}\approx 66\arcsec$ (10.6 kpc) in this filter.

Assuming that the galaxy has an oblate axisymmetric shape and is inclined at the same angle as the molecular gas disk ($i\approx 48\degr$; see Section~\ref{sec:dynamical}), we deprojected the $H$-band MGE model and numerically integrated the resulting stellar luminosity densities assuming an initial $H$-band $M/L$ ratio $\upsh=1$ to determine the stellar contribution to circular velocity (\vcst) as a function of radius. During gas-dynamical modeling, these \vcst\ values are then scaled by $\sqrt{\upsh}$ that is a free parameter of the fits. We do not include contributions from a dark matter halo, as these are negligible within the central few kpc of the galaxy \citep{bru04}.

\subsection{Disk Extinction Modeling}
\label{sec:extinction}

Modeling the extinction from the circumnuclear dust disk is essential in order to derive accurate models of the host galaxy stellar mass profile from the \hst\ images. The $H$-band nuclear morphology seen in Figure \ref{fig:mosaic} is suggestive of a ring-like obscuring structure at $R \approx 0\farcs5$ in which the extinction is most pronounced on the southern (near) side of the disk. At somewhat larger radii, ring-like obscuration also appears in optical \hst\ images \citepalias[see \citealt{cap05} and][]{boi17}. The $H$-band surface brightness profile (Figure \ref{fig:mge}) shows an apparent break at $R \approx 2\arcsec$ to a nearly flat inner slope at smaller radii, suggesting substantial extinction of the $H$-band light by the disk within the galaxy's inner arcsecond.

The \hst\ $B-I$ color map reveals that the region of largest optical color excess (relative to the host galaxy color of $B-I\approx2.30$ mag outside the dust disk region) is confined to a ring located at $R \approx 1\arcsec$. On the southern side of the ring, the $B-I$ color is $\sim0.3-0.4$ mag redder than the host galaxy outside the dust disk.  At radii $<1\arcsec$, the color map shows a patchy structure with multiple concentric ringlets and a greater overall color excess of $(B-I) \approx 0.1 - 0.15$ mag relative to the host galaxy on larger scales.

\begin{figure}
\begin{center}
\includegraphics[trim=0mm 0mm 0mm 0mm, clip, width=\columnwidth]{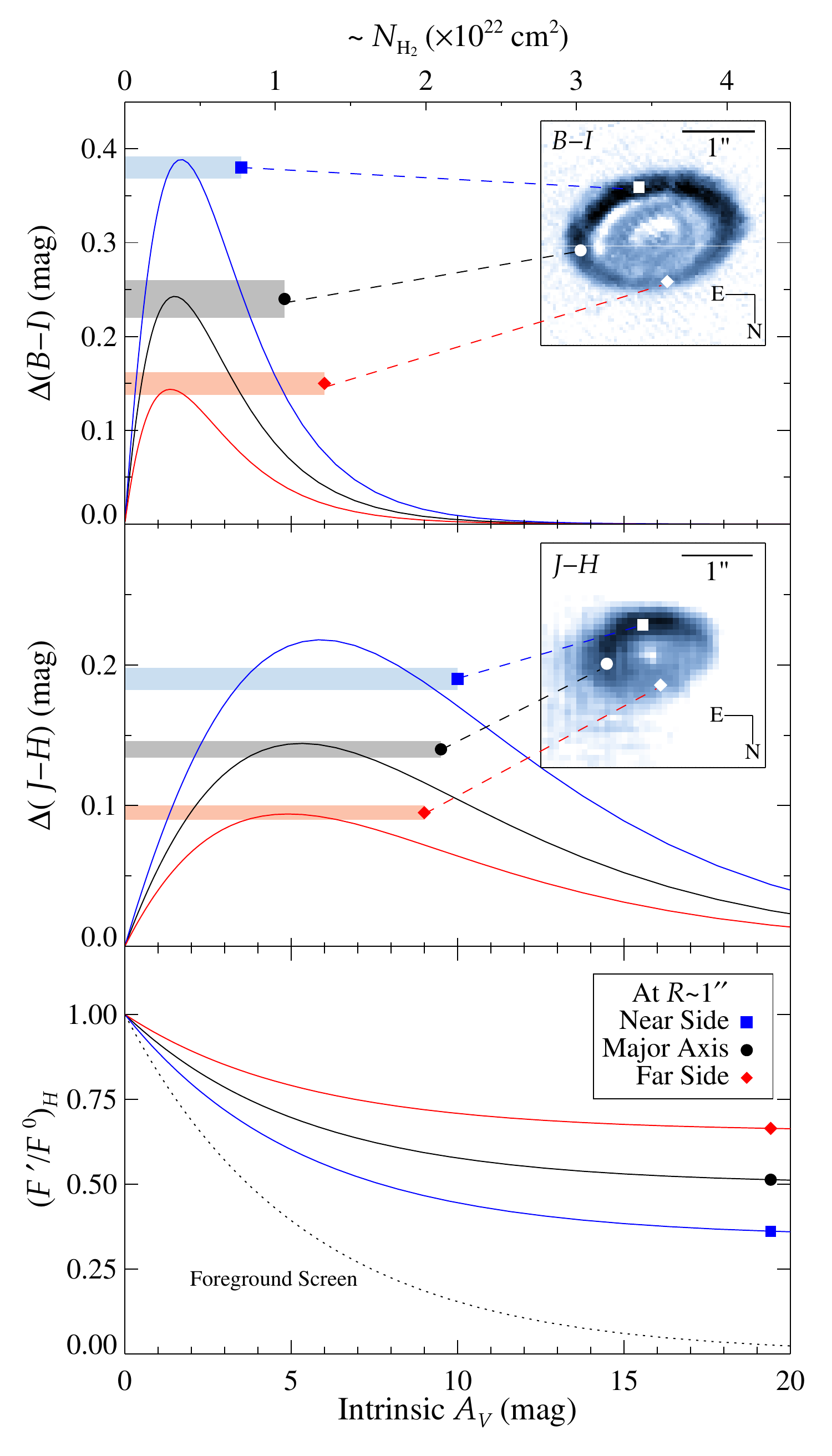}
\begin{singlespace}
  \caption{Modeled color excess (\textit{top and middle panels}) and integrated line-of-sight $H$-band intensity (\textit{bottom panel}) as functions of intrinsic $V$-band extinction $A_V$ for the inclined, embedded-screen dust disk model with $i=48\degr$ (see Equations~\ref{eqn:1} and \ref{eqn:2}). Results were calculated for three disk locations each for the $B-I$ and $J-H$ maps, at points within the ring of maximum color excess for each of the two color maps. Horizontal bars illustrate the ranges of $B-I$ and $J-H$ colors at each of these positions for comparison with model predictions. The $J-H$ color reaches maximal values at smaller radii than the $B-I$ color, indicating that the disk becomes increasingly opaque towards the center. The bottom panel also illustrates the integrated line-of-sight $H$-band intensity for the case of a foreground rather than embedded dust screen (\textit{dotted curve}); in this case, the observed flux falls to zero in the limit of high disk extinction.}
\end{singlespace}
\label{fig:dust}
\end{center}
\end{figure}

If interpreted as due to a foreground screen of extinction in front of NGC 3258, these color excesses would indicate modest extinction values reaching $A_V \sim 0.3$ mag (or $A_H \sim 0.06$ mag) in a ring and that decreases by a factor of two towards the central portion of the disk \citep[assuming a standard Galactic extinction law;][]{car89}. However, such small extinctions would not be sufficient to create the observed absorption feature in the nuclear $H$-band light (Figure~\ref{fig:mosaic}). Furthermore, the CO(2$-$1) surface brightness of the disk suggests an average $V$-band extinction as high as 5$-$10 mag over the disk surface \citepalias{boi17}. This seeming discrepancy is the result of the disk's location in the midplane of the host galaxy: interpreting the observed color excess with foreground-screen models would greatly underpredict the disk's true optical depth \citep{tra01}.

In this situation, the obscuring structure is an inclined, dusty disk in the midplane of the galaxy. Starlight originating from in front of the disk is unobscured, while light from within and behind the disk is attenuated. In the limit of very high optical depth in a very thin disk, light from the far side of the disk would be completely obscured, and a $B-I$ color map would not reveal any color excess. The maximum observed $B-I$ color excess would occur for some moderate value of disk optical depth that would permit some reddened starlight to pass through the disk. For an inclined, embedded dust disk, the near side of the disk would be expected to show a larger color excess than the far side as the near side of the disk obscures a greater fraction of the host galaxy's starlight \citep{elm99a}.

To examine the relationship between disk optical depth and observed color excess, we employed a simple embedded-screen model following the method described by \citet{via17}. In this model, the obscuring structure is treated as a thin, inclined disk bisecting the galaxy. Along a given line of sight, the fraction $b$ of total stellar light originating \textit{behind} the disk is obscured by simple screen extinction, while the fraction $f$ in front remains unaffected. For full generality in the case of a thick disk, a small fraction ($w=1-f-b$) of the total light may  originate within the disk and therefore experience ``mixed'' attenuation. Rewriting Equation 6 of \citet{via17} in terms of the extinction $A_\lambda$, the wavelength-dependent ratio $F^\prime/F^0$ of observed to intrinsic integrated stellar light takes the form
\begin{equation} \label{eqn:1}
    \left(\frac{F^\prime{}}{F^0}\right)_\lambda\approx f+w \left[\frac{1-10^{-A_\lambda/2.5}}{0.921A_\lambda}\right]+b \left[10^{-A_\lambda/2.5}\right]\,.
\end{equation}
We used the same $R_V=3.1$ extinction law to characterize $A_\lambda$, and for simplicity assumed a very thin ($w\rightarrow0$) disk. To determine fractions $f$ and $b$ across the arcsecond-scale disk, we populated a model galaxy cube with stellar densities deprojected from the $H$-band MGE solution and adopted $i=48\degr$ for the dust disk based on initial gas-dynamical modeling results in Section~\ref{sec:dynamical}. We evaluated Equation~\ref{eqn:1} at the pivot wavelengths of the ACS and WFC3 filters to generate predictions for the opacity-dependent color excess at each spatial location:
\begin{equation} \label{eqn:2}
    \Delta(B-I)=-2.5\log_{10}\left[\left(\frac{F^\prime}{F^0}\right)_B\left(\frac{F^\prime}{F^0}\right)_I^{-1}\right],
\end{equation}
with a similar form for $\Delta(J-H)$.

In Figure~\ref{fig:dust}, we show the modeled color excesses $\Delta(B-I)$ and $\Delta(J-H)$ as a function of the intrinsic extinction $A_V$ of the obscuring disk, extracted at three locations each in order to illustrate the effect of the disk inclination on the color excess at different locations in the disk. These major and minor axis positions coincide with the elliptical rings of maximal color excess observed at $R \sim 1\farcs1$ and $\sim 0\farcs5$ for the $B-I$ and $J-H$ color maps, respectively. As expected, the color excess predicted by the model is very small for both very low and very high disk optical depth, and reaches a maximum value at intermediate extinction. The predicted $B-I$ color excess peaks at a disk extinction of $A_V \approx 1.5-2$ mag, while the $J-H$ color excess peaks at $A_V \approx 5-6$ mag. Away from these extinctions corresponding to peak color excesses, the observed color excess no longer corresponds to a unique $A_V$ value. The color excesses on the near side of the disk are predicted to be more than twice the value of the color excesses on the far side.

Remarkably, this simple model predicts maximum color excess values that are in very close agreement with both the $B-I$ and $J-H$ color maps of NGC 3258, as seen in Figure \ref{fig:dust}. This consistency indicates that the disk optical depth rises from $A_V \sim 1.5$ mag near the disk edge to at least 5 mag at $R \sim 0\farcs5$. Such high intrinsic extinction corresponds to a substantial attenuation of $H$-band light within the galaxy's inner arcsecond, as illustrated in the bottom panel of Figure \ref{fig:dust}.

\begin{figure}
\begin{center}
\includegraphics[trim=0mm 0mm 0mm 0mm, clip, width=\columnwidth]{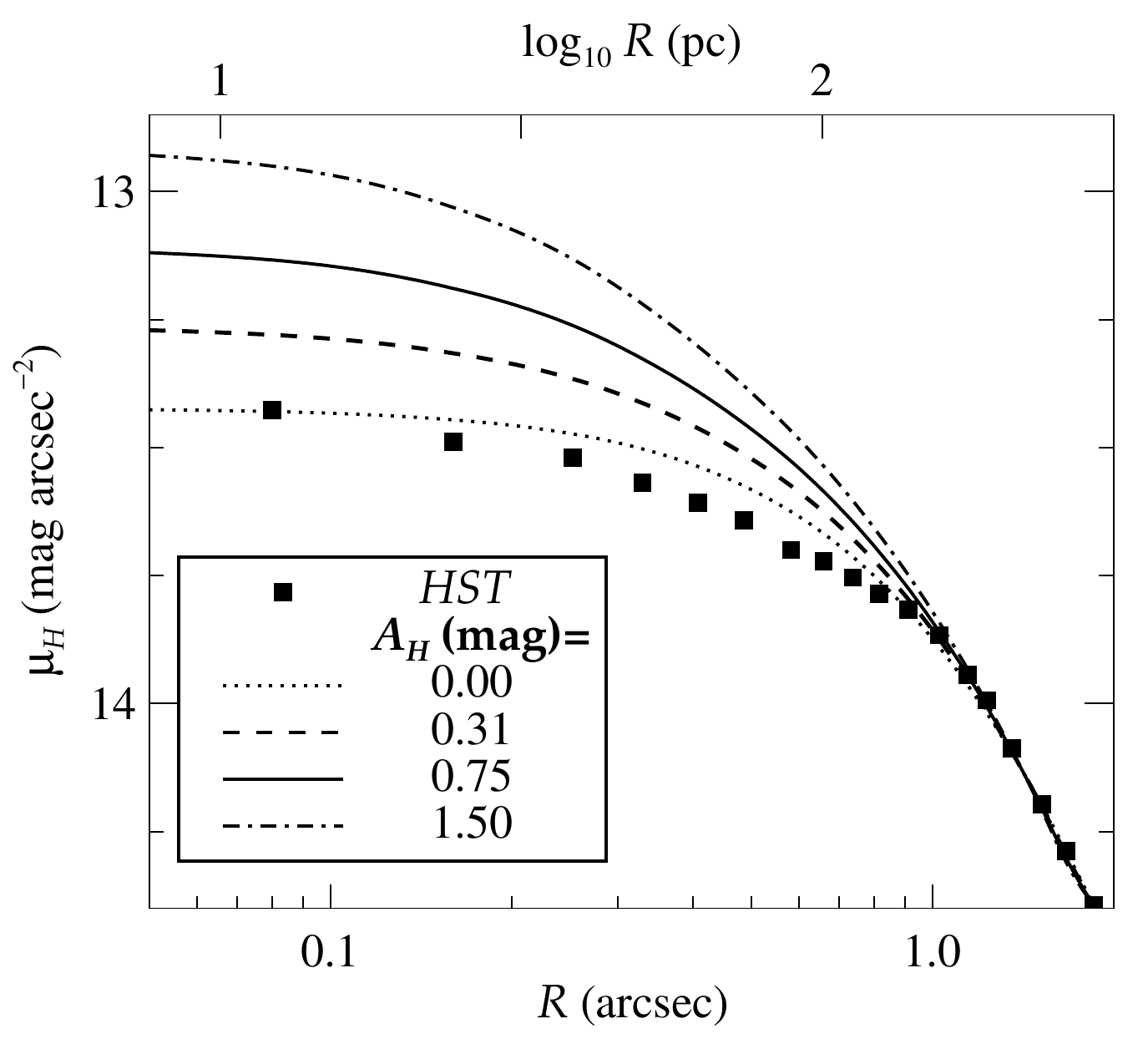}
\begin{singlespace}
  \caption{Nuclear $H$-band surface brightness profile of NGC 3258, showing an abrupt flattening of the stellar slope that coincides with the increasing $J-H$ color inward of $R \sim 1\arcsec$. After masking out the circumnuclear regions where dust obscuration appears highest, we model the $H$-band mosaic with an MGE (labeled as the $A_H=0$ case; see Table~\ref{tbl:mge} and Figure~\ref{fig:mge}). For comparison, we include model surface brightness profiles constructed to approximately correct the central $H$-band measurements for dust obscuration. We selected intrinsic $A_H=0.31$, 0.75, and 1.50 mag extinction, corresponding to loss of \textonequarter, \textonehalf, and \textthreequarters\ of the innermost stellar light behind this dusty disk.}
\end{singlespace}
\label{fig:mgex4}
\end{center}
\end{figure}

These model results imply that the central region of the dust disk is sufficiently opaque to absorb a significant fraction of the $H$-band galaxy light originating from behind the disk, and we conclude that extinction is responsible for some of the central flattening in the $H$-band radial profile. However, there is no straightforward method to correct the observed $H$-band radial profile for extinction based on the color excess maps. Recovering the intrinsic stellar surface brightness via spectral energy distribution measurements at each spatial location would require realistic radiative transfer modeling \citep[e.g.,][]{cam15} that accounts for the disk geometry and thickness, dust scattering, and extinction within the disk. Possible additional contributions of light from recent star formation in the disk or a weak active nucleus would further complicate any extinction correction method based on the observed color excess maps. Such modeling is beyond the scope of this work.

Instead, we adopted a simpler approach to examine the impact of extinction on the inferred \vcst\ profile by adjusting the central $H$-band surface brightness profile to correct for three fiducial values of disk extinction that bracket the likely range. The inner $H$-band stellar surface brightness follows a double power-law profile, so we fit the central $R \lesssim 10\arcsec$ of the mosaic with a PSF-convolved 2D Nuker function \citep[using \texttt{GALFIT};][]{pen02}, which yields an inner cusp slope $\gamma=0.01$ and a break radius $r_\mathrm{b}=1.5\arcsec$ (corresponding to $\sim$230 pc) that extends slightly beyond the dust disk radius. This $r_\mathrm{b}$ is consistent with those measured for other massive ETGs \citep[e.g.,][]{fab97,lau05}. After fixing all other Nuker parameters, we adjusted $\gamma$ to 0.09, 0.17, and 0.26 to approximately correct for absorption of \textonequarter, \textonehalf, and \textthreequarters\ of the integrated stellar light originating \textit{behind} the disk (for $R \lesssim 0\farcs25$), respectively, corresponding to disk intrinsic optical depths of $A_H=0.31$, 0.75, and 1.50 mag (or $A_V=1.67$, 4.04, and 8.09 mag). The maximum $\gamma$ we use is within the range generally associated with core galaxies \citep[$\gamma\leq 0.3$; e.g.,][]{fab97}. For each $A_H$ case, we created a new model image by seamlessly replacing the dust-obscured region (out to $R=1\farcs5$) with the associated \texttt{GALFIT} product. These dust-corrected $H$-band surface brightness profiles are shown in Figure \ref{fig:mgex4}. We parameterized each new model image using the MGE method and used the results to derive three additional, ``dust-corrected'' circular velocity profiles. In Section~\ref{sec:dynamical}, we employ all four \vcst\ profiles (the original and the three dust-corrected profiles) in gas-dynamical models to quantify the impact of dust obscuration on the final \mbh\ measurement.

\section{ALMA Data}
\label{sec:almadata}

\begin{figure}
\begin{center}
\includegraphics[trim=0mm 0mm 0mm 0mm, clip, width=0.85\columnwidth]{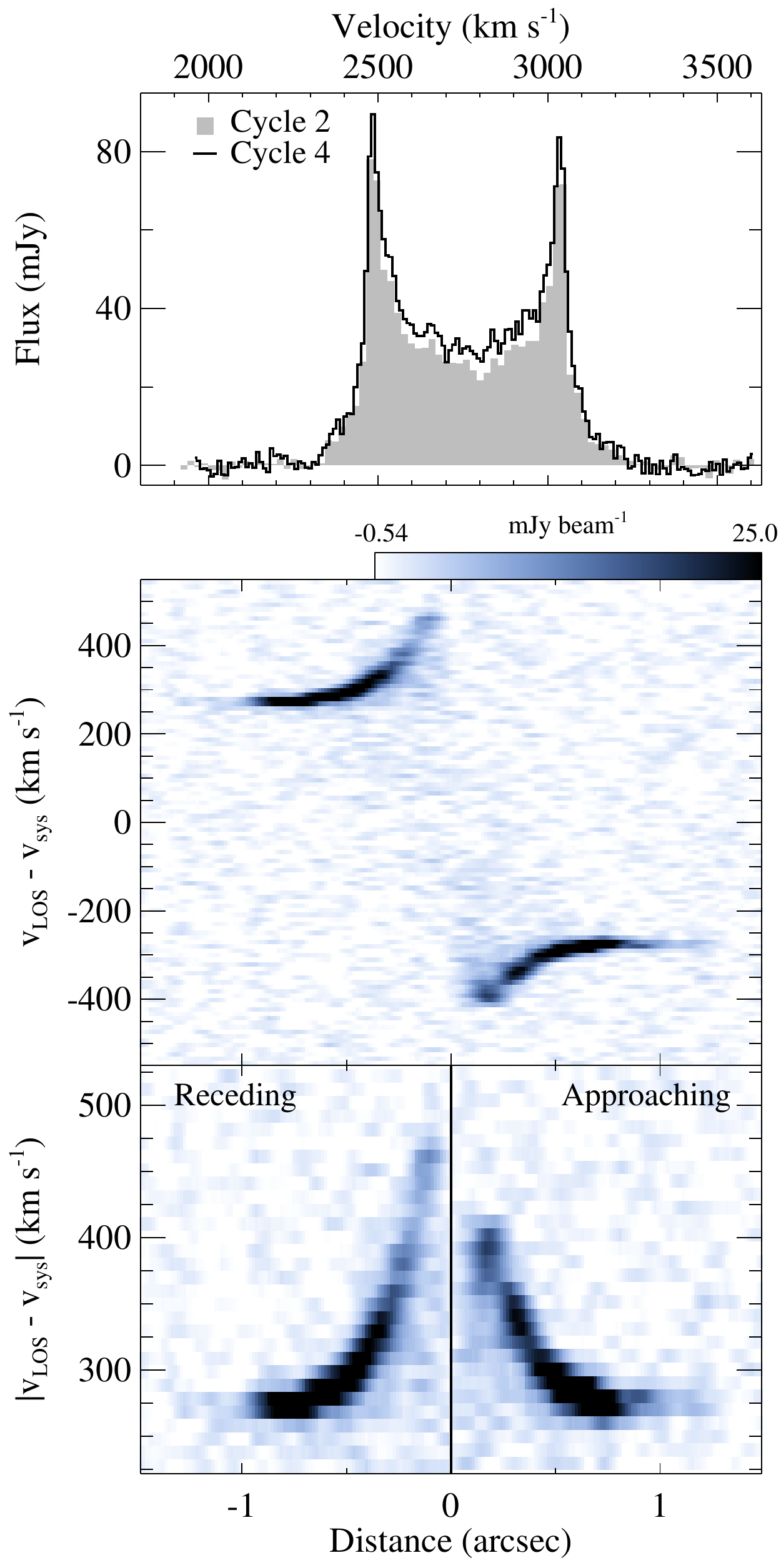}
\begin{singlespace}
  \caption{Velocity profile (\textit{top}) and PVD (\textit{middle}) from the Cycle 4 CO(2$-$1) observation. Flux densities were integrated in each channel within an elliptical area with semimajor and semiminor axes of 1\farcs25 and 0\farcs84, oriented at PA=77\degr. The Cycle 2 profile is included for comparison. The PVD was extracted along the disk major axis (PA=77\degr) with an extraction width equal to the geometric average of their beam FWHM; line-of-sight velocities are relative to the galaxy systemic velocity $\vsys=2761$ \kms. The data reveal smooth and well-ordered quasi-Keplerian disk rotation on the whole, with a deficit of central emission and a velocity asymmetry in the inner 0\farcs2 on the approaching side of the disk (\textit{bottom}).}
\end{singlespace}
\label{fig:pvd}
\end{center}
\end{figure}

\subsection{Observations and data processing}
\label{sec:almaobs}

The new Cycle 4 data were obtained in ALMA Program 2016.1.00854.S during 7-8 August 2017 in the C40-7 configuration, which had baselines ranging from 21 to 3696 m. Observations consisted of a single pointing with three $\sim$2 GHz$-$bandwidth spectral windows, one of which was centered on the redshifted $^{12}$CO(2$-$1) 230.538 GHz line while the remaining two measured the continuum at average (sky) frequencies of 228.4 and 243.0 GHz. Three execution blocks were carried out in good weather conditions (precipitable water vapor of 0.3$-$1.0 mm) for a total on-source integration time of 135 minutes. Line and continuum spectral windows were sampled using channel widths of 3.91 MHz (after $8\times$ online channel averaging) and 15.6 MHz, respectively. The data were flux calibrated using ALMA quasar standards J1037$-$2934 and J1107$-$4449, which have absolute flux calibration accuracies of $\sim$10\% \citep{fom14}. We have propagated this uncertainty into all subsequent flux and flux density measurements.

Prior to line and continuum imaging, we flagged and calibrated the Cycle 4 visibilities using version 4.7.2 of the \texttt{Common Astronomy Software Applications (CASA)} pipeline. \texttt{CASA} \texttt{TCLEAN} deconvolution with Briggs \citep[$r=0.5$;][]{bri95} weighting results in a synthesized beam with FWHM $\theta_\mathrm{FWHM}=0\farcs11 \times 0\farcs08$ at PA=88\degr. We first imaged the line-free channels (with a 5.2 GHz total bandwidth) to produce a continuum map with a point-source sensitivity of $\sim$11 $\mu$Jy beam\per. After $uv$-plane continuum subtraction, we imaged the primary spectral window into a line cube with 7.81 MHz channels (corresponding to rest-frame velocity widths of $\sim$10.2 \kms) that have typical rms sensitivities of $\sim$0.27 mJy beam\per.

\begin{figure}
\begin{center}
\includegraphics[trim=0mm 0mm 0mm 0mm, clip, width=0.85\columnwidth]{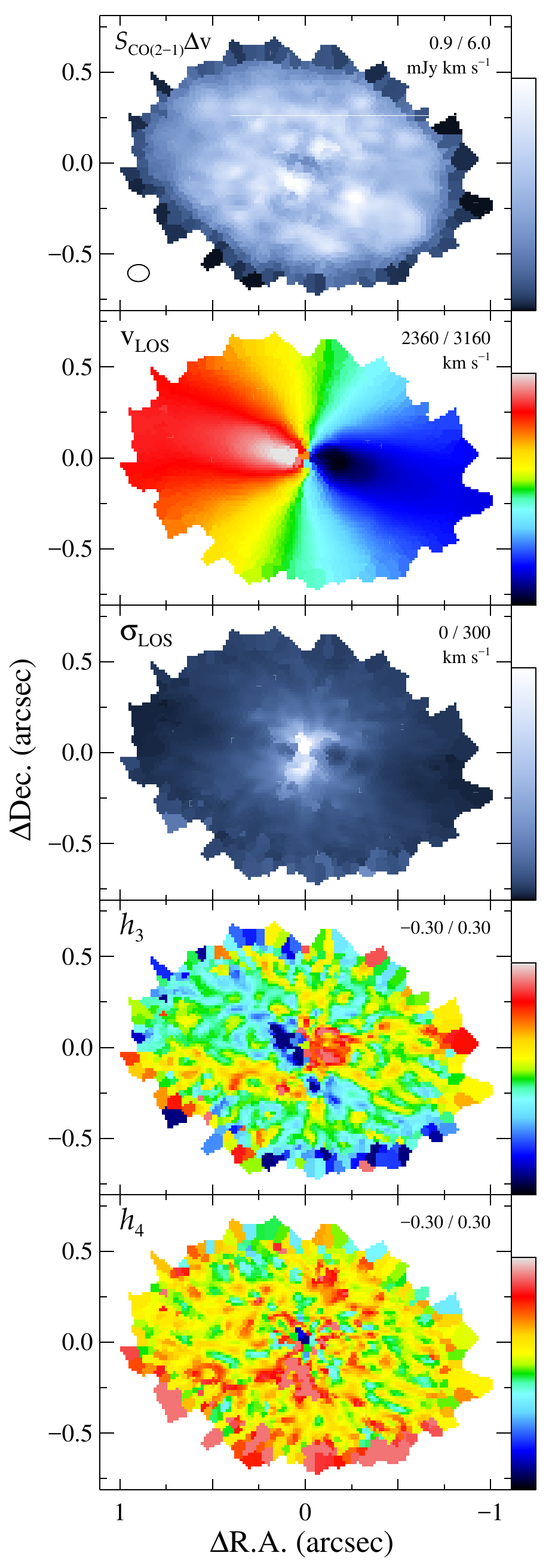}
\begin{singlespace}
  \caption{Maps of CO(2$-$1) flux and kinematic quantities (\vlos, \sigmalos, $h_3$, and $h_4$) measured from the ALMA Cycle 4 data cube. Ranges in each frame indicate the minimum and maximum data values mapped to the color tables shown at the right side of the figure. The ellipse in the top panel shows the FWHM size of the ALMA synthesized beam.}
\end{singlespace}
\label{fig:c4gh}
\end{center}
\end{figure}

These ALMA Cycle 4 data are a significant improvement in both angular resolution and sensitivity over the Cycle 2 observations of this target from Program 2013.1.00229.S, which are described in \citetalias{boi17}. However, the sparse central $uv$-plane coverage of the C40-7 configuration results in a $\sim$1\farcs2 maximum recoverable scale that may resolve out some smoothly-distributed emission in the 2\farcs4$-$wide disk. We therefore simultaneously imaged together the Cycle 2 and 4 visibilities using a multiscale deconvolution. After natural weighting of the visibilities, we obtained a continuum map with $\theta_\mathrm{FWHM}=0\farcs14\times0\farcs11$ at PA=$-$82\degr\ and an rms level of 9.8 $\mu$Jy beam\per. Briggs ($r=0.5$) weighting produced a line cube with $\theta_\mathrm{FWHM}=0\farcs12\times0\farcs09$ at PA=89\degr\ with $\sim$0.23 mJy beam\per\ sensitivities in $\sim$10.2 \kms\ channels at the 0\farcs015 pixel\per\ scale. Although incorporating these shorter-baseline data does slightly expand the synthesized beam major and minor axes to a geometric mean of $\sim$0\farcs10, we recovered more CO line and extended continuum emission than from imaging of the Cycle 4 data set alone. For the remainder of this paper, we refer to the results of our Cycle 2+4 multiscale deconvolution of continuum and spectral line data simply as Cycle 4 imaging.

\begin{figure}
\begin{center}
\includegraphics[trim=0mm 0mm 0mm 0mm, clip, width=\columnwidth]{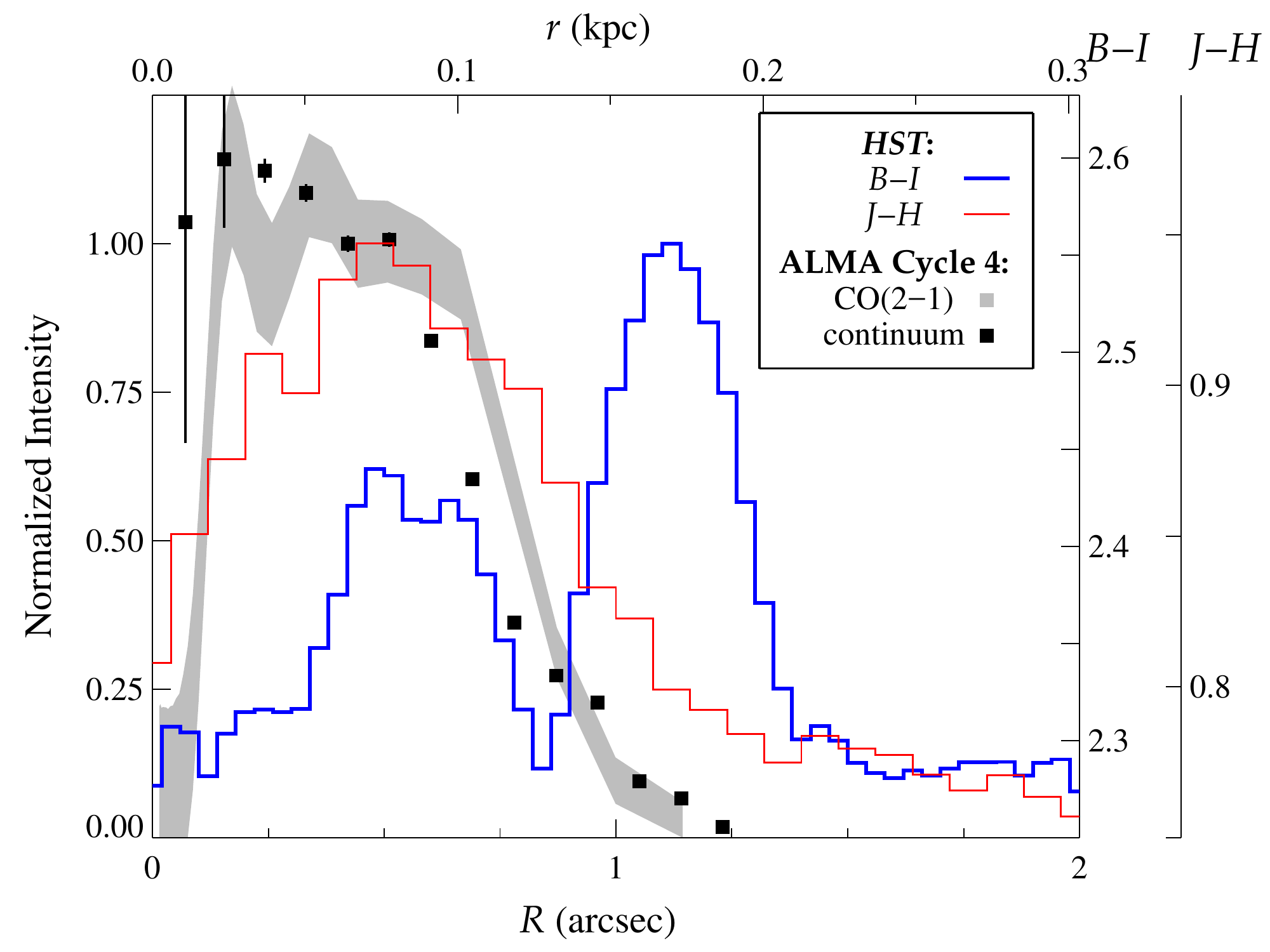}
\begin{singlespace}
  \caption{Radial profiles showing both the ALMA Cycle 4 CO(2$-$1) and continuum emission (averaged on elliptical annuli) and the optical and NIR \hst\ colors (extracted along the major axis). The left ordinate labels refer to the ALMA measurements (normalized at $R\sim0\farcs5$) while the right labels indicate the observed colors. The $B-I$ color excess reaches its maximum value at a radius where the dust optical depth becomes small enough to permit substantial optical light to pass through the disk. In contrast, the $J-H$ profile more closely follows the CO and continuum emission profiles because the maximum $J-H$ color excess occurs at higher values of the disk optical depth, as illustrated in the model calculations shown in Figure \ref{fig:dust}.}
\end{singlespace}
\label{fig:hstalma_axis}
\end{center}
\end{figure}

\subsection{Emission Line Properties}
\label{sec:almaproperties}

In the Cycle 4 line cube, we detect CO(2$-$1) emission out to $R\sim1\farcs05$ and in channels spanning 900 \kms. The highest velocity line emission (relative to the disk systemic velocity $\vsys\approx 2761$ \kms) is directly adjacent to the galaxy nucleus. We integrated the cube flux densities in each channel over the elliptical disk area to determine its velocity profile (Figure~\ref{fig:pvd}). The double-horned profile shape is very similar to that seen in the Cycle 2 data, while the total line flux of $S_{\rm CO(2-1)}\Delta v=27.40\pm0.15\pm2.74$ Jy \kms\ (statistical and systematic uncertainties, respectively) is slightly higher than the Cycle 2 value of $23.89\pm2.39$ Jy \kms\ reported in \citetalias{boi17}.

\begin{figure*}
\begin{center}
\includegraphics[trim=0mm 0mm 0mm 0mm, clip, width=0.95\textwidth]{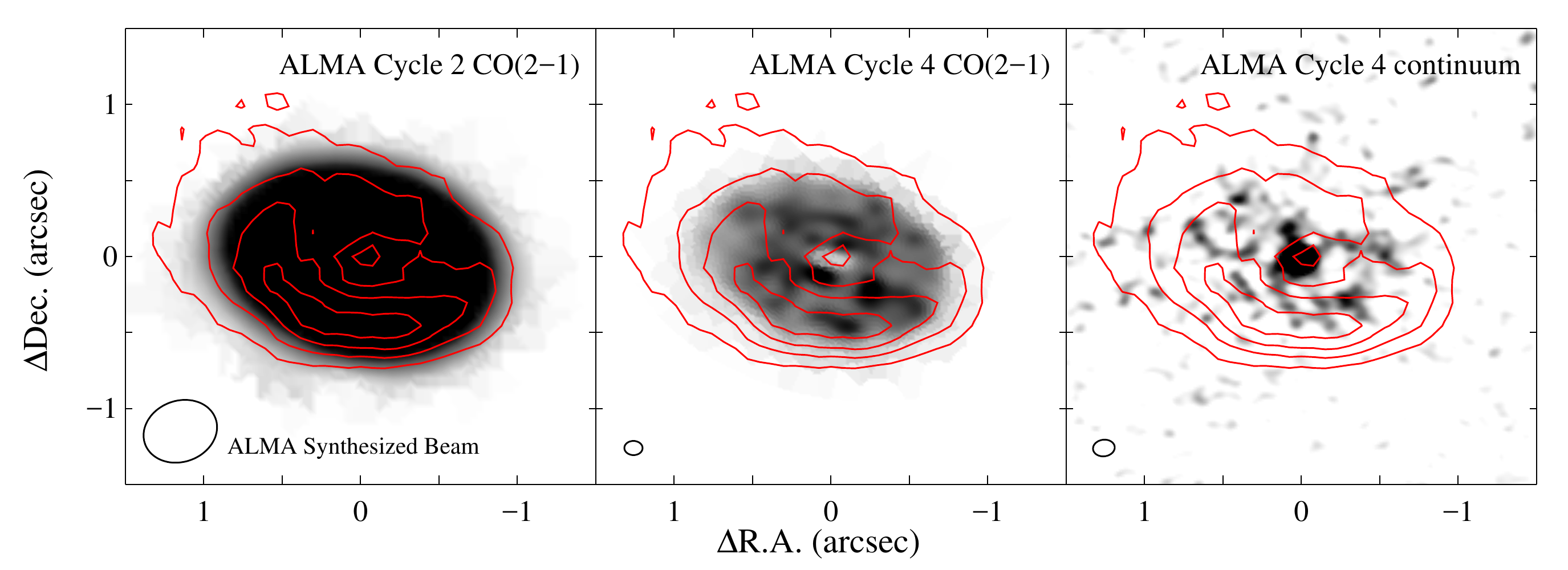}
\begin{singlespace}
  \caption{CO(2$-$1) line flux maps measured from ALMA Cycle 2 (\textit{left panel}) and Cycle 4 (\textit{middle panel}) imaging, revealing clumpy emission substructure in the latter case. Red contours show the \hst\ $J-H$ color map.  The right panel shows a map of the high-resolution $\sim$236 GHz continuum that presumably arises from thermal dust emission \citepalias[see][]{boi17}. Both CO and continuum emission are concentrated in a region that extends out to the ring of maximal NIR color.}
\end{singlespace}
\label{fig:hstalma_maps}
\end{center}
\end{figure*}

We extracted a position-velocity diagram (PVD) from the Cycle 4 cube along a PA = 77\degr\ with a spatial extraction width equal to the average synthesized beam FWHM of 0\farcs10 (see Figure~\ref{fig:pvd}). The CO line-of-sight velocities span the same range as in the Cycle 2 PVD, but in the Cycle 4 data the CO emission is resolved into a tight locus of quasi-Keplerian rotation arising from the point mass BH and extended galaxy mass distribution. These CO emission-line velocities rise to a peak on either side of the nucleus, tracing gas rotation to within $\sim$20 pc of the galaxy center. This remarkable PVD structure resolves the central rise in rotation velocity far better than any published ALMA observation of circumnuclear gas in any other galaxy. In contrast to the Cycle 2 data, spatial blurring of high-velocity and low-velocity emission due to beam smearing in the inner disk is almost completely eliminated. The line-of-sight velocity $|\vlos-\vsys|$ of this innermost CO emission reaches $\sim$480 \kms. Assuming a regularly rotating disk inclined by $i\approx48\degr$, the corresponding circular velocity of $v_{\rm c} \approx 650$ \kms\ at this radius would suggest $\mbh \approx 2\times10^9$ \msun. This value of \mbh\ implies $\rg \approx 0\farcs9$, which in turn indicates that nearly all of the dust disk lies within \rg.

As described in Section~\ref{sec:dynamical}, we fit gas-dynamical models directly to both the Cycle 2 and 4 CO line cubes. For visualization purposes, we parameterized the line-of-sight velocity distributions using Gauss-Hermite functions \citep[GH;][]{mar93}. For low S/N regions at the disk center and near the edge, adjacent spectra were combined together prior to line profile fitting \citep[using a Voronoi tessellation of a preliminary CO flux map;][]{cap03}. We display GH moment maps for the Cycle 2 data in \citetalias{boi17} and for Cycle 4 in Figure~\ref{fig:c4gh}, which includes the integrated CO(2$-$1) line flux, \vlos, and velocity dispersion \sigmalos. Due to beam smearing, both the Cycle 2 and 4 moment maps show high $|h_3|$ and $|h_4|$ values of up to $\sim 0.25-0.30$ for radii $\lesssim 0\farcs2$. For the lower-resolution data set, these non-Gaussian coefficients remain elevated in coherent patterns out to $R \sim 1\arcsec$.

Making the same assumptions about the CO-to-H$_2$ conversion factor $\alpha_{\rm CO}$ as in \citetalias{boi17}, the CO flux measured from the Cycle 4 data implies a total H$_2$+He mass $M_{\rm gas}=(1.0\pm0.3)\times10^8$ \msun\ for the gas disk (including uncertainties in galaxy distance and flux calibration). The Cycle 2 and 4 data both show a centrally concentrated CO flux distribution (Figure~\ref{fig:hstalma_axis}), and the Cycle 4 imaging with a beam size corresponding to 17 pc partially resolves the CO(2$-$1) emission into large, cloud-like knots (Figure \ref{fig:hstalma_maps}). Clumpy emission-line structure appears to be common for molecular gas disks in ETGs when observed at similar physical resolutions \citep{uto15,bar16a,dav17b,dav18}. We identify a central hole in CO surface brightness with a radius of $\sim 0\farcs13$ that corresponds to the innermost emission detected in the PVD.

The gas kinematics are nearly Keplerian close to the disk center, flattening out to $\vlos \sim 280$ \kms\ for $R>0\farcs6$ due to the increasing contributions of host galaxy mass at larger radii. Examination of the Cycle 4 PVD shows an asymmetry in the peak velocities on either side of the nucleus, which reach $+483$ and $-414$ \kms\ relative to \vsys\ on the receding and approaching sides of the disk, respectively. For $R<0\farcs2$, the approaching (western) emission appears to show sub-Keplerian rotation velocities.

\begin{figure}
\begin{center}
\includegraphics[trim=0mm 0mm 0mm 0mm, clip, width=\columnwidth]{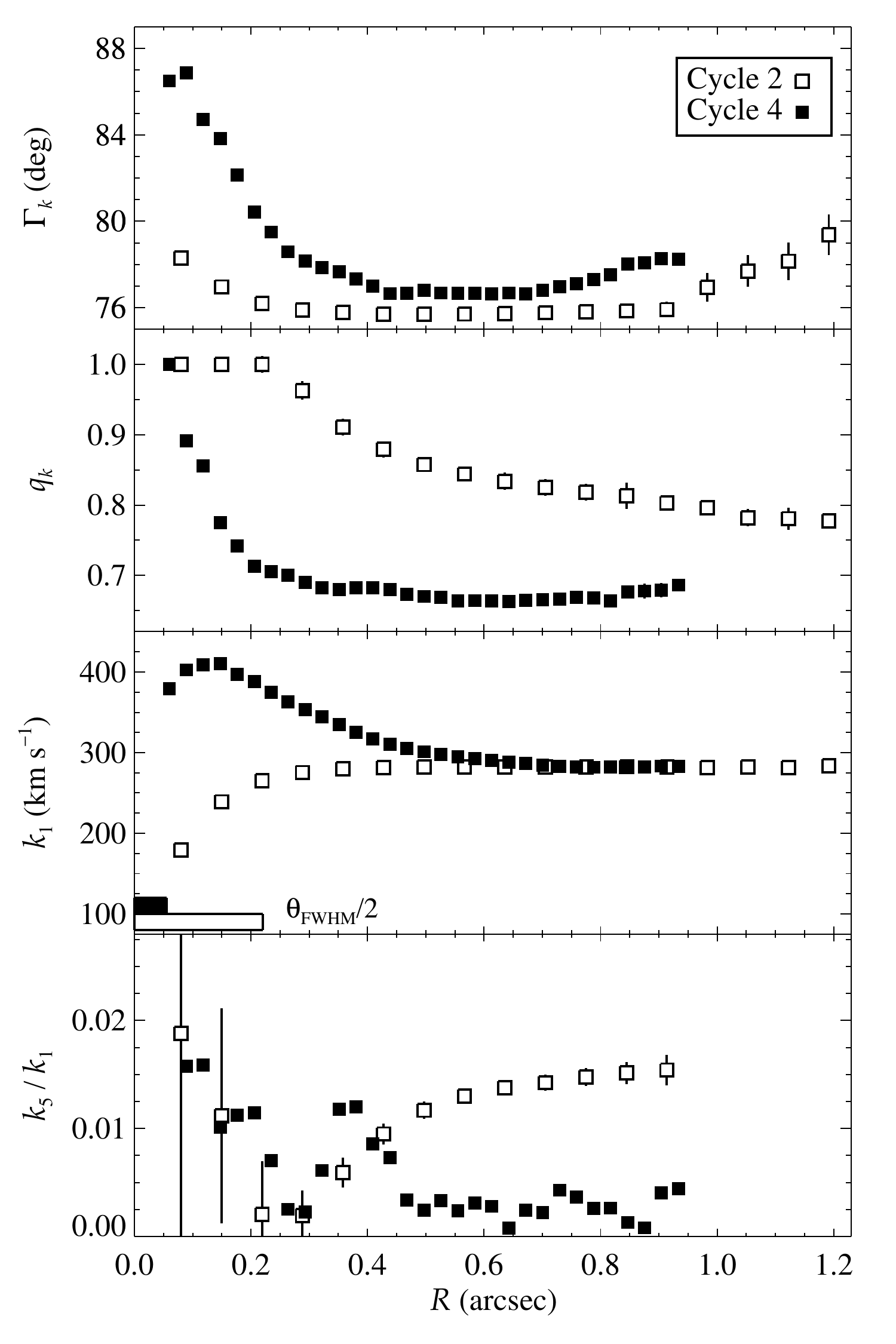}
\begin{singlespace}
  \caption{Kinemetry decomposition of the Cycle 4 \vlos\ field, showing the major axis PA $\Gamma_k$, kinematic ellipse flattening $q_k$, line-of-nodes velocity coefficient $k_1$, and ratio $k_5/k_1$. Cycle 2 results from \citetalias{boi17} are shown for comparison. The $k_1$ coefficient derived from the Cycle 2 observations agrees to within $\sim$10\% with the higher resolution results at an angular distance of the Cycle 2 synthesized beam FWHM from the disk center. The lower resolution $q_k$ values converge much more slowly and remain discrepant out to the edge of this molecular disk \citep[see][Figure A2]{kra08}.}
\end{singlespace}
\label{fig:kin}
\end{center}
\end{figure}

The observed velocity field also exhibits minor kinematic warping, most noticeably at radii $\lesssim 0\farcs25$. To characterize deviations from coplanar rotation, we decomposed the $v_{\rm LOS}$ map using kinemetry \citep{kra06} to measure the kinematic PA $\Gamma_k$ and axis ratio $q_k$, as well as circular ($k_1$) and non-circular ($k_5$) velocity components, as a function of radius. Results are shown in Figure~\ref{fig:kin}. While the primary kinematic twist $\Delta \Gamma_k \sim 10\degr$ occupies the inner half arcsecond, the disk remains slightly warped out to the edge of the detected CO(2$-$1) emission. Beam smearing reduces the velocity amplitude along the line of nodes for $R \lesssim 0\farcs2$, while at greater radii $k_1 \approx  v_{\rm c} \sin i$. The measured $q_k$ values show a central rise to unity that may in part be the result of finite angular resolution (i.e., circularization of the nuclear velocity field; see \citetalias{boi17}). For thin disk rotation $q_k \approx \cos i$, so $q_k \rightarrow 0.67$ with increasing radius suggests an outer disk inclination angle of $\sim$48\degr. Similar to the Cycle 2 kinemetry results, the coefficient ratio $k_5/k_1 \lesssim 0.02$ at all radii, suggesting only negligible deviations from circular rotation despite the evident warping of the disk.

The observed CO(2$-$1) line dispersion ranges from $\sim$7$-$415 \kms. The highest values found around the nucleus and on either side of the major axis can be attributed primarily to beam smearing and intrapixel velocity gradients \citep{bar16b}. However, the \sigmalos\ field reaches its maximum not at the disk center as expected but $\sim$0\farcs05 northward. The central $\sigmalos \sim300$ \kms\ may in part be lower due to the coincident hole in CO(2$-$1) flux, with some broad, low S/N line profile wings buried beneath the noise. Along the disk major axis, the measured line dispersion rapidly decreases to $\lesssim50$ \kms\ for $R\gtrsim0\farcs07$ and falls below 7 \kms\ near the disk edge.

\section{Dynamical Modeling}
\label{sec:dynamical}

In this section, we present results from dynamical modeling of the circumnuclear disk in NGC 3258. We begin with models for the simple case of a geometrically flat disk, and then consider a tilted-ring model designed to provide a better fit to the disk's warped geometry. We employ two different methods to constrain the mass distribution of the host galaxy: first, the standard approach of using the dust-corrected MGE models that is more widely applicable to data that do not highly resolve gas rotation within \rg, and second, a method using only the ALMA CO kinematics to determine the extended mass profile. We also consider models with different prescriptions for the spatial variation of the turbulent velocity dispersion of the molecular gas. Models are fit to the ALMA Cycle 4 data, and we also describe model fits to the lower-resolution Cycle 2 data to illustrate the effect of angular resolution on the \mbh\ determination. For clarity, the various models used in this paper are labeled and described in Table~\ref{tbl:mods}. Models A and B are fit to the Cycle 2 data, while models C$-$F are fit to the Cycle 4 data. We adopt model F1 as our final best result: this includes the tilted-ring disk structure, a spatially varying turbulent velocity dispersion, and the extended mass profile determined from the ALMA kinematic data. We also conduct additional tests based on variants of model F1 to estimate the systematic uncertainties in the BH mass. Unless otherwise specified, all modeling results described below refer to the Cycle 4 data; fits to the Cycle 2 data are described in \S\ref{sec:thinmodel_results_c2}.

\begin{deluxetable}{ccccc}
\tabletypesize{\scriptsize}
\tablecaption{Dynamical Model Properties\label{tbl:mods}}
\tablewidth{0pt}
\tablehead{
\colhead{Model} & \colhead{Cycle} & \colhead{Mass Model} & \colhead{Disk Structure} & \colhead{$\sigmaturb(r)$}
}
\startdata
A1 & 2 & MGE; $A_H=0$ & Flat & Uniform \\ 
B1--B4 & 2 & MGE; $A_H=0,0.31,0.75,1.50$ & Flat & Gaussian \\
C1 & 4 & MGE; $A_H=0$ & Flat & Uniform \\
D1--D4 & 4 & MGE; $A_H=0,0.31,0.75,1.50$ & Flat & Gaussian \\
E1 & 4 & MGE; $A_H=0.75$ & Tilted ring & Gaussian \\
F1 & 4 & $v_{\rm ext}$ & Tilted ring & Gaussian
\enddata
\begin{singlespace}
  \tablecomments{Properties of the dynamical models. Models A--B were fit to the ALMA Cycle 2 data cube, while Models C--F were fit to the Cycle 4 data. Contributions from the galaxy's extended mass distribution to the circular velocity were included either by using the extinction-corrected MGE deprojection of the host galaxy luminosity profile measured from the \hst\ $H$-band image (after incorporating a spatially uniform mass-to-light ratio \upsh), or by allowing the circular velocity due to spatially extended mass [$\vext(r)$] to vary independently within 10 radial bins as described in \S\ref{sec:vext_method}. The $H$-band extinction is listed (in magnitudes) for each MGE-based model; this refers to the extinction due to the inclined dust disk embedded in the galaxy midplane, which attenuates light originating from the far side of the disk.}
\end{singlespace}
\end{deluxetable}

\subsection{Initial flat-disk models}
\label{sec:thinmodel}

\subsubsection{Method}
\label{sec:thinmodel_method}

We first describe the basic flat-disk modeling procedure, which builds on methods developed for the analysis of \hst\ ionized-gas kinematics and uses forward modeling of line profiles from a rotating disk \citep[e.g.,][]{mac97,van98,bar01}. A major difference is that we fit models directly to the observed ALMA data cube, making use of all available information, rather than fitting models to velocity and velocity dispersion curves extracted from the data, as was done for \hst\ gas-dynamical measurements.  Our flat-disk modeling method was previously used to measure the black hole mass in NGC 1332 \citep{bar16a, bar16b}, and is similar to procedures used by other groups to measure black hole masses from molecular gas kinematics \citep[e.g,][]{dav17b, oni17}. \citet{bar16b} present a detailed description of the method, which we summarize here.

The model calculation starts by determining the circular velocity as a function of radius for a thin, flat disk orbiting in the combined gravitational potential of a central black hole and the spatially extended mass distribution of the host galaxy. Line-of-sight projections of the disk rotation velocity are determined at each point on the sky for a given disk inclination and major axis position angle and for an assumed distance to the galaxy. Then, a spectral cube is generated by assuming an intrinsically Gaussian line profile at each point in the disk, with some specified turbulent velocity dispersion. The model cube is constructed to match the observed frequency spacing of the ALMA data cube for direct comparison.  At each spatial grid point, the total flux in the modeled line profile is determined using a map of the CO surface brightness distribution determined from the ALMA observation. Each velocity channel of the model is convolved with the ALMA synthesized beam (an elliptical Gaussian).  In order to capture details of sub-pixel gradients in rotation velocity near the disk center, the model calculation and beam convolution are carried out on an oversampled spatial grid (relative to the ALMA data cube) and the modeled cubes are then downsampled to match the ALMA pixel scale. Models are optimized by \chisq\ minimization using a downhill simplex minimization method \citep{pre92} by fitting the calculated cubes directly to the ALMA data cube. Further details of these steps are described below.

These basic models employ at least nine free parameters: the black hole mass \mbh, the stellar $H$-band $M/L$ ratio \upsh, the pixel location of the disk's dynamical center ($x_\mathrm{c}, y_\mathrm{c}$), the disk inclination angle $i$ and major-axis position angle $\Gamma$ of the receding side of the disk, the systemic velocity \vsys, the turbulent velocity dispersion \sigmaturb, and a flux-scaling factor $f_0$ to correct for possible flux normalization mismatch between the data and model. The gas velocity dispersion \sigmaturb\ can be set to a uniform (but freely varying) value over the disk surface, or allowed to vary as a function of radius with the introduction of additional free parameters. The models are calculated on a pixel grid that is oversampled by a factor of $s$ relative to the ALMA data cube pixel size of 0\farcs015 pixel\per. In other words, each ALMA spatial pixel is subdivided into an $s\times s$ grid of sub-pixel elements. For NGC 3258, we calculated initial models for values of $s$ ranging from 1 to 14.

\begin{figure*}
\begin{center}
\includegraphics[trim=0mm 0mm 0mm 0mm, clip, width=0.9\textwidth]{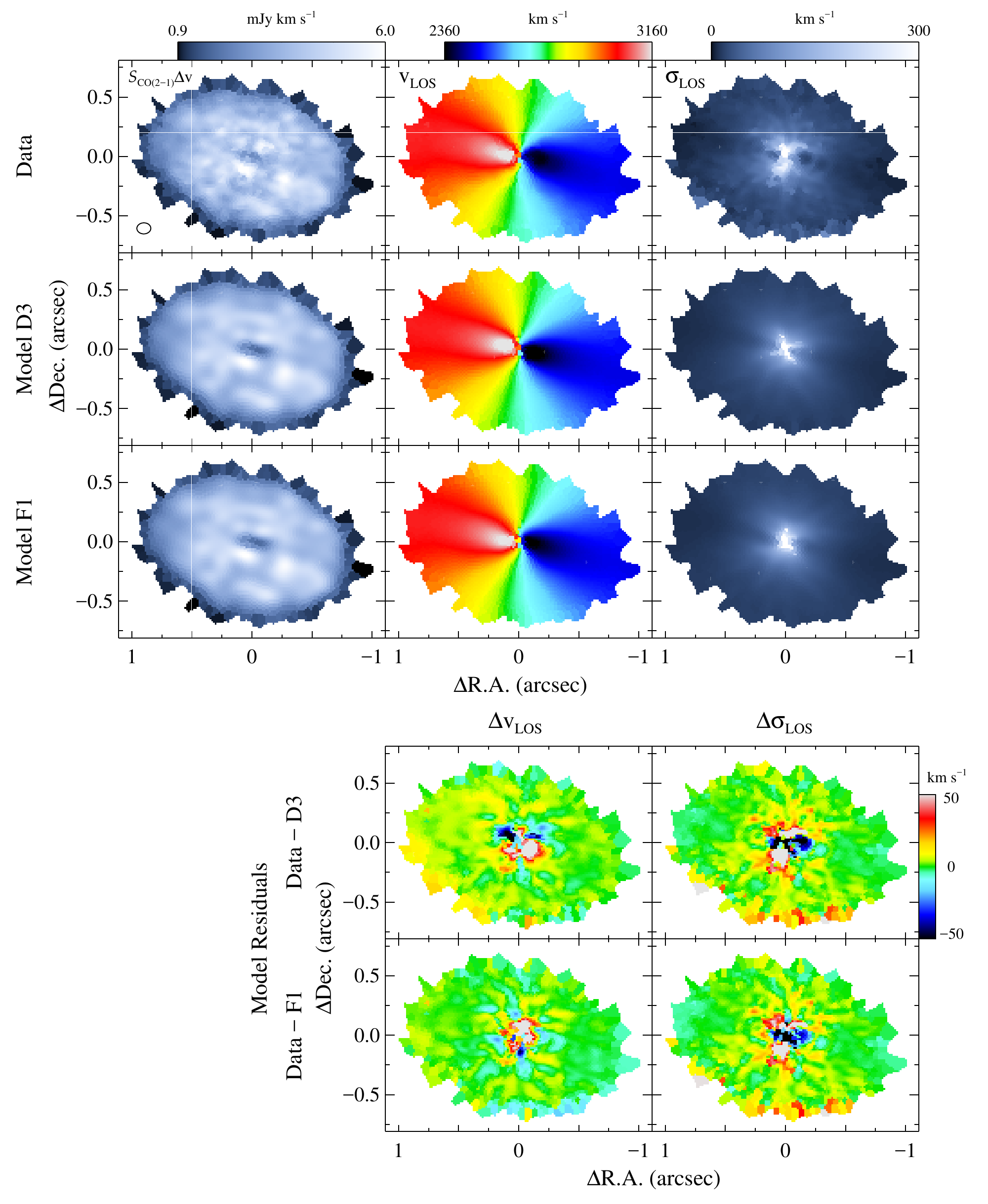}
\begin{singlespace}
  \caption{Maps of CO(2$-$1) flux and kinematic quantities (\vlos\ and \sigmalos) measured from the ALMA data cube (\textit{top row}) and from flat-disk model D3 (\textit{second row}) and tilted-ring model F1 (\textit{third row}). Ranges in each frame indicate the minimum and maximum data values mapped to the color tables shown above the figures. The model CO flux map used for models D3 and F1 was formed by collapsing the data cube regions that show emission above the $2\sigma$ sensitivity level. Residual maps (data$-$model; \emph{lower panels}) of line-of-sight velocity and line dispersion show generally small deviations between models and data over most of the disk surface. At the disk center these deviations become much larger, with, e.g., the models D3 and F1 $\Delta \vlos$ ranging from $-80$ to +180 \kms\ and $-25$ to +50 \kms, respectively. We retain the $\pm50$ \kms\ color scale ranges to highlight the better overall fit of model F1. Note that our models are fitted to the full three-dimensional data cube, while these kinematic maps represent quantities extracted from the data and model.}
\end{singlespace}
\label{fig:c4ghm}
\end{center}
\end{figure*}

A required input to this calculation is a model map of the disk's CO surface brightness distribution prior to convolution by the ALMA synthesized beam. To generate this map, we collapsed the ALMA Cycle 4 data cube to form an image (see Figure~\ref{fig:c4ghm}), and applied ten iterations of the IRAF STSDAS Richardson-Lucy deconvolution \citep{ric72,luc74} task \texttt{LUCY} using the elliptical Gaussian synthesized beam.

The disk's circular velocity $v_{\rm c}(r)$ is calculated as a function of radius for rotation in the combined gravitational potential of the BH (a point source at $r=0$) and the host galaxy. The host galaxy contribution $\vcstar(r)$ to the circular velocity is determined using the host galaxy luminosity profiles derived from the dust-corrected MGE models, with these velocity values scaled by $\sqrt{\upsh}$. We assume a spatially uniform $M/L$ ratio in our model calculations. The optically thick dust disk within the inner kpc of NGC 3258 makes it difficult to constrain any possible $M/L$ gradient, but the three dust-corrected stellar luminosity profiles described in \S\ref{sec:stellarprofile} correspond to a range of different central mass profile slopes that collectively encompass the possible effect of stellar $M/L$ variations. We do not include the gas disk itself in the mass model. In \S\ref{sec:almaproperties}, we estimate the disk's H$_2$+He gas mass to be $\sim$10$^8$ \msun, and this gas mass is distributed in a disk extending out to $r\sim150$ pc, within which the total enclosed mass is $\sim5\times10^9$ \msun. In effect, the gas disk's small contribution to the $M(r)$ profile will be subsumed into the $M/L$ parameter, although there will be a small residual error since the disk's radial mass profile differs from the stellar $M(r)$ profile. Because our final dynamical model (described in \S\ref{sec:detailedmodel_method} below) determines the spatially extended mass profile directly from the kinematic data, independent of the host galaxy surface brightness profile measurements, that method incorporates all gravitating mass contributions that may be present.

For the turbulent velocity dispersion within the molecular gas disk, we adopt either a spatially uniform value across the disk surface ($\sigmaturb = \sigma_1$), or an axisymmetric model allowing for radial variation in \sigmaturb\ with a Gaussian radial profile: $\sigmaturb = \sigma_0 \exp[-(r-r_0)^2/2\mu^2)] + \sigma_1$, where $\sigma_0$, $\sigma_1$, $\mu$, and $r_0$ are free parameters.  We use \sigmaturb\ to represent the combination of processes contributing to the emergent line width of the disk: internal turbulence and rotation of individual clouds, as well as radial velocity variations between clouds contained within a given grid element, whether due to rotational shear in the disk or random cloud-to-cloud velocity variations. The molecular gas kinetic temperature in ETG circumnuclear disks is very low, $\sim$10$-$20 K \citep{bay13}, so gas temperature makes a negligible contribution to the CO line widths. In the ALMA data cube, the minimum observed line dispersion is just $\sim$7 \kms, while the central rise to $\sim$300 \kms\ is likely almost entirely the result of beam smearing at small radii (see \citealp{bar16b} for a detailed discussion of this effect). The Gaussian \sigmaturb\ model allows for the possibility that some portion of this central increase in line width is intrinsic.

\begin{figure*}
\begin{center}
\includegraphics[trim=0mm 0mm 0mm 0mm, clip, width=\textwidth]{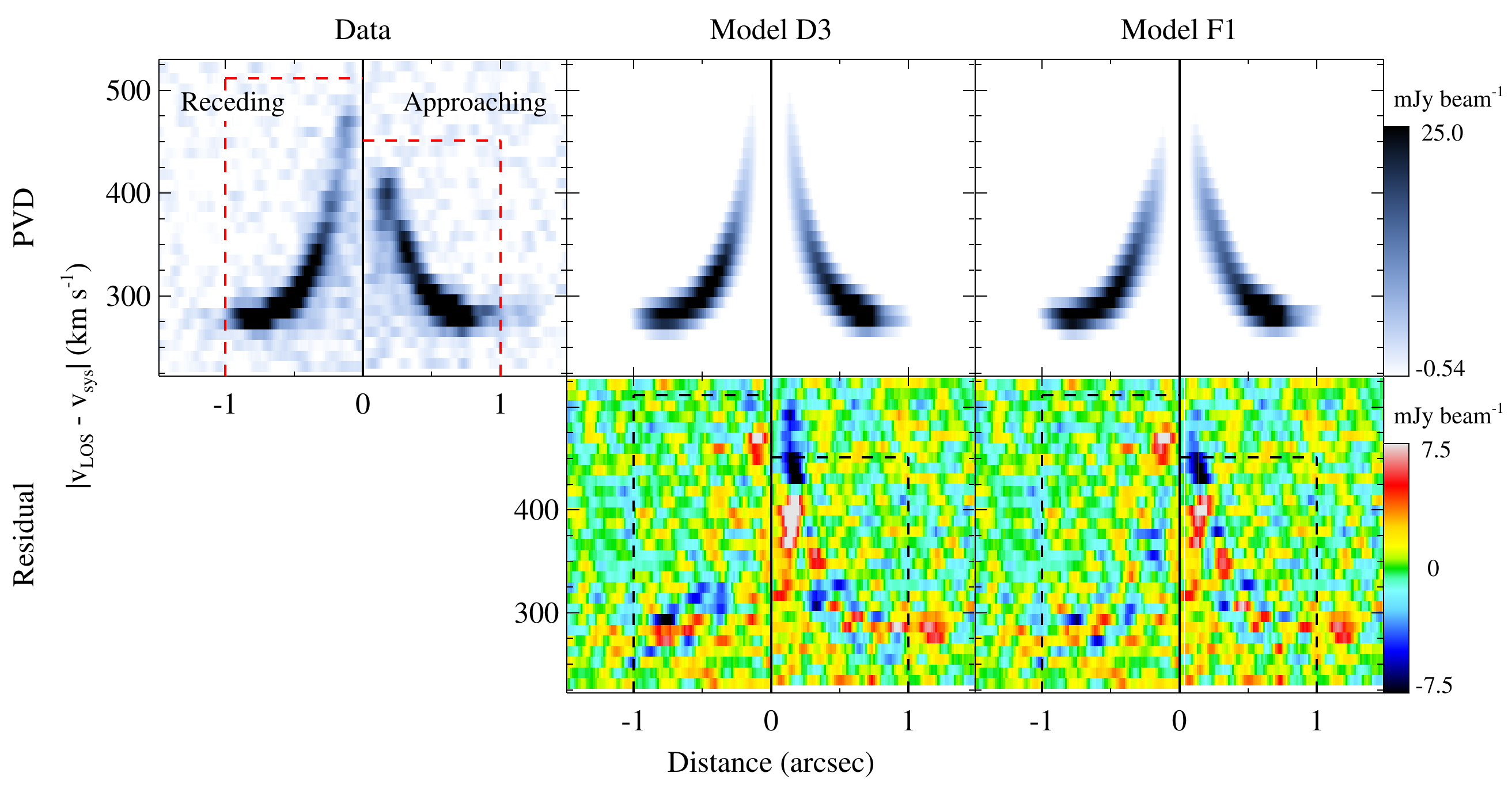}
\begin{singlespace}
  \caption{Model D3 and F1 PVDs (\textit{upper central and right panels}) extracted in the same manner as the data (\textit{upper left panel}), with data-model PVD residuals (\textit{lower panels}) that highlight discrepancies in the full-cube model fits. We demarcate (\textit{dashed lines}) the fitting region projected onto the PVD plane. The detailed disk model F1 shows better agreement with the data at all radii, although large residuals near the kinematic center remain due to the asymmetric CO velocities on the approaching side of the disk.}
\end{singlespace}
\label{fig:c4pvdm}
\end{center}
\end{figure*}

We populate the model cube at each spatial location with Gaussian emergent line profiles, defined by the projected line-of-sight velocity and \sigmaturb\ value at each oversampled grid point. The model cube spectral axis is observed frequency, and we transform rest-frame projected velocities and \sigmaturb\ maps to observed frequencies prior to creating the line profiles \citep{mey17}. The line profile flux at each spatial location is determined using the model flux map. As the CO surface brightness distribution is not known on subpixel scales, each line profile within an $s\times s$ block corresponding to a single ALMA pixel is equally weighted in flux, such that the total $s\times s$ region contains the same total flux as the deconvolved CO flux model at that pixel location.

The two final steps of the model calculation are the convolution of each model cube channel with the synthesized beam, and averaging of each $s\times s$ block of oversampled pixels into a single pixel matching the scale of the ALMA data cube. In principle, for highest fidelity the beam convolution would be computed on the oversampled pixel grid. Beam convolution is the most time-consuming portion of the model calculation procedure, and for large values of $s$ this would become prohibitively slow. In fact, we found that the modeling results do not appreciably change if the model cube is first rebinned to the original pixel scale of the ALMA data prior to the beam convolution step since the synthesized beam is already oversampled by the chosen pixel size. We adopted this method in order to minimize the computational time required for model optimization.

\begin{deluxetable*}{lccccccccccccr}
\tabletypesize{\scriptsize}
\tablecaption{Modeling Results\label{tbl:mod_res}}
\tablewidth{0pt}
\tablehead{
Model & \colhead{\mbh} & \colhead{\upsh} & \colhead{$i$} & \colhead{$\Gamma$} & \colhead{$\sigma_1$} & \colhead{$\sigma_0$} & \colhead{$r_0$} & \colhead{$\mu$} & \colhead{$x_\mathrm{c}$} & \colhead{$y_\mathrm{c}$} & \colhead{\vsys} & \colhead{$f_0$} & \colhead{$\chisqnu$} \\
 & ($10^9$ \msun) & (\msun/\lsun) & (\degr) & (\degr) & (\kms) & (\kms) & (pc) & (pc) & (\arcsec) & (\arcsec) & (\kms) &  &  
}
\startdata
A1 & 2.386 & 3.16 & 46.0 & 76.8 & 7.44 & \nodata & \nodata & \nodata & $-$0.022 & $-$0.020 & 2760.76 & 1.02 & 1.783 \\
B1 & 2.386 & 3.16 & 46.0 & 76.8 & 7.42 & 4.75 & 0.14 & $-$0.46 & $-$0.022 & $-$0.020 & 2760.76 & 1.02 & 1.784 \\
B2 & 2.309 & 2.88 & 45.9 & 76.8 & 7.74 & 16.9 & 0.62 & \phantom{$-$}0.53 & $-$0.022 & $-$0.020 & 2760.76 & 1.02 & 1.806 \\
B3 & 2.216 & 2.65 & 46.0 & 76.8 & 7.71 & 4.11 & 1.04 & \phantom{$-$}4.53 & $-$0.021 & $-$0.021 & 2760.79 & 1.02 & 1.858 \\
B4 & 2.087 & 2.42 & 46.1 & 76.8 & 7.86 & 0.14 & 1.54 & \phantom{$-$}9.78 & $-$0.021 & $-$0.021 & 2760.79 & 1.01 & 1.935 \\ \hline
C1 & 2.280 & 2.72 & 49.0 & 77.0 & 10.5 & \nodata & \nodata & \nodata & $-$0.001 & $-$0.003 & 2760.84 & 1.06 & 1.229 \\
D1 & 2.276 & 2.73 & 49.0 & 77.0 & 6.83 & 8.58 & \phantom{$-$}20.3 & 65.9 & $-$0.001 & $-$0.003 & 2760.83 & 1.07 & 1.219 \\
D2 & 2.215 & 2.46 & 49.0 & 77.0 & 7.14 & 8.44 & \phantom{$-$}25.0 & 59.5 & $-$0.001 & $-$0.003 & 2760.87 & 1.07 & 1.217 \\
D3 & 2.144 & 2.26 & 49.0 & 77.0 & 7.74 & 8.02 & \phantom{$-$}36.44 & 46.9 & $-$0.001 & $-$0.003 & 2760.91 & 1.07 & 1.217 \\
D4 & 2.059 & 2.04 & 49.0 & 77.0 & 8.72 & 7.07 & \phantom{$-$}52.4 & 29.9 & $-$0.001 & $-$0.003 & 2760.97 & 1.07 & 1.219 \\
E1 & 2.203 & 2.18 & 24.2$-$49.8 & 76.2$-$96.4 & 6.54 & 22.5 & $-$53.2 & 83.6 & $-$0.002 & $-$0.003 & 2760.83 & 1.07 & 1.180 \\
F1 & 2.249 & \nodata & 27.5$-$49.3 & 76.2$-$93.6 & 6.32 & 21.9 & $-$51.3 & 84.7 & $-$0.002 & $-$0.003 & 2760.82 & 1.07 & 1.179 \\
  & (0.004) & \nodata & \nodata & \nodata & (0.16) & (0.40) & (0.47) & (0.39) & (0.001) & (0.001) & (0.07) & (0.002) & \nodata \\
\enddata
\begin{singlespace}
  \tablecomments{Best-fit parameter values obtained from model fits to the Cycle 2 (Models A$-$B, \textit{top}) and Cycle 4 (Models C$-$F, \textit{bottom}) data cubes. See Table~\ref{tbl:mods} for a description of each model. Model F1 is the final best-fitting model. The major axis position angle $\Gamma$ is measured east of north for the receding side of the disk. The position of the disk kinematic center ($x_\mathrm{c}$, $y_\mathrm{c}$) is given in terms of right ascension and declination offsets from the nuclear continuum source centroid at 10$^{\rm h}$28$^{\rm m}$53\fs550, $-$35\degr36\arcmin19\farcs78 (J2000). In these models, the disk systemic velocity  \vsys\ is taken to be the recessional velocity $cz_\mathrm{obs}$ in the barycentric frame that is used to transform the models to observed frequency units. For tilted-ring disk models E1 and F1, $\Gamma$ gives the range of ring major-axis PA and $i$ corresponds to the range in $q$ values determined at the ring radial positions. The last row of the table lists 1$\sigma$ parameter uncertainties on Model F1 determined after Monte Carlo resampling and re-fitting of the best-fit model cube.}
\end{singlespace}
\end{deluxetable*}


Within each frequency channel, the background noise in the ALMA data cube contains strong pixel-to-pixel correlations on scales comparable to and smaller than the angular scale of the synthesized beam. Constructing the full covariance matrix to account for correlated errors in these data remains very challenging \citep[e.g.,][]{hez16,oni17}. Instead, we mitigate the impact of correlated errors in the \chisq\ calculation by first spatially block-averaging the data in $5\times5$ pixel regions to form roughly beam-sized cells. We then measure the rms noise levels in line-free regions in each frequency channel of the block-averaged data cube. The channel-dependent background noise is somewhat lower than measured at the original pixel scale, averaging to 0.18 mJy beam\per\ in $\sim$10.2 \kms\ channels. For each model iteration, the beam-convolved model cube is block-averaged in the same way as the data. After rebinning, we calculate \chisq\ within an elliptical spatial fitting region that includes nearly the entire disk area with major axis radius $\rfit=1\arcsec$, PA=77\degr, and axis ratio $q=0.67$, and a spectral fitting region $2310\leq \vlos \leq 3273$ \kms\ that spans a slightly larger velocity range than does the CO(2$-$1) emission (see Figure~\ref{fig:c4pvdm}). The block-averaged fitting region used to compute \chisq\ contains 415 spatial pixels and 94 frequency channels, for a total of 39010 data points.

\begin{figure}
\begin{center}
\includegraphics[trim=0mm 0mm 0mm 0mm, clip, width=\columnwidth]{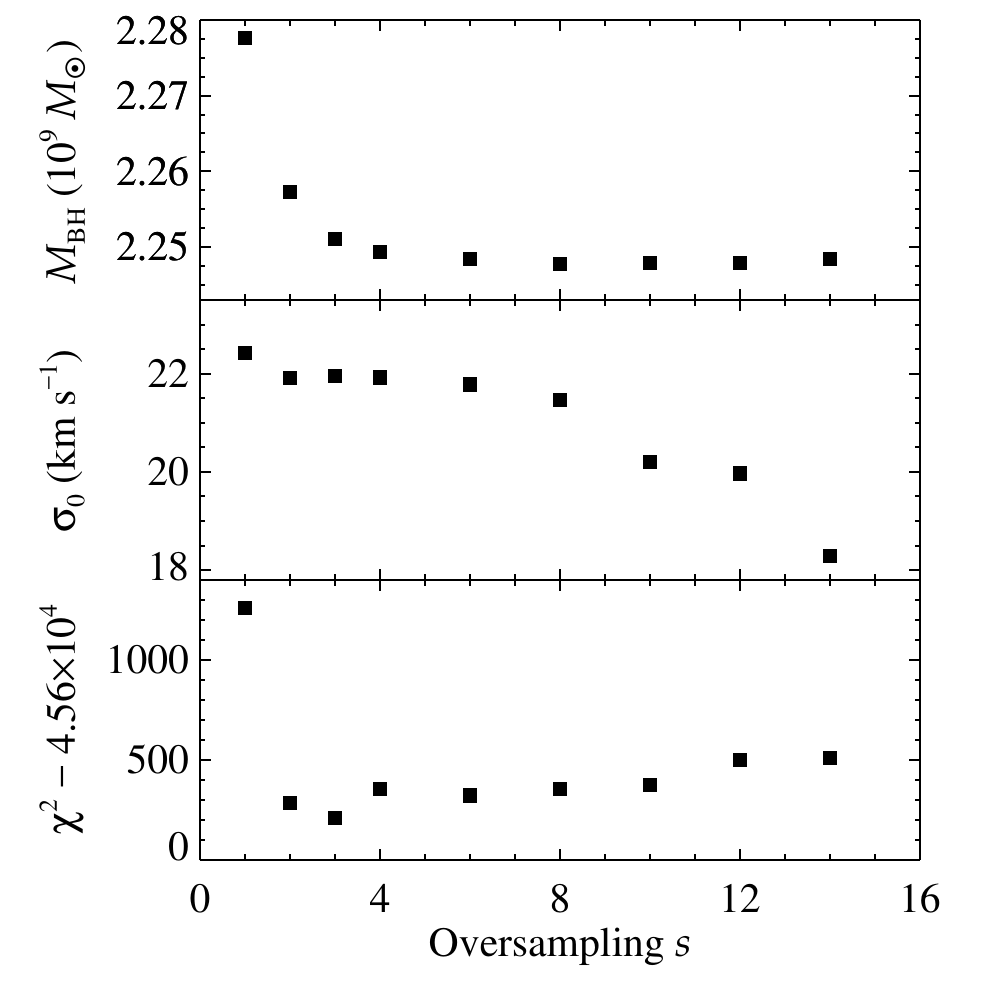}
\begin{singlespace}
  \caption{Model optimization results as a function of spatial oversampling factor $s$. This is illustrated here for model F1 (our final best-fitting model), and we find qualitatively similar results for flat-disk and MGE-based models.}
\end{singlespace}
\label{fig:subsamp}
\end{center}
\end{figure}

\subsubsection{Flat disk model results}
\label{sec:thinmodel_results}

In initial modeling trials, we tested how parameter values change with increasing oversampling factor $s$. As shown in Figure~\ref{fig:subsamp}, we find that the best-fitting BH mass converges to stable values for $s\geq 4$, while for smaller values of $s$ the best-fitting \mbh\ increases by at most $\sim$1\%. Computation time increases dramatically for $s>4$ with very little change in the resulting BH mass, so for the remainder of this work all model calculations use $s=4$.

Our initial model C1 incorporated a flat disk and a spatially uniform turbulent velocity dispersion $\sigmaturb(r) = \sigma_1$. The extended mass distribution was characterized by the initial $\vcst(r)$ profile assuming no extinction in the disk ($A_H=0$). After optimizing to the CO(2$-$1) data cube, we obtained best-fit parameters $\mbh=2.280\times10^9$ \msun, $\upsh=2.72$ \msun/\lsun, and $\sigmaturb=10.5$ \kms\ (see Table~\ref{tbl:mod_res} for the complete results). The total $\chisq=47929.2$ results in $\chisqnu=1.229$ over $\ndof=39001$ degrees of freedom. This basic dynamical model reproduces the general CO kinematic behavior moderately well, although quantitatively it does not constitute a good fit to the data. For this \ndof, a formally acceptable fit should achieve $\chisqnu \leq 1.012$.

At high angular resolution, the CO line structure in each channel map forms a tight locus of emission with a characteristic "V" shape. We constructed a residual cube by subtracting the model from the data cube, and in each channel identify regions where the model is mismatched with the data: line structure discrepancies between data and model channel slices can be separated into those that arise from neglect of disk kinematic warping (i.e., rotational components that warp the "V" shape) and those that stem from an inadequate host galaxy mass model (i.e., that shift and dilate the locus in the radial direction). In most channels, we find the discrepancies to be primarily rotational.

\begin{figure}
\begin{center}
\includegraphics[trim=0mm 0mm 0mm 0mm, clip, width=\columnwidth]{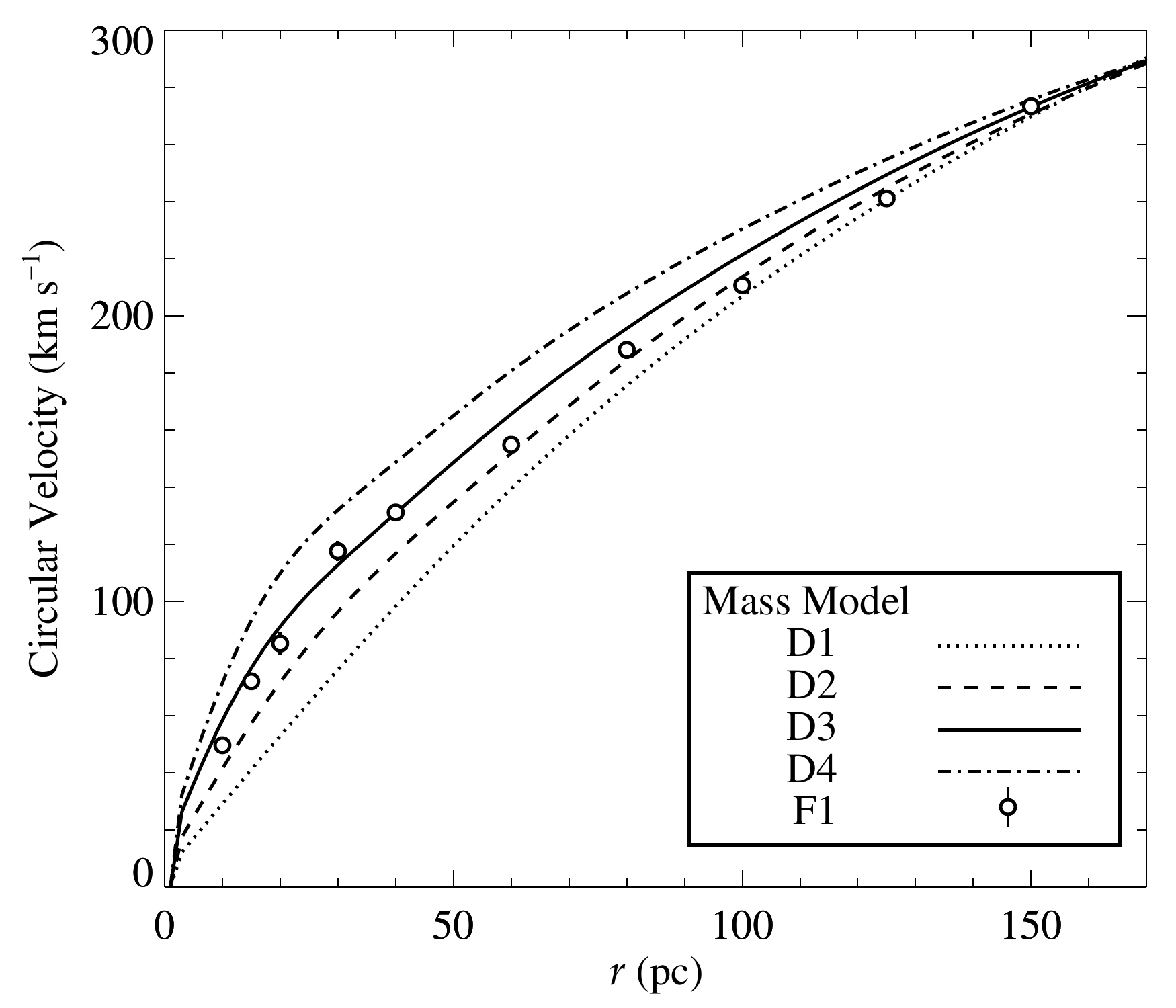}
\begin{singlespace}
  \caption{Plot comparing the \vcst\ profiles after scaling by the \upsh\ values obtained in models D1--D4. The best-fit radial circular velocity profile \vext\ (corresponding to model F1) lies within the envelope of these MGE-derived \vcst\ solutions, albeit with a different dependence on radius. Uncertainties in \vext\ are based on the Monte Carlo error analysis procedure described in \S\ref{sec:montecarlo} and are listed in Table \ref{tbl:disk}.}
\end{singlespace}
\label{fig:stellradprof}
\end{center}
\end{figure}

In models D1--D4, we adopted the more flexible Gaussian $\sigmaturb(r)$ function and used each of the extinction-corrected \vcst\ profiles in turn to explore the impact of central dust extinction on \mbh. The parameters for their best-fit model cubes converge to a range of values $\mbh=(2.059-2.276)\times10^9$ \msun\ and $\upsh=2.04-2.73$ \msun/\lsun. These best fits obtain minimum $\chisq=47458.1-47547.7$ over $\ndof=38998$, corresponding to an average $\chisqnu \approx 1.22$ with slight preference for model D3 (corresponding to a central disk extinction of $A_H=0.75$ mag). Including a radially-varying \sigmaturb\ does improve the overall fit without significantly affecting the BH mass, as demonstrated by the decrease in \chisq\ from model C1 to D1. In model D3, $\sigmaturb(r)$ reaches a peak of 18.0 \kms\ at $r=12.2$ pc, decreasing to 8.3 \kms\ at the disk edge.

As the extinction correction increases from $A_H=0$ to 1.50 mag, the corresponding \vcst\ profiles reflect increasing stellar luminosity density at all disk radii with a bias towards increasing nuclear contributions (see Figure~\ref{fig:stellradprof}). Since the total enclosed mass is tightly constrained by velocities at the outer edge of the disk, a more cuspy central stellar surface brightness slope arising from a higher assumed extinction has the effect of slightly lowering both the best-fit \mbh\ and \upsh\ values. The highest of these \upsh\ values measured using models D1 and D2 are elevated when compared to typical dynamical $H$-band $M/L$ ratios in other ETGs \citep[e.g.,][]{oni17,yil17} while remaining consistent with those derived from stellar population synthesis modeling \citep[e.g.,][]{zib09}.

To visualize the model results, we show GH moments in Figure~\ref{fig:c4ghm} and the PVD in Figure~\ref{fig:c4pvdm} that are measured from the best-fit model D3 cube in the same manner as the data. The flat disk model velocities closely agree with the observed velocity field for most of the disk, with typical residuals $|\Delta \vlos| \lesssim3$ \kms. The velocity peaks in the flat disk model are offset from the observed locations by nearly 0\farcs05 (in a clockwise direction about the disk center) with large associated residuals ranging between $-80$ and $+180$ \kms, demonstrating the limitations of a flat disk formalism when modeling even mildly warped disks. In \S\ref{sec:add_tests} below, we also explore the possibility that a non-circular component of the gas velocity may account for the central kinematic twists.

\begin{figure*}
\begin{center}
\includegraphics[trim=0mm 0mm 0mm 0mm, clip, width=\textwidth]{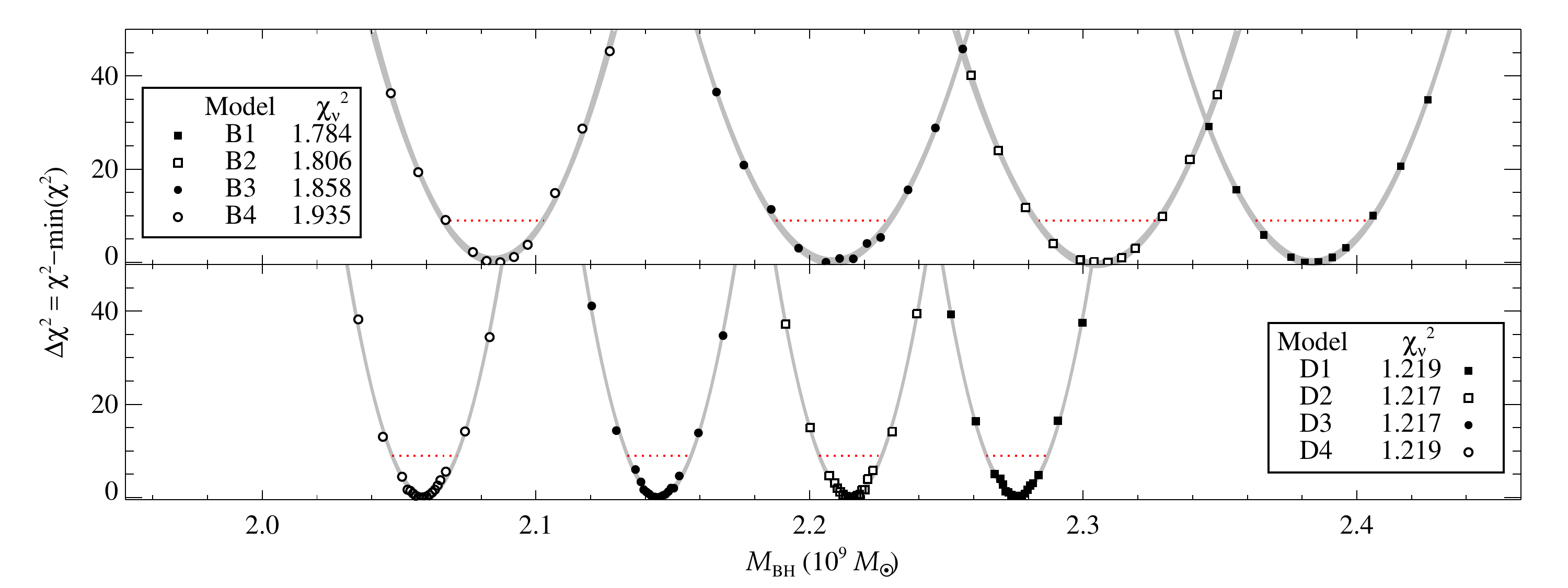}
\begin{singlespace}
  \caption{Results from \chisq\ minimization as a function of BH mass using models that assume flat disk rotation and MGE-based host galaxy mass profiles, for the Cycle 2 (upper panel) and Cycle 4 (lower panel) ALMA data. Shaded regions are Gaussian fits (with uncertainties) to these $\Delta \chisq = \chisq-\min(\chisq)$ values. The nominal $3\sigma$ uncertainty ranges $\Delta \chisq (\mbh) \leq 9$ (\textit{dotted lines}) are demarcated in each case. The B1$-$B4 (Cycle 2) and D1$-$D4 (Cycle 4) model fits incorporate the same set of four extinction-corrected host galaxy models. The $\Delta \chisq (\mbh)$ curves indicate narrow statistical uncertainties for an individual mass model.  The range in best-fit \mbh\ values shows that the uncertainty in \mbh\ due to the extinction correction applied to the MGE model is substantially larger than the model-fitting uncertainty on \mbh\ for a given dust-corrected MGE profile.}
\end{singlespace}
\label{fig:datamodchi2_c2c4}
\end{center}
\end{figure*}

In Figure~\ref{fig:datamodchi2_c2c4} we show $\Delta\chisq$ curves as a function of fixed BH mass for models D1$-$D4. Assuming the usual $\Delta \chisq \leq 9$ criterion, the $3\sigma$ (statistical)  uncertainties in \mbh\ for a given host galaxy model would be less than 1\% of \mbh. For the preferred model D3, the nominal $1\sigma$ uncertainty obtained by $\Delta \chisq \leq 1$ is estimated to be less than 0.2\% of its best-fit \mbh\ value. The range in BH mass of $\Delta \mbh =2.2 \times 10^8$ \msun\ (nearly 10\% of the BH mass) for these four models with different \vcst\ profiles far exceeds the statistical bounds on any one of the four. This range is representative of the systematic uncertainty introduced by dust extinction.

We note that \chisqnu\ values from these fits to the data cube do not faithfully characterize the model fidelity, because block-averaging does not fully mitigate the correlations between neighboring pixels. 
Thus, we do not use the $\Delta \chisq$ curves when determining the error budget on \mbh; instead, we adopt a Monte Carlo resampling procedure (described in \S\ref{sec:montecarlo}) to calculate the final statistical uncertainty.

For models D1$-$D4, the radius of the BH sphere of influence (defined as the radius within which the enclosed stellar mass is equal to \mbh) is 131--143 pc, projecting to an angular size of 0\farcs86--0\farcs94.

\subsubsection{Cycle 2 Comparison}
\label{sec:thinmodel_results_c2}

Our Cycle 2 CO(2$-$1) imaging of NGC 3258 with $\theta_\mathrm{FWHM}=0\farcs48 \times 0\farcs40$ provides $\sim$2 resolution elements across the BH radius of influence, so this initial data set should also allow for a confident BH mass measurement, although it will still be subject to the same uncertainty in the dust-disk extinction correction. Comparison with the Cycle 4 models provides an opportunity to test the impact of angular resolution on the best-fit \mbh.

\begin{figure}
\begin{center}
\includegraphics[trim=0mm 0mm 0mm 0mm, clip, width=\columnwidth]{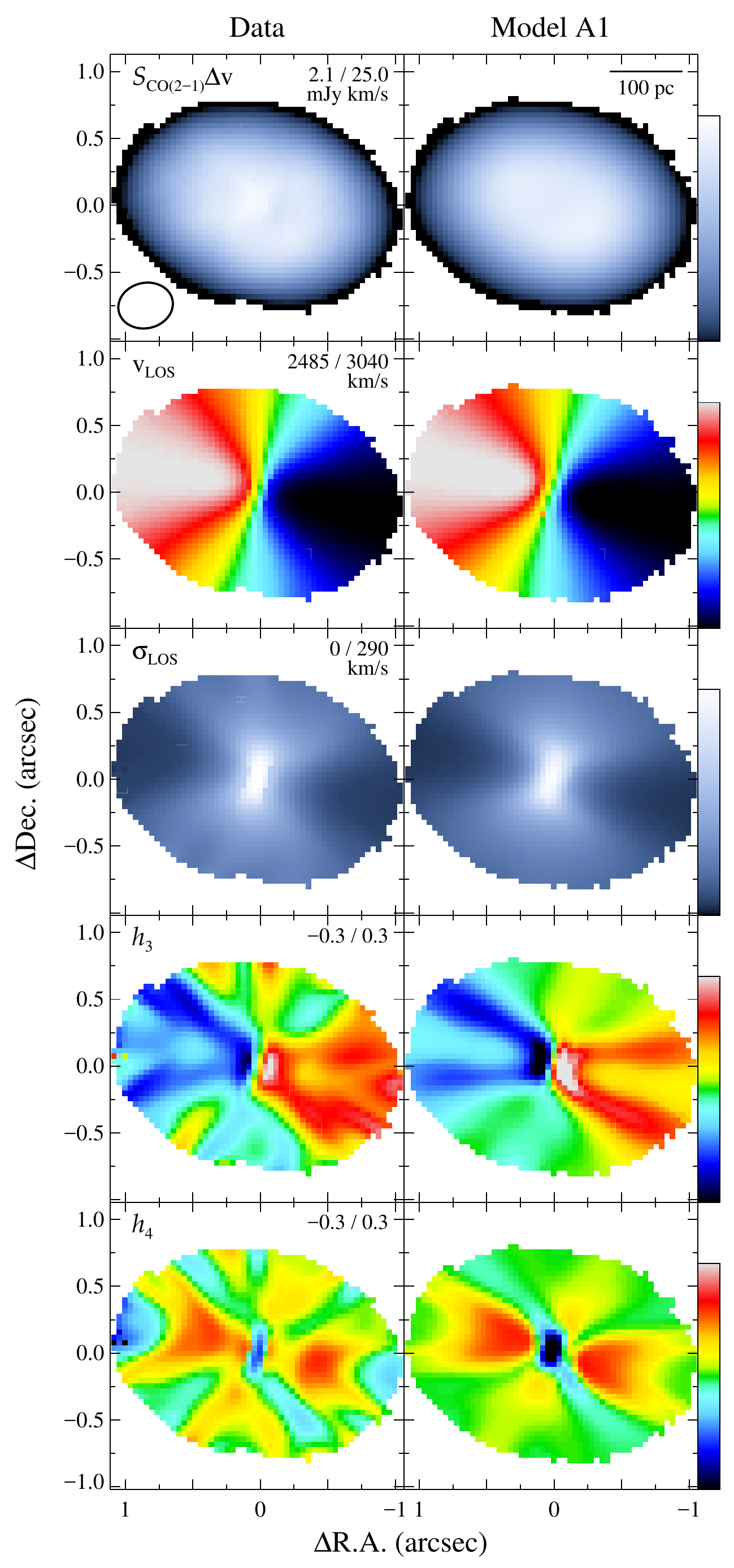}
\begin{singlespace}
  \caption{Comparison between flux and kinematic moments measured from the Cycle 2 CO(2$-$1) data cube (\textit{left}) and those measured from the best-fitting model A1 (\textit{right}). Ranges in each data frame indicate the intensity/color scale extremes. Due to strong central beam smearing of this $\sim0\farcs44-$resolution data, the kinematic signature of the BH is primarily found in higher-order (especially $h_3$) moments.}
\end{singlespace}
\label{fig:c2gh}
\end{center}
\end{figure}

As described in \citetalias{boi17}, the Cycle 2 data cube has a pixel scale of 0\farcs04 and a rest-frame velocity channel width of $\sim$20.3 \kms\ for CO(2$-$1) emission redshifted to the systemic velocity of NGC 3258. We fit the Cycle 2 data in models A1 and B1$-$B4 using procedures that correspond to Cycle 4 models C1 and D1$-$D4. We treated the Cycle 2 modeling in a self-contained manner by using a Richardson-Lucy deconvolution of the smoother Cycle 2 CO distribution to weight the model line profiles. We block-averaged both data and model cubes in $4 \times 4$ pixel regions prior to calculating model goodness-of-fit. These new cell sizes are significantly smaller than the synthesized beam area but allow for many spatial cells across the disk. At this more coarse angular resolution, the slightly larger $\rfit = 1\farcs2$ fitting region contains 124 spatial cells and 46 frequency channels, for a total of 5704 data pixels. Results of the Cycle 2 model fits are listed in Table \ref{tbl:mod_res}, and Figure \ref{fig:datamodchi2_c2c4} shows the $\Delta \chisq$ curves for models B1$-$B4 for comparison with the analogous Cycle 4 models D1$-$D4.

Overall, the Cycle 2 model fits yield \mbh\ values within a few percent of the values obtained from the analogous Cycle 4 models, and \upsh\ values that are $\sim$20\% greater than those from the corresponding Cycle 4 models. GH moments measured from the best-fitting model B1 cube show good agreement with those obtained from the data (see Figure~\ref{fig:c2gh}). The Cycle 2 model fits also find low \sigmaturb\ with Gaussian line width coefficients similar to those obtained from the Cycle 4 data. From examination of fitting residuals in the data cube, we find large residuals near the disk center, which we attribute in part to insufficient resolution in the flux map used in the modeling procedure. The Cycle 2 data do not recover the central hole in CO(2$-$1) surface brightness, and as a result the model assigns too much flux to pixels at LOS velocities $>500$ \kms\ in the innermost region of the disk, producing line profiles that exceed the maximum observed $|\vlos-\vsys|$. The worsening \chisq\ from models B1 to B4 stems from the additional central stellar mass that is introduced by the increasingly dust-corrected \vcst\ mass models, thereby increasing the model rotation speed near the BH. For an individual host galaxy mass model, the $\Delta \chisq$ curve is wider for the Cycle 2 data than for the corresponding Cycle 4 model fit, implying statistical uncertainties that are larger by a factor of $\sim$2 for the same host galaxy radial profile. (This analysis does not consider the larger \chisqnu\ values obtained for the Cycle 2 modeling due to significant correlated noise between block-averaged cells. However, even if we were to inflate the background rms noise to drive \chisqnu\ to unity, the $\Delta \chisq$ criterion would not yield significantly broader confidence intervals for \mbh.)

Despite these issues, the close agreement in \mbh\ between the Cycle 2 and Cycle 4 flat-disk model fits demonstrates that the Cycle 2 data already provide a good determination of \mbh. For a fixed host-galaxy mass model, the improvement in ALMA angular resolution (from resolving \rg\ by a factor of $\sim$2 to a factor of $\sim$10) results in a relatively modest improvement in precision on \mbh. In either case, the dominant uncertainty when using MGE-based mass models stems from the uncertainty in the dust extinction correction rather than from the model-fitting precision. It is important to note that these model fits are carried out over a spatial region that is almost entirely contained within \rg\ for NGC 3258. As a result, the uncertainty in the central stellar mass profile slope only results in a modest (several percent) uncertainty in \mbh\ even for the Cycle 2 data. In many other ALMA gas-dynamical targets, the molecular disk extends to radii well beyond \rg. In such cases, if the model fits are carried out over the entire dust disk, the fitting results will tend to be dominated by the influence of the large fraction of spatial pixels well outside of \rg, in which case the uncertainty in the dust extinction correction will likely lead to a far larger range of uncertainty in \mbh\ than what we find for NGC 3258.

\begin{figure*}
\begin{center}
\includegraphics[trim=0mm 0mm 0mm 0mm, clip, width=\textwidth]{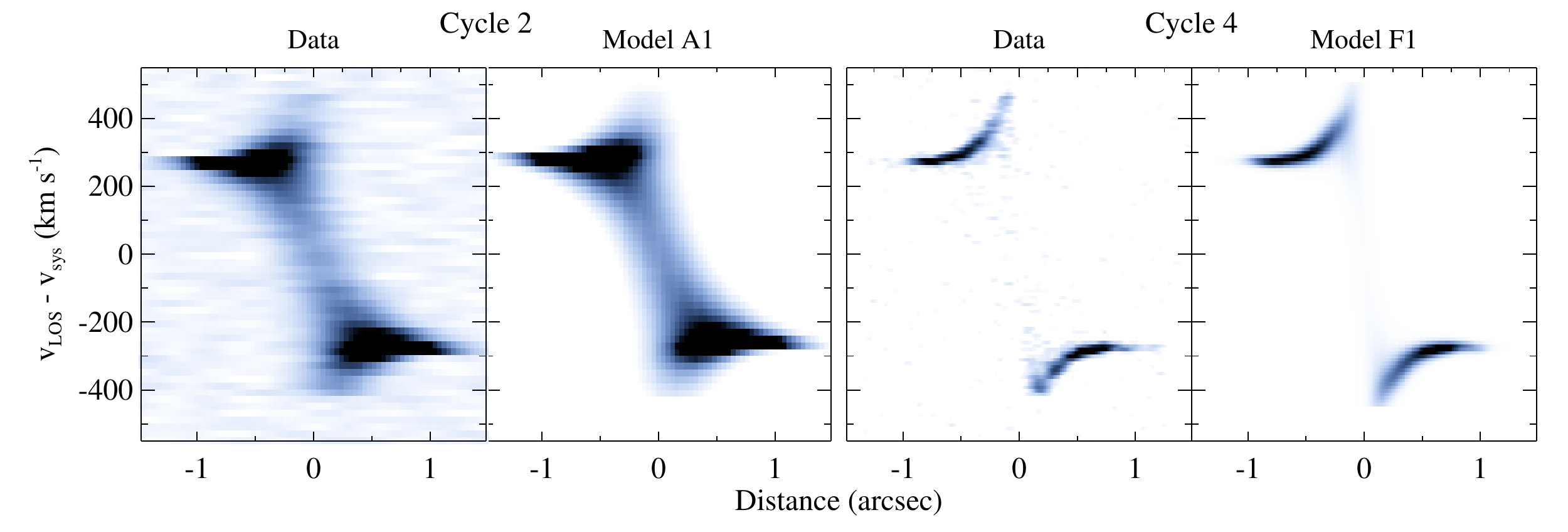}
\begin{singlespace}
  \caption{Comparison between the Cycle 2 (\textit{left}) and Cycle 4 (\textit{right}) data and model PVDs. Line-of-sight velocities are relative to the galaxy systemic velocity. These PVDs were extracted at a PA of 77\degr with a width equal to the geometric average of the beam major and minor FWHMs.}
\end{singlespace}
\label{fig:pvdmod}
\end{center}
\end{figure*}

\subsection{Detailed Dynamical Modeling}
\label{sec:detailedmodel_method}

In this section, we introduce more general descriptions for the disk structure and host galaxy mass distribution, with the addition of two additional features to the modeling procedure described above. First, we incorporate a tilted-ring model that fits the disk's mildly warped structure more accurately than flat-disk models. Second, we employ a method to constrain the host galaxy's spatially extended mass profile solely through fitting to the ALMA CO kinematics, rather than relying on the \hst\ imaging data (and an uncertain extinction correction) to constrain the host galaxy model. These two improvements are made possible by the angular resolution of the ALMA Cycle 4 observations, which fully resolve the rotational structure of the disk without significant blurring of the central kinematics by beam smearing.

\begin{figure}
\begin{center}
\includegraphics[trim=0mm 0mm 0mm 0mm, clip, width=\columnwidth]{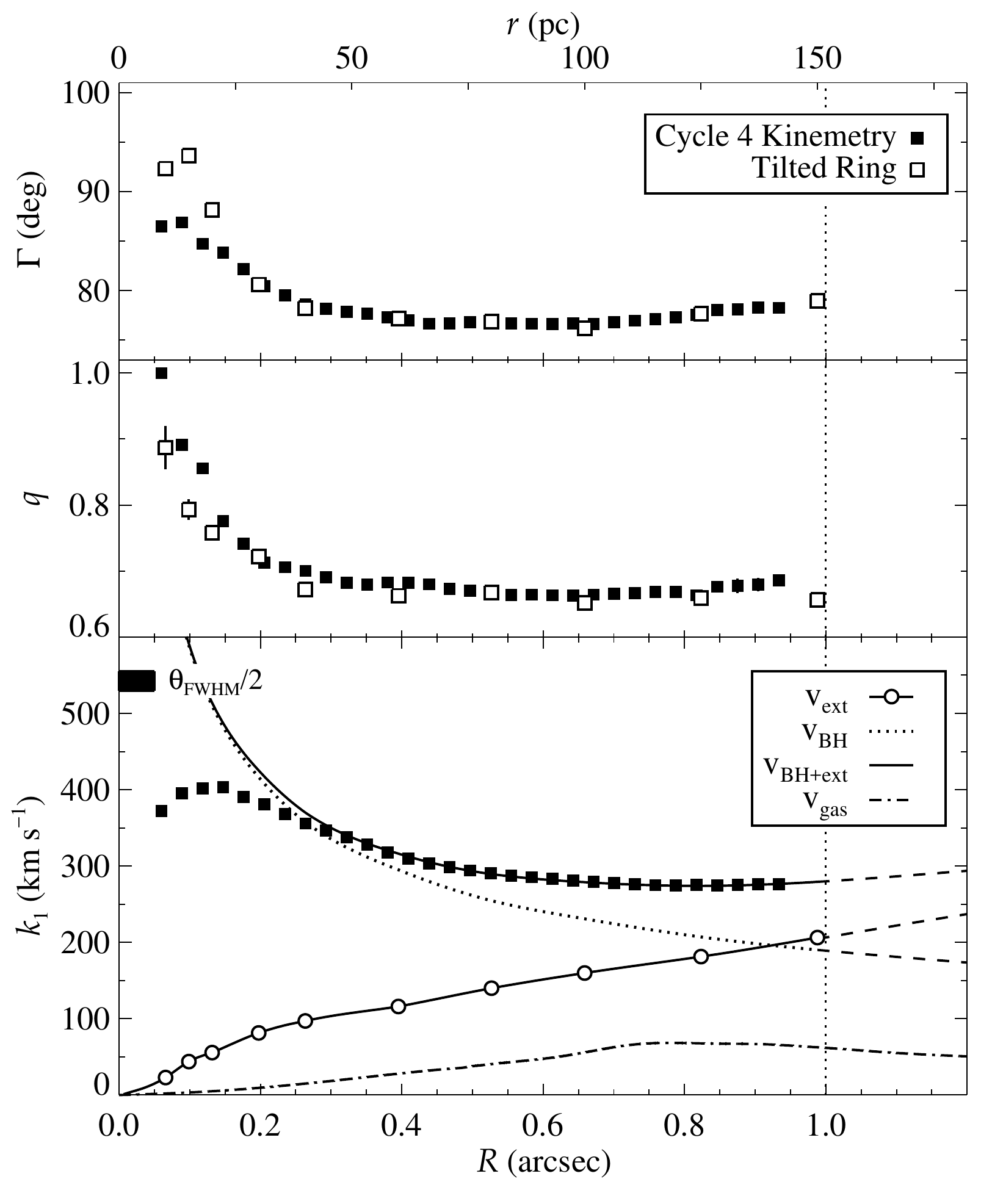}
\begin{singlespace}
  \caption{Comparison between the Cycle 4 \vlos\ kinemetry and best-fitting parameters from model F1. Except for the central couple of beam FWHMs (within $\sim$0\farcs2), position angles and axis ratios agree well with the non-parametric, freely-varying tilted-ring $\Gamma$ and $q$ parameters. Good agreement is likewise found between the measured line-of-nodes velocities ($k_1$) and model F1 LOS velocities that include contributions from both the BH and \vext\ due to the extended mass distribution. For comparison, this plot also includes the expected LOS velocity profile $v_{\rm gas}$ in the galaxy midplane that arises from the molecular gas disk assuming standard CO-to-H$_2$ conversion (see \S\ref{sec:almaproperties}). Parameter error bars are estimates derived from Monte Carlo resampling of the best-fitting model cube (see Table~\ref{tbl:disk}).}
\end{singlespace}
\label{fig:kinplots}
\end{center}
\end{figure}

\begin{figure*}
\begin{center}
\includegraphics[trim=0mm 0mm 0mm 0mm, clip, width=\textwidth]{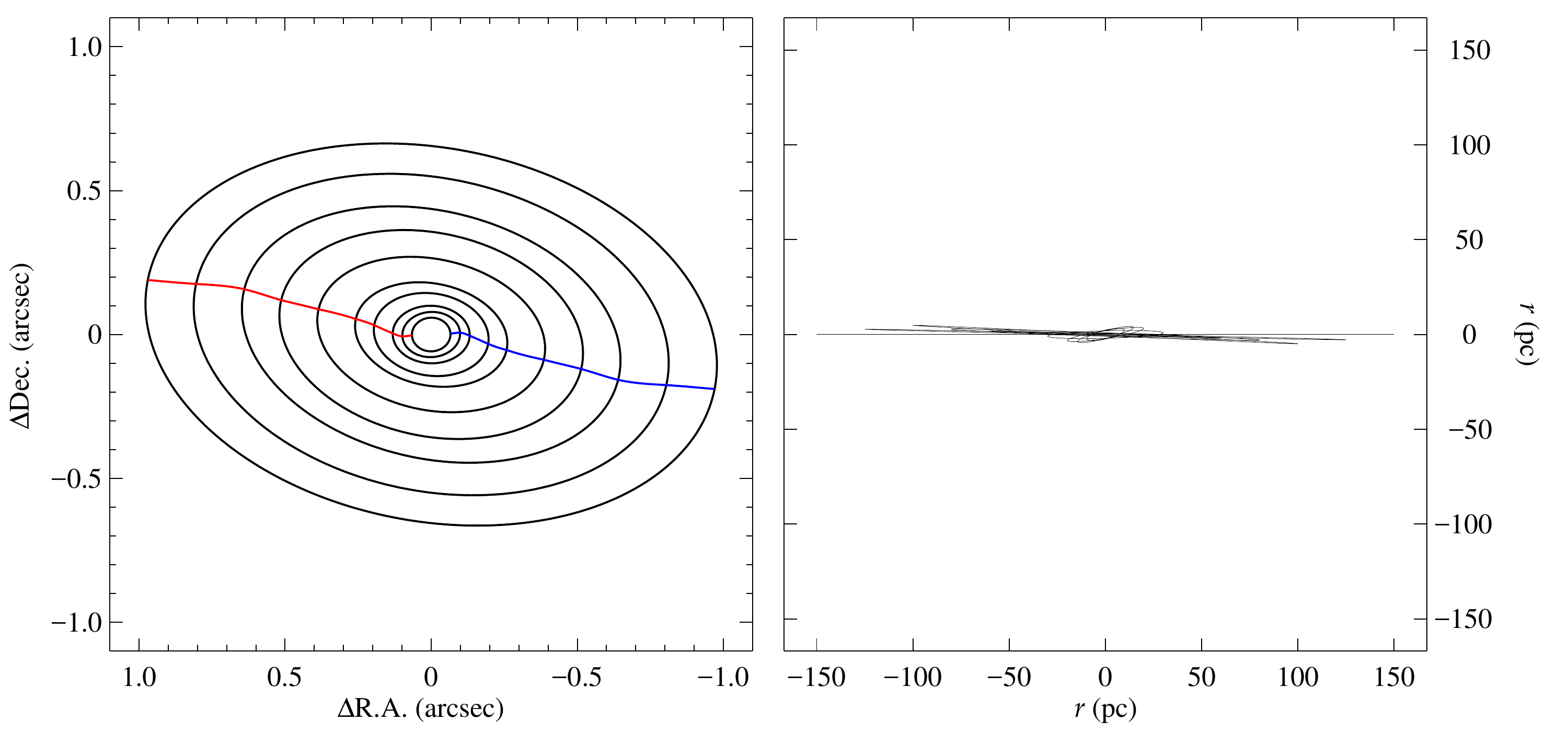}
\begin{singlespace}
  \caption{The best-fit tilted-ring structure from model F1, shown both in projection (\textit{left}, with the line of nodes delineated) and in the plane of the outermost ring (\textit{left}, with -x corresponding to east), demonstrating that the warped disk remains relatively flat.}
\end{singlespace}
\label{fig:kinmod}
\end{center}
\end{figure*}

\subsubsection{Tilted-ring model}
\label{sec:detailedmodel_twist}

As shown in Figure~\ref{fig:c4ghm}, the \vlos\ map for flat-disk model D3 does not fully reproduce the observed velocity field due to the mild kinematic twist near the disk center discussed in \S\ref{sec:almaproperties}. Given the high precision enabled by the resolution of our Cycle 4 observation, it is important to determine how the disk's warp may affect the inferred BH mass. In models E1 and F1, we implement a tilted-ring model that characterizes a warped (but still intrinsically thin) disk using a series of concentric rings \citep[e.g.,][]{rog74}, with each ring allowed to have an arbitrary PA $\Gamma$ and inclination angle $i =$ arccos q. The model comprises ten rings spanning the disk's radius (see Table~\ref{tbl:disk} for the ring radii), approximately matching the number of beam widths across \rfit. The non-linear spacing between annuli was chosen to better characterize the more abrupt increases in $\Delta \Gamma$ and $\Delta q$ toward the disk center.

For each model iteration, we form continuous $\Gamma(r)$ and $q(r)$ functions by a cubic spline interpolation of these ring parameters at intermediate radii and construct 2D maps of intrinsic radius and inclination at each projected disk location on the plane of the sky. One-to-one correspondence between a projected and physical disk location is not in general preserved for warped disks \citep[especially for rapid shifts in $\Gamma$ and $q$ shifts and for more edge-on disks; see][]{cor97,joz07,dav13a}. However, this approximation is suitable for the moderately inclined and warped gas disk in NGC 3258. We proceed to generate model line profile cubes using the maps of intrinsic \vlos\ and \sigmalos. Beam smearing and subsequent model fitting steps are applied as described previously. We allow the values of $\Gamma$ and $q$ for each ring to vary freely while simultaneously optimizing other disk parameters. Model E1 incorporates the tilted-ring disk structure and employs the same MGE-based host galaxy profile as model D3 (with $A_H=0.75$ mag), while model F1 incorporates both the tilted-ring disk and the new host galaxy mass modeling procedure described below.

\subsubsection{Host galaxy mass profile from CO kinematics}
\label{sec:vext_method}

The four MGE-derived \vcst\ profiles are corrected for a plausible range of central disk extinction levels. In models D1$-$D4, we adopted each profile in turn to explore the impact on the derived \mbh\ of the range in possible host galaxy mass distributions. The optimized models are very similar in a statistical sense yet yield best-fit \mbh\ values that span a mass range of about 10\%. We cannot determine the correct (average) extinction level using these \vcst\ profiles alone, and the associated systematics would make the Cycle 4 modeling results nearly as uncertain as those from the Cycle 2 data set.

Fortunately, our Cycle 4 data are so well resolved that we can constrain the galaxy mass profile directly by modeling the CO kinematics, without reference to the NIR imaging data. This is the only method that can potentially reduce the systematic uncertainties to the percent level (or better) for such dusty disks, because host galaxy models based on image deprojections will always suffer from systematic uncertainty in the extinction correction. We refer to the circular velocity profile arising from the extended mass distribution as $\vext(r)$. This velocity profile primarily reflects the stellar mass distribution but also includes any other gravitating mass, including the gas disk itself (see Figure~\ref{fig:kinplots}) as well as the expected very small contribution of dark matter \citep[extrapolation from observations and simulations of luminous ETGs suggests a dark matter mass of less than $\sim 10^7$ \msun\ enclosed within NGC 3258's central dust disk region; e.g.,][]{new13,wan18}.

In model F1, we describe the extended mass distribution in terms of a circular velocity profile with ten free parameters, where the free parameters correspond to the values of \vext\ at the same set of ten ring radii used to generate the tilted-ring model.  We create the model velocity field by cubic spline interpolation of \vext\ between the rings to determine its value at each disk location, afterwards calculating the disk rotation speed at each position resulting from both the BH mass and the spatially extended mass, and finally projecting the rotation speed along the line of sight using the tilted-ring model to calculate \vlos\ at each spatial pixel in the model. The ten free \vext\ parameters were optimized simultaneously with the tilted-ring angular parameters and the other disk parameters, for a total of 39 free parameters in the final model. The only constraints we applied to the circular velocity model is that $\vext(r)$ was required to be an increasing function of radius and that $\vext =0$ at $r=0$. This method of determining \vext\ is largely equivalent to allowing a radially-varying $M/L$ ratio when scaling the stellar luminosity profile \citep{dav17a,dav18}. However, our method eliminates any dependence on the luminosity profile derived from imaging data, instead allowing nearly complete freedom in the $M(r)$ profile to match the kinematic data.

\begin{deluxetable}{ccccc}
\tabletypesize{\scriptsize}
\tablecaption{Tilted-ring Model Parameters and Host Galaxy Circular Velocity Profile\label{tbl:disk}}
\tablewidth{0pt}
\tablehead{
\multicolumn{2}{c}{Disk Radius} & \colhead{$\Gamma$} & \colhead{$q$} & \colhead{\vext} \\
\colhead{(arcsec)} & \colhead{(pc)} & \colhead{(\degr)} & \colhead{ } & \colhead{(\kms)}
}
\startdata
0.066 & 10 & 92.31 (0.22) & 0.887 (0.033) & \phantom{ }49.7 (2.5) \\
\rule{0pt}{4ex}0.099 & 15 & 93.61 (0.24) & 0.793 (0.016) & \phantom{ }72.0 (2.8) \\
\rule{0pt}{4ex}0.131 & 20 & 88.14 (0.21) & 0.758 (0.007) & \phantom{ }85.2 (4.1) \\
\rule{0pt}{4ex}0.197 & 30 & 80.60 (0.22) & 0.722 (0.004) & 117.6 (3.6) \\
\rule{0pt}{4ex}0.264 & 40 & 78.21 (0.17) & 0.672 (0.002) & 131.1 (1.6)  \\
\rule{0pt}{4ex}0.395 & 60 & 77.17 (0.09) & 0.662 (0.001) & 154.9 (1.3) \\
\rule{0pt}{4ex}0.527 & 80 & 76.88 (0.07) & 0.668 (0.001) & 188.1 (0.8) \\
\rule{0pt}{4ex}0.659 & 100 & 76.17 (0.05) & 0.652 (0.001) & 210.7 (0.4) \\
\rule{0pt}{4ex}0.823 & 125 & 77.68 (0.07) & 0.659 (0.001) & 241.1 (0.4) \\
\rule{0pt}{4ex}0.988 & 150 & 78.95 (0.09) & 0.656 (0.002) & 273.4 (0.9)
\enddata
\begin{singlespace}
  \tablecomments{Best-fitting model F1 parameters $\Gamma$ and $q$ for each ring when employing the tilted-ring geometry, and circular velocities \vext\ that arise from the spatially extended host galaxy mass distribution. These parameters were allowed to freely vary at each of the ten fixed radial locations, the only restriction being that \vext\ was required to be a strictly increasing function of radius. The corresponding ring physical distances in parcsecs are shown assuming $1\arcsec=151.8$ pc. Statistical uncertainties (in parentheses) were derived from Monte Carlo resampling of the optimized model cube.}
\end{singlespace}
\end{deluxetable}

\subsection{Detailed modeling results}
\label{sec:detailedmodel_results}

We first optimize model E1, which includes a tilted-ring geometry and an extinction-corrected ($A_H=0.75$ mag) galaxy mass distribution. Aside from the flexible disk structure, this scenario is identical to the D3 case, making it possible to isolate the impact of disk warping on the derived BH mass. The optimized model converges to $\mbh=2.203\times10^9$ \msun\ and $\upsh=2.18$ \msun/\lsun. The total $\chisq=46009.9$ yields $\chisqnu=1.180$ over $\ndof=38980$ and represents the most substantial fit improvement for the Cycle 4 gas-dynamical models; in contrast, model D3 achieved $\chisqnu = 1.217$. The tilted-ring angular parameters smoothly increase by $\Delta \Gamma \approx 20\degr$ and $\Delta q \approx 0.27$ (corresponding to an inclination angle decrease $\Delta i \approx -26\degr$) towards the disk center. The shift to a more face-on disk orientation at small radii projects circumnuclear nuclear rotation away from the line of sight and results in a moderate $\Delta \mbh =5.9\times10^7$ \msun\ (or $\sim$3\%) increase in \mbh\ relative to the otherwise analogous flat-disk model D3.

We go on to investigate the host galaxy mass profile in model F1, which is identical to E1 except that it employs the freely varying \vext\ method to represent the host galaxy mass distribution instead of the MGE-based host galaxy model. In this case, the BH mass converges to $2.249\times10^9$ \msun, and the best fit achieves $\chisq=45956.4$ over $\ndof=38971$ for $\chisqnu=1.179$. The $\chisqnu$ statistic decreases only slightly from model E1 to F1, indicating that between models D3 and F1 most of the improvement in fit quality was the result of including the tilted-ring disk geometry (detailed in Figures~\ref{fig:kinplots} and \ref{fig:kinmod}) rather than allowing additional freedom in the host galaxy model. However, the primary advantage of the freely varying \vext\ host galaxy model is that it removes the systematic uncertainty in \mbh\ resulting from dust extinction that is inherent in the MGE-based host galaxy models. Model F1 attains the lowest \chisqnu\ value of any of our trial models, and we consider it to be our final preferred model for the NGC 3258 disk.

As a result of beam smearing of the central disk kinematics, the kinemetry measurements $\Gamma_k$ and $q_k$ do not trace the intrinsic disk structure as faithfully within the inner couple of beam widths. In particular, within the innermost 0\farcs2, the strong intrinsic change in line-of-nodes PA implied by the tilted-ring model exceeds the $\Gamma_k$ rise, and the axis ratio $q_k$ approaches unity near the nucleus while the tilted-ring model axis ratio reaches a central value of $\approx0.89$ for model F1 (see Figure~\ref{fig:kinplots}). At $R>0\farcs2$, beam smearing has less impact on the observed kinematics, and $\Gamma_k$ and $q_k$ more closely trace the $\Gamma(r)$ and $q(r)$ profiles of the tilted-ring model.

In Figure~\ref{fig:c4ghm} we show GH moment and residual maps measured from this best model F1 cube. We find the velocity residuals $\Delta \vlos$ are generally small and centered about zero, with $\sim$60\% of the spatial pixels in the model falling within $\pm2$ \kms\ of the observed velocity in the data. The tilted-ring disk model alleviates most of the large central discrepancies apparent in the flat-disk velocity map generated from model D3, although the $|\vlos-\vsys|$ asymmetry between the quasi-Keplerian peaks still leads to $\sim$50 \kms\ velocity residuals for $R < 0\farcs15$ in model F1.

The data$-$F1 $\Delta \sigmalos$ residuals are also substantial at pixels near the nucleus, and we find the Gaussian $\sigmaturb$ profile underpredicts the line widths directly north and south of the nucleus by $\sim$100 \kms. These locations are coincident with bright clumps of CO emission, and we expect these non-axisymmetric excess line width features either to be correlated with gas cloud size \citep[e.g.,][]{she12} or to result from strong tidal shear within $\sim$20 pc of the BH. At the nucleus, this model overpredicts the data $\sigmalos$ by over 200 \kms, and in \S\ref{sec:add_tests} we discuss how this feature may be the result of an inadequate CO surface brightness map.

For the best-fitting model F1, the BH radius of influence (defined as the radius within which the enclosed stellar mass is equal to \mbh) is $\rg = 143$ pc, or 0\farcs 94.

\subsection{Monte Carlo error analysis}
\label{sec:montecarlo}

\begin{figure}
\begin{center}
\includegraphics[trim=0mm 0mm 0mm 0mm, clip, width=\columnwidth]{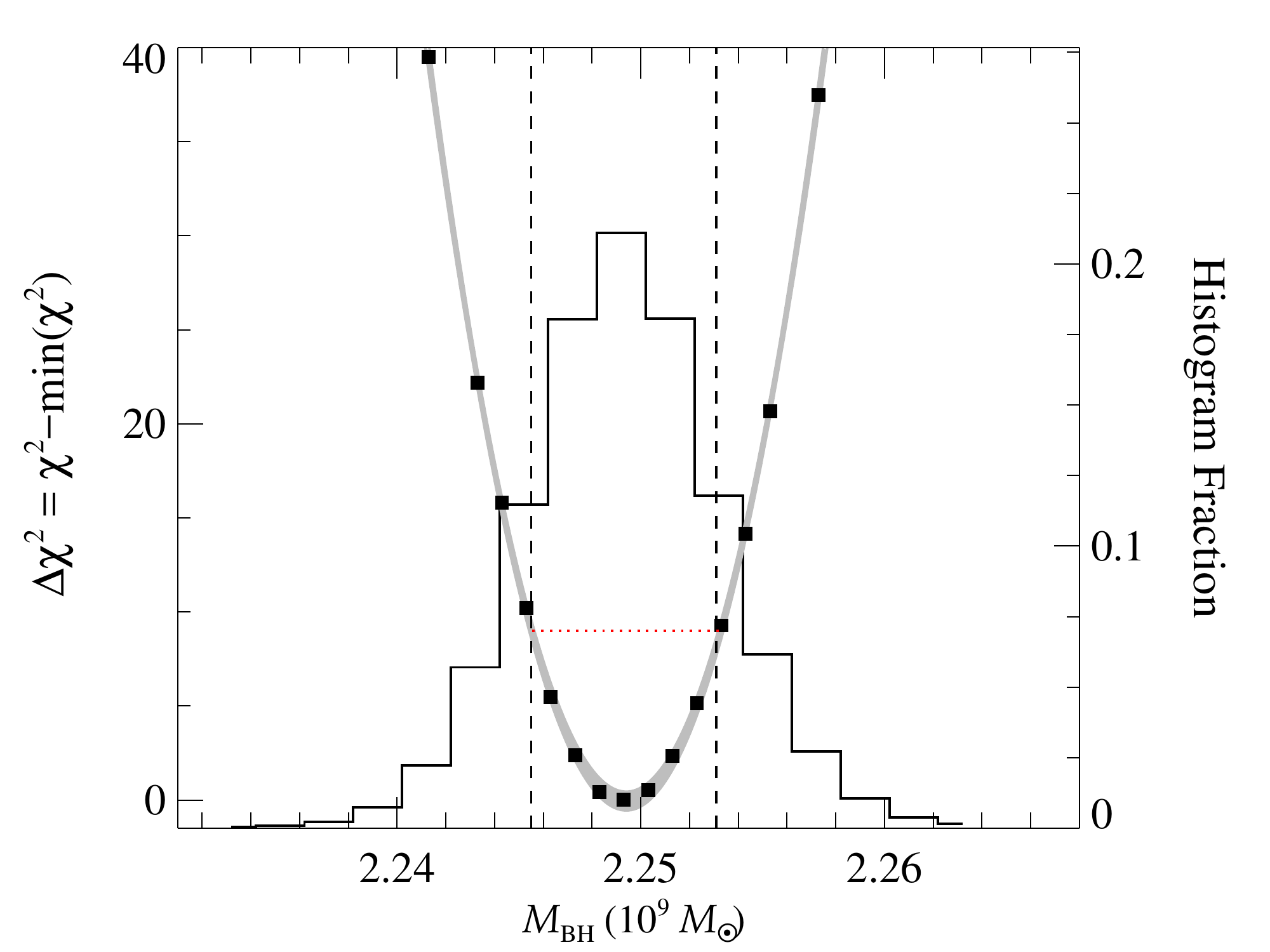}
\begin{singlespace}
  \caption{$\Delta \chisq (\mbh)$ minimization curve for model F1, which includes both a tilted-ring geometry and a radial circular velocity profile \vext. Shaded regions are Gaussian fits (with uncertainties) to these $\Delta \chi^2$ values. The nominal $3\sigma$ uncertainty range $\Delta\chi^2=9$ (\textit{dotted line}) in \mbh\ is much more narrow than found for models D1$-$D4 in Figure~\ref{fig:datamodchi2_c2c4}. The histogram shown here contains the set of BH masses determined after Monte Carlo resampling the best model F1 cube, with the $1\sigma$ statistical uncertainty range (\textit{dashed lines}) demarcated.}
\end{singlespace}
\label{fig:datamodchi2_c4}
\end{center}
\end{figure}

Although our models match the overall kinematic structure of the disk well in general, the final $\chisqnu = 1.179$ for model F1 indicates that the model is not formally an acceptable fit to the data for 38971 degrees of freedom. In this situation, determining the statistical uncertainty on \mbh\ and other parameters by examining contours in $\Delta\chisq$ would tend to underestimate the true uncertainty range. For example, measuring $\chi^2$ as a function of fixed \mbh\ while allowing all the other model parameters to freely vary results in a very narrow $\Delta\chi^2$ curve, with the $\Delta\chisq=9$ range (the nominal $3\sigma$ uncertainty range) corresponding to $\pm0.2\%$ of the best-fitting BH mass (see Figure~\ref{fig:datamodchi2_c4}), and $\Delta\chisq=1$ (for a $1\sigma$ uncertainty range) corresponding to just $\sim 0.06\%$ of \mbh, although the bottom of the $\Delta\chisq$ curve for model F1 is slightly irregular.

To obtain a more robust measure of the statistical model-fitting uncertainties in this situation, we carried out 100 Monte Carlo realizations of the best-fit model F1 cube. To add realistic noise to this model cube, we used noise from line-free channels in the continuum-subtracted ALMA data cube itself. We extracted nearly 100 line-free channels from the data cube where $|\vlos-\vsys|>500$ \kms, and randomly assigned and added these noise slices to the model cube channels at each resampling iteration. After incorporating this realistic noise, we carried out complete model fits to each noise-randomized model cube following the same procedure as for model F1, including both the tilted-ring model and the flexible \vext\ description. All model parameters were allowed to freely vary. From this suite of Monte Carlo realizations, we determined $1\sigma$ uncertainties on each of the thirty-nine model parameters by taking the standard deviation of the set of their best-fit values.

We include the histogram of \mbh\ values determined from this procedure in Figure~\ref{fig:datamodchi2_c4} to compare to the $\Delta \chisq$ results. While somewhat broader than the \chisq\ bounds, the distribution of \mbh\ values remains very narrow. We adopt the standard deviation $3.8\times10^6$ \msun\ (corresponding to $\sim$0.17\% of the BH mass) of these Monte Carlo results as the final $1\sigma$ statistical uncertainty on \mbh. Tables \ref{tbl:mod_res} and \ref{tbl:disk} list the full set of parameter uncertainties for model F1 based on these Monte Carlo simulations.

\subsection{Additional tests and error budget}
\label{sec:add_tests}

We now describe additional tests conducted to estimate the systematic uncertainties on \mbh. In each test, we modified aspects of model F1 to explore the sensitivity of our model-fitting results to various details of the model construction.

\textit{Pixel oversampling and block averaging:} We adopted an oversampling factor $s=4$ for the model fits described above, based on the results shown in Figure \ref{fig:subsamp}. Ionized gas disk dynamical modeling has demonstrated a typical scatter in derived \mbh\ values of a few percent for different $s$ values \citep[e.g.,][]{bar01}, behavior that may also apply to ALMA data \citep{bar16b}. Model F1 tests show that oversampling factors $s<4$ do not adequately sample the velocity field, resulting in a $\sim$1.3\% decrease in BH mass from $s=1$ to 4 that we do not include in the final error budget. The \mbh\ results are very stable for $s\geq 4$, with the best-fit BH mass decreasing by $\Delta \mbh=-1.5\times 10^6$ \msun\ (corresponding to $\sim$0.07\% in BH mass; see Figure~\ref{fig:subsamp}) as $s$ increases to 14.

As described in \S\ref{sec:thinmodel_method}, the Cycle 4 data and model cubes were spatially block-averaged into $5\times5$ pixel bins prior to computing $\chi^2$ for each model iteration, in order to mitigate the impact of correlated noise on spatial scales smaller than the ALMA beam width. We also explored block-averaging the data and model cubes using pixel regions ranging from $1\times1$ (no averaging) to $10\times10$ and found only a negligible impact on the derived \mbh, with the best-fit models converging to a narrow range of BH masses with a scatter of $\Delta \mbh=\pm 1.1\times 10^6$ \msun\ about the model F1 value.

\begin{figure}
\begin{center}
\includegraphics[trim=0mm 0mm 0mm 0mm, clip, width=\columnwidth]{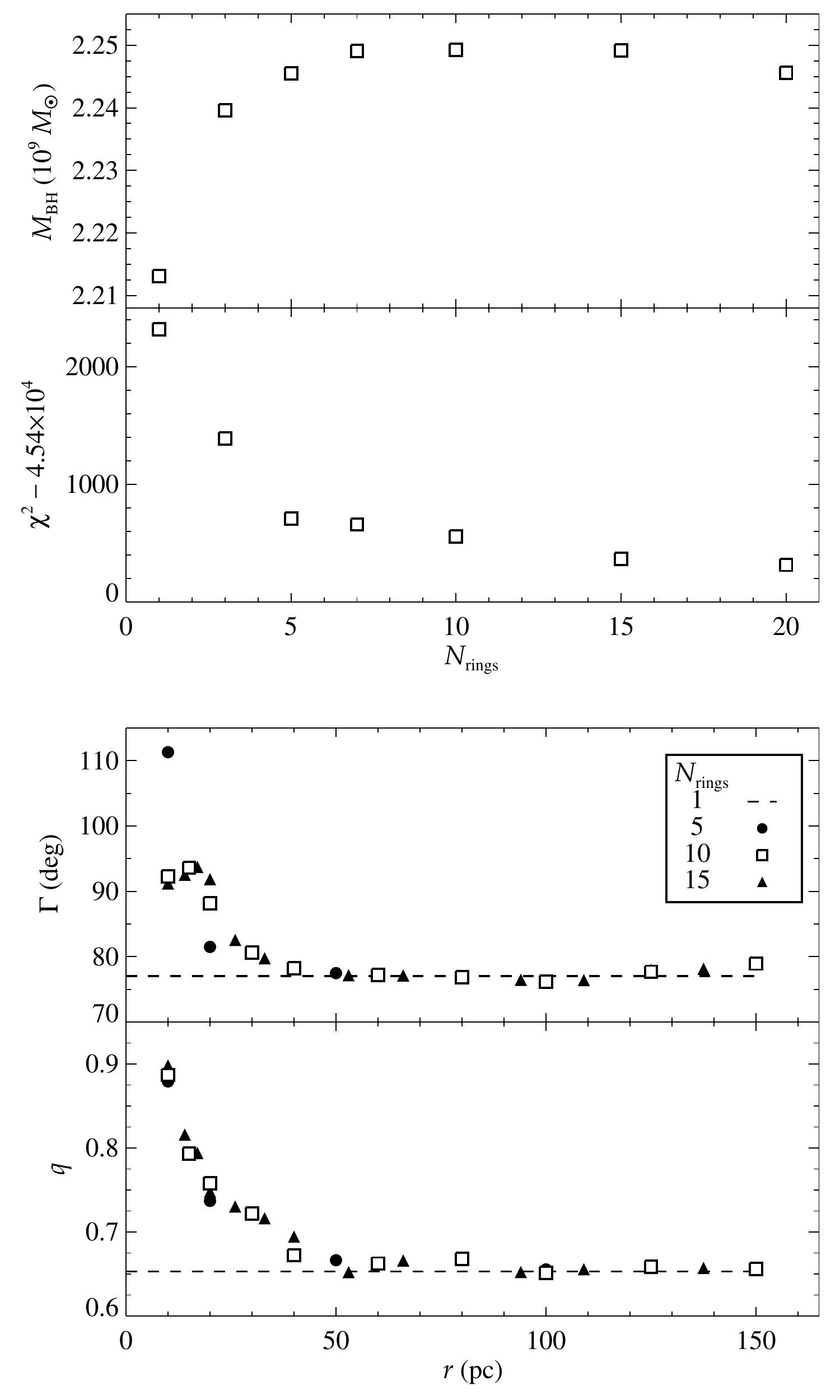}
\begin{singlespace}
  \caption{Comparison of model F1 results after changing the number of rings $N_\mathrm{rings}$ in the tilted-ring model from one (a flat disk) to twenty. The top panels show that, while the overall fit improves with larger $N_\mathrm{rings}$, the BH mass is essentially unaffected for values above ten. The bottom panels show best-fit tilted-ring parameters as a function of disk radius.}
\end{singlespace}
\label{fig:nrings}
\end{center}
\end{figure}

\textit{Tilted-ring model:} The choice of ten rings to anchor our tilted-ring model was somewhat arbitrary but appears sufficient to recover the disk structure. Model parameters may be sensitive to the number of rings $N_\mathrm{rings}$ that define the warped disk, so we explored different annular spacing from single $\Gamma$ and $q$ values (a flat disk) to $N_\mathrm{rings}=20$. To isolate the effect of changing the number and spacing of rings in the warped disk model, \vext\ is still optimized at the initial ten radial locations. As shown in Figure~\ref{fig:nrings}, increasing the number of rings does improve the overall fit but the BH mass is not significantly impacted for $N_\mathrm{rings}\geq 5$. When using between ten and twenty rings, the best-fit BH masses span a range of only $3.7\times 10^6$ \msun\ ($\sim$0.16\% in BH mass). For $N_\mathrm{rings}\geq 10$, the tilted-ring solutions return consistent, small-amplitude oscillations in $\Gamma(r)$ and $q(r)$ (of $\sim$2\degr\ in both PA and $i$; see Figures~\ref{fig:kinmod} and \ref{fig:nrings}) for radii $r>50$ pc.

\textit{Fitting region:} In the models described in Tables~\ref{tbl:mods} and \ref{tbl:mod_res}, we measured $\chi^2$ by fitting to essentially the entire disk. However, our symmetric models cannot fully account for local irregularities in the velocity and velocity dispersion fields. Here, we highlight the most apparent discrepancies and, by adjusting the model fitting region, estimate their potential impact on our dynamical modeling results.

The fitting region for model F1 (and all other models in Table~\ref{tbl:mods}) gives roughly equal weight to the red and blueshifted portions of the inner disk, even though the molecular gas within $R<0\farcs2$ on the approaching side of the disk appears to be in sub-Keplerian rotation.  Assuming that the approaching velocity peak represents one of these local irregularities, we explored its impact on model results by excluding the affected data: we restricted the fitting region on the approaching side to channels where $|\vlos-\vsys|<350$ \kms. The fitting region is otherwise unchanged, and this test retains the full generality of model F1. After optimizing to the data cube, we find only a small BH mass increase $\Delta \mbh=2.6\times10^6$ \msun\ relative to the model F1 results. Excluding channels with obviously asymmetric gas rotation reduces the number of data points by nearly 11\% while only decreasing the number of cells containing CO emission by just 2\%. As a result, this adjustment to the fitting region produces only a small change in $\mbh$.

Due to the abundance of data points at larger radii, the full fitting region gives greater weight to data near the disk edge than near the BH. We explored the impact of our choice of fitting radius by calculating a model with $\rfit=0\farcs5$ ($\sim$75 pc), and fitting to the same range of velocity channels as model F1. This spatial region extends to the edge of the observed central upturns in CO rotation speed and includes gas that is maximally sensitive to the dynamical influence of the BH. We optimized the tilted-ring and \vext\ models only out to the first ring location beyond the new \rfit\ (at $r\sim 0\farcs54$). The final BH mass increases by $\Delta \mbh =3.9\times 10^6$ \msun\ ($\sim$0.17\%) relative to model F1. This change in BH mass is so small in part due to the radial flexibility of \vext. Using model D3 with an MGE-derived host-galaxy mass profile for comparison, adopting this same $\rfit=0\farcs 5$ during model optimization induces a larger $\sim$0.5\% relative increase in its best-fit BH mass.

The central CO(2$-$1) line widths in the best-fit model F1 cube are significantly discrepant with the data, as seen in the large $\Delta \sigmalos$ values adjacent to the nucleus along the disk minor axis (Figure~\ref{fig:c4ghm}). We considered the impact of these local line width excesses on modeling results by excluding spatial locations where $\Delta \sigmalos > 25$ \kms\ across all channels. Not surprisingly, this $\sim$3\% decrease in \ndof\ produces a much improved overall model fit with $\chisqnu=1.146$. However, the BH mass only increases by about 0.02\%, so we do not expect these local line-width irregularities to cause significant error in \mbh.

The first two adjustments to the fitting region both produce $\Delta \mbh$ changes that are commensurate with the model F1 BH mass statistical uncertainty. To understand the significance of these shifts, we applied the same Monte Carlo error analysis to the best-fit test model cubes, subject to the respective changes to the fitting region. The resulting $1\sigma$ statistical uncertainties are $\sim$1.2$\times 10^7$ \msun\ (roughly 0.5\%) in \mbh\ for each test. In the first case, the larger BH mass statistical uncertainty is driven by more poorly constrained $\Gamma$, $q$, and \vext\ values for $r<30$ pc; in the second case, it arises due to less certainty in \vsys\ and the \sigmaturb\ parameters. These tests demonstrate that, irrespective of the elevated \chisqnu\ values, our model fits to the totality of the disk yield an \mbh\ measurement that is insensitive to locally irregular kinematics. Figure~\ref{fig:c4pvdm} illustrates the good agreement between the observed and modeled PVDs everywhere except the approaching velocity peak for $|\vlos-\vsys|\sim400$ \kms.

\textit{Central CO hole:} Model F1 overpredicts the CO line widths at the nucleus, with data-model $\Delta \sigmalos$ residuals falling below $-200$ \kms\ in the central pixel. The most simple explanation is that low S/N nuclear CO emission may produce high-velocity line wings that remain buried beneath the noise and are therefore not reflected in the observed line widths. Another plausible explanation is that the deconvolved model CO flux map contains excess surface brightness at the disk center, overproducing unresolved high-velocity emission at the nucleus that translates to high model \sigmalos\ values. To test this possibility, we set the intrinsic model CO flux to zero within the synthesized beam area centered on the nucleus before again optimizing the model cube. In this case, the model cube line widths measured from the best-fit model decrease by $\sim$50 \kms\ with a slight overall improvement in the fit (to $\chisqnu\sim 1.177$). However, setting the central CO hole surface brightness to zero only increases the best-fit \mbh\ by 0.01\%.

\begin{figure}
\begin{center}
\includegraphics[trim=0mm 0mm 0mm 0mm, clip, width=\columnwidth]{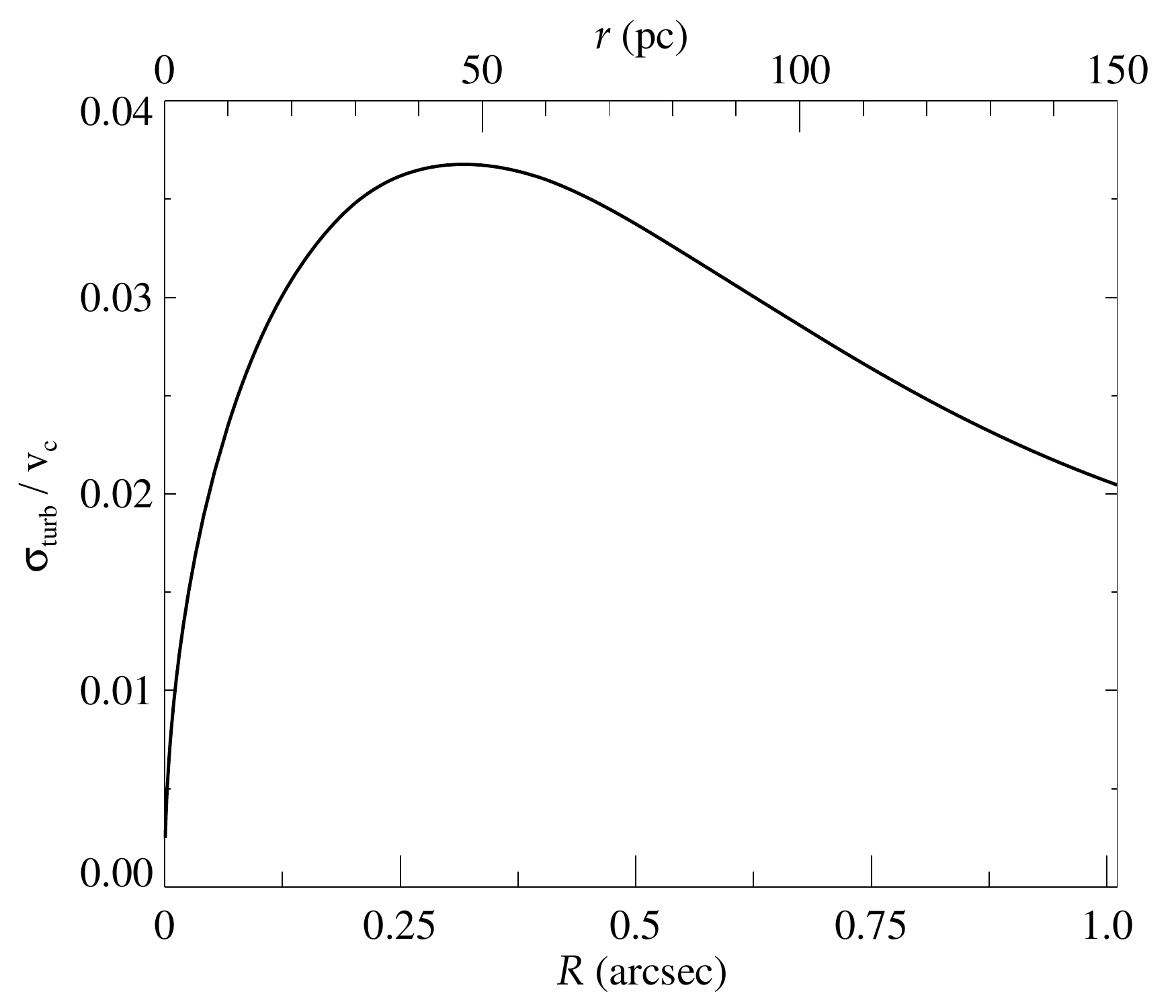}
\begin{singlespace}
  \caption{From our best-fit model F1, the ratio of intrinsic line dispersion to the disk circular velocity as a function of radius.}
\end{singlespace}
\label{fig:sigvc}
\end{center}
\end{figure}

\textit{Radial motion:} Regardless of their formation method \citep[e.g., see][]{lau05,dav11b,martini13}, circumnuclear disks experience both secular evolution \citep{dav18} and ongoing gas accretion \citep[albeit perhaps at very low levels;][]{vande15} that may result in detectable deviations from purely circular rotation. The relaxed molecular gas kinematics and regular dust disk morphology do not suggest any recent disruptions to the NGC 3258 molecular gas disk, although its mildly warped structure may indicate an ongoing settling process or perturbations arising from a triaxial galaxy potential \citep[e.g.,][]{ems11}.

We first adapted model F1 to include a spatially-uniform radial velocity term \vrad\ as a free parameter to simulate either bulk gas inflow or outflow. The radial flow component is projected along the line of sight and added to the projected tangential speed, which we approximate with the circular speed for the assumed \mbh\ value and the host-galaxy mass model. While not a self-consistent model of disk rotation, we simultaneously optimized \vrad\ with the other free parameters in this toy model to see how much radial flow is kinematically allowed by the data. This initial test favors bulk inflow with a speed of just 0.85 \kms\ while the BH mass converges to the exact model F1 value.

Radial flows introduce twists in the line of nodes of otherwise circularly-rotating disks \citep[e.g., see the analogous protoplanetary disk modeling of][Figure A4]{walshc17}. Because the kinematic twists appear strongest within the inner $\sim$40 pc (see Figure~\ref{fig:kin}), we tested whether the kinematics within this region might be consistent with flat-disk rotation and a higher inflow speed. After adding a radial flow component to model D3 with \vrad\ applying just to pixels at $R<0\farcs26$, we find that \vrad\ converges to an inflow speed of 41 \kms. This test largely reproduces the kinematics within this central region and achieves an overall $\chi^2=47335.7$ with $\ndof=38997$ for $\chi^2_\nu=1.214$, which is a modest improvement over the original $\chi^2_\nu=1.217$ for model D3. While the above \vrad\ value approaches 10\% of the circular rotation speed at these radii, the best-fit BH mass of $2.213\times10^9$ \msun\ is only about 0.2\% lower than the corresponding best-fit \mbh\ in Table~\ref{tbl:mod_res}.

We then adopted \vrad\ as a free parameter for $R<0\farcs26$ in our model F1 framework. After simultaneously optimizing all 40 free parameters, we find a global minimum with $\vrad\sim0$ \kms\ while the remaining model parameters converge to the fiducial values given in Tables~\ref{tbl:mod_res} and \ref{tbl:disk}. Since a radial flow component can reproduce some of the apparent kinematic twists that arise from a warped disk, we anticipated significant degeneracy between \vrad\ and the $\Gamma$ and $q$ parameters for at least the inner ring positions. After setting the initial inflow speed guess to 40 \kms, the model F1 variant settles on a local minimum where $\vrad=26$ \kms\ and the $\Gamma$ and $q$ parameters remain below 80\degr\ and 0.74, respectively. This local minimum achieves a slightly worse $\chi^2_\nu$ of 1.180 and returns a BH mass of $2.236\times10^9$ \msun\ that is only 0.7\% lower than reported for the original model F1 in Table~\ref{tbl:mod_res}.

Finally, to rule out any significant impact of radial gas motion on the BH mass measurement, we again incorporated a bulk flow term \vrad\ into model F1 but only fit the model to points where $R>0\farcs26$, thereby focusing on the region with the lowest disk warping to minimize possible degeneracies. The $\Gamma$, $q$, and \vext\ parameters for the first four ring positions are fixed to the values in Table~\ref{tbl:disk}. We find that \vrad\ settles on an inflow speed of 0.86 \kms\ while the BH mass converges to $2.247\times10^9$ \msun, which corresponds to a mass difference of $\sim$0.1\% from the fiducial value. After Monte Carlo resampling the resulting best-fit model cube, the distribution of \vrad\ values suggests that the possible detection of bulk radial inflow in the outer disk region is not particularly significant, being only $1.3\sigma$ removed from the $\vrad\sim0$ \kms\ case. Since the kinematic twists in the CO velocity field appear to arise almost entirely from an intrinsically warped inner disk and not from gas inflow, we do not include any $\Delta\mbh$ from this radial flow analysis in the final error budget.

Our conclusion of a low inflow rate within the CO disk is consistent (modulo an assumption of a \textit{steady} flow) with evidence of a low inflow rate on smaller scales. If we assume an average inflow speed of just 1 \kms\ (a level that is dynamically unimportant for our BH mass measurement), the entire circumnuclear disk with a radius of $\sim$150 pc would accrete onto the BH in about 150 Myr. For a total gas mass of $\sim$10$^8$ \msun, the average mass accretion rate over this accretion timescale is about 0.7 \msun\ yr$^{-1}$. This in turn translates to a ratio of BH mass accretion to the Eddington limit of $\dot{M}/\dot{M}_\mathrm{EDD}\sim0.014$ \citep[assuming a standard radiative efficiency $\epsilon=0.1$;][]{vande10}, which would imply an accretion luminosity of $L_\mathrm{bol}\sim10^{45.6}$ erg s\per. We do not see evidence for luminous AGN activity in \hst\ imaging, optical spectroscopy \citep{jon09}, or molecular gas outflows, suggesting that any modest inflow of molecular gas within the CO disk neither reaches the nucleus nor is directed out -- consistent with negligible if any inflow at all.

\textit{Turbulence:} For gas-dynamical modeling of some ionized gas disks, the intrinsic line widths are a substantial fraction of the disk rotation speed, suggesting significant local turbulence that is generally presumed to provide pressure support to the disk \citep{ver00,bar01,wal13}. In these cases, models based on purely circular rotation will underestimate the true BH masses, because the disk rotation velocity will lag behind the circular velocity (analogous to asymmetric drift in stellar dynamics). In thin-disk models that neglect this asymmetric drift effect, the fractional bias in \mbh\ is expected to be of order $(\sigmaturb/v_\mathrm{c})^2$. 

Our gas-dynamical models assume a perfectly thin and dynamical cold disk within NGC 3258, and do not account for the dynamical effect of turbulent pressure support. For the best-fit model F1, $\sigmaturb/v_\mathrm{c}$ reaches a maximum of 0.037 at $\sim$50 pc from the disk center (see Figure~\ref{fig:sigvc}) and a mean value $\langle\sigmaturb/v_\mathrm{c}\rangle = 0.030$ averaged over the disk surface. This molecular gas disk is truly dynamically cold. Since the fractional change to the BH mass resulting from turbulent pressure support scales as $(\sigmaturb/v_\mathrm{c})^2$, we expect an upward correction to \mbh\ of order $\sim3\times 10^6$ \msun\ (corresponding to $\sim$0.14\%) that is similar to the statistical model-fitting uncertainty.

\textit{Distance Uncertainty:} Since the enclosed mass in the rotating disk model scales as $M(r)=rv_\mathrm{c}^2/G$, the inferred BH mass should in principle be directly proportional to the assumed angular size distance, although in practice other modeling details such as beam smearing may slightly modify this dependence. We anticipate that the uncertainty in the galaxy's adopted distance $D_L=31.9$ Mpc of slightly more than 10\% will introduce a commensurate systematic uncertainty in BH mass. We quantify this uncertainty by calculating two test models with the luminosity distance shifted by $\pm 1\sigma$ from the assumed value (i.e., $D_L=35.8$ and 28.0 Mpc, corresponding to angular scales of 170 and 133 pc arcsec\per, respectively). After optimizing over all free parameters, we obtain best-fit BH masses that are $\Delta \mbh = \pm 2.7\times10^8$ \msun\ (or about 12\%) removed from the fiducial model F1 value. We note that the uncertainty in our assumed NGC 3258 $D_L$ value does not include any systematic contributions that arise from Cephied period-luminosity metallicity corrections or uncertainties in the zero point \citep[that are of order $\sim$0.1 mag; e.g.,][]{mei05,bla10}.

Some estimates of NGC 3258's distance disagree with the ground-based SBF measurent from \citet{ton01} by more than its quoted $1\sigma$ errors. Using \hst\ observations to measure SBFs in this galaxy, \citet{can05} determined $m-M=33.00\pm0.15$ mag, corresponding to $D_L=39.8\pm2.8$ Mpc, although their analysis lacked empirical calibration of the SBF method in the F814W filter \citep{bla10}. Using an angular scale of 189.5 pc arcsec\per\ derived for this second SBF distance, the best-fit BH mass increases by $\Delta \mbh =5.6\times 10^8$ \msun\ (or about 25\%) from the model F1 case. Other distance measurement techniques yield distance modulii between $32.42\pm 0.19$ \citep[or $D_L=30.5\pm 2.8$ Mpc, using the globular cluster luminosity function;][]{bas08} and $33.73\pm 0.41$ mag \citep[or $D_L=55.7\pm 11.6$ Mpc, using the Fundamental Plane;][]{bla02}, with respective ($\Delta \mbh/\msun$) of $-5.9\times 10^7$ and $1.6\times 10^9$ from our model F1 results. We report a $\pm 12\%$ systematic distance uncertainty in the BH mass based on the reported SBF distance uncertainty from \citet{ton01}, but the \mbh\ may plausibly lie in the range $(2.0-3.8)\times 10^9$ \msun\ based on these other distance estimates. Thus, while our model fits provide a highly precise determination of \mbh\ given an assumed distance to NGC 3258, the galaxy distance uncertainty dominates the total \mbh\ error budget.

As a final note on distance uncertainties, the preceding calculations have not accounted for source or observer peculiar velocities. Ideally, line-of-sight velocities and line width maps are transformed into observed frequency units assuming separate cosmological and peculiar redshifts in the relationship $(1+z_\mathrm{obs})=(1+z_\mathrm{cos})(1+z_\mathrm{pec})$, with the angular size distance depending on $z_\mathrm{cos}$ and not $z_\mathrm{obs}$. To investigate the impact on our \mbh\ determination from this neglect of peculiar motion, we first removed the Sun's peculiar velocity contributions by transforming the Cycle 4 data into the cosmic microwave background (CMB) frequency reference frame wherein $z_\mathrm{obs}=0.010283$. Our adopted $D_L$ for this galaxy corresponds to $z_\mathrm{cos}=0.007745$ \citep[][using $H_0=73.24$ \kms\ Mpc\per; \citealp{rie18a}]{wri06}, which translates to $1\arcsec=152.3$ pc and a line-of-sight Doppler shift $v_\mathrm{pec}\approx 760$ \kms\ for NGC 3258 in the CMB frame. Then, we fixed this $z_\mathrm{cos}$ value in a test model while allowing NGC 3258's peculiar velocity  $v_\mathrm{pec}$ to vary as a free parameter in place of \vsys. This test converges to $v_\mathrm{pec}=753$ \kms\ with a BH mass decrease of $\Delta \mbh=-1.3\times 10^7$ \msun\ from our model F1 result. In light of this galaxy's disparate distance estimates, we did not attempt to separate out its cosmological and peculiar redshift contributions in models A$-$F, and we do not consider peculiar velocity systematics in the final BH mass error budget.

\textit{Final error budget:} The statistical uncertainties on \mbh\ are equivalent to the largest model-dependent systematic terms, while the distance uncertainties are much larger than either of these other terms. Given the wide range of relative contributions, we separated these into distinct statistical (stat), model systematic (mod), and distance systematic (dist) terms in the final BH mass error budget. To estimate the total model systematic uncertainty, we separately combined in quadrature the positive and negative $\Delta \mbh$ contributions listed above, with the largest of these (non-distance) systematics being at the 0.2\% level. Our final BH mass with $1\sigma$ uncertainty ranges is then $(\mbh/10^9\,\msun)=2.249\pm0.004$ (stat) $^{+0.007}_{-0.004}$ (mod) $\pm0.270$ (dist).

\section{Discussion}
\label{sec:discussion}

\subsection{BH Mass}
\label{sec:disc_bhmass}

NGC 3258 has no previous BH mass measurement to compare with our gas-dynamical modeling results. Using this galaxy's $\sigma_\star$ and $M_K$ values and uncertainties listed in Section~\ref{sec:intro}, standard $\mbh-\sigma_\star$ and $\mbh-L_K$ relations for classical bulges and elliptical galaxies \citep{kor13} predict ($\mbh/10^9$ \msun) values of $(0.62^{+0.43}_{-0.23})$ and $(1.00^{+0.18}_{-0.16})$, respectively. Our NGC 3258 BH mass of $2.249 \times 10^9$ \msun\ is more than a factor of two larger than these predictions and lies on the extreme edge of measurements populating the $\mbh-\sigma_\star$ and $\mbh-L_K$ relations. Significant tension between prediction and measurement remains when employing a different univariate correlation \citep[see also][]{van16,sag16}, or after accounting for the impact of distance uncertainty on \mbh\ and $L_K$.

Quiescent BCGs and BGGs often exhibit cored surface brightness profiles \citep{lau07b,rus13b}, presumably formed through scouring by massive binary BHs \citep[e.g.,][]{rav02,tho14}. Even a partial depletion of their stellar core will suppress $\sigma_\star$ measurements for these luminous galaxies relative to $\mbh-\sigma_\star$ extrapolations for normal ETGs \citep{lau07a}. A more reliable indicator for cored galaxies is the break radius $r_\mathrm{b}$, which is found to scale with both \mbh\ and \rg\ \citep{tho16}. While certainly not an extreme example \citep[e.g., see][]{dul17}, $H$-band surface brightness profile modeling of NGC 3258 described in \S\ref{sec:extinction} suggests a break radius of $\sim$230 pc. Circumnuclear dust extinction that acts on similar scales makes it difficult to confidently determine $r_{\rm b}$ from the NIR imaging alone. Based on our measured BH mass and sphere of influence, the $\rg-r_\mathrm{b}$ and $\mbh-r_\mathrm{b}$ relations of \citet{tho16} return a predicted $r_{\rm b}$ between 130-160 pc, which is slightly lower than the measured $r_\mathrm{b}$ but remains consistent within the scatter of these relationships.

\subsection{The impact of angular resolution on BH mass measurement precision}
\label{sec:disc_res}
 
In general, the most precise extragalactic BH mass measurements are those derived from H$_2$O megamaser disk observations. These maser BH mass measurements typically have statistical and systematic uncertainties of at least a few percent \citep[e.g.,][]{kuo11,gao17,zha18}.  However, the BH mass measurement we present here for NGC 3258 has higher precision than many maser BH measurements (apart from distance uncertainties). Here, we discuss the impact of angular resolution on \mbh\ determination as well as various limiting factors that have affected other gas-dynamical modeling efforts.

\begin{figure}
\begin{center}
\includegraphics[trim=0mm 0mm 0mm 0mm, clip, width=\columnwidth]{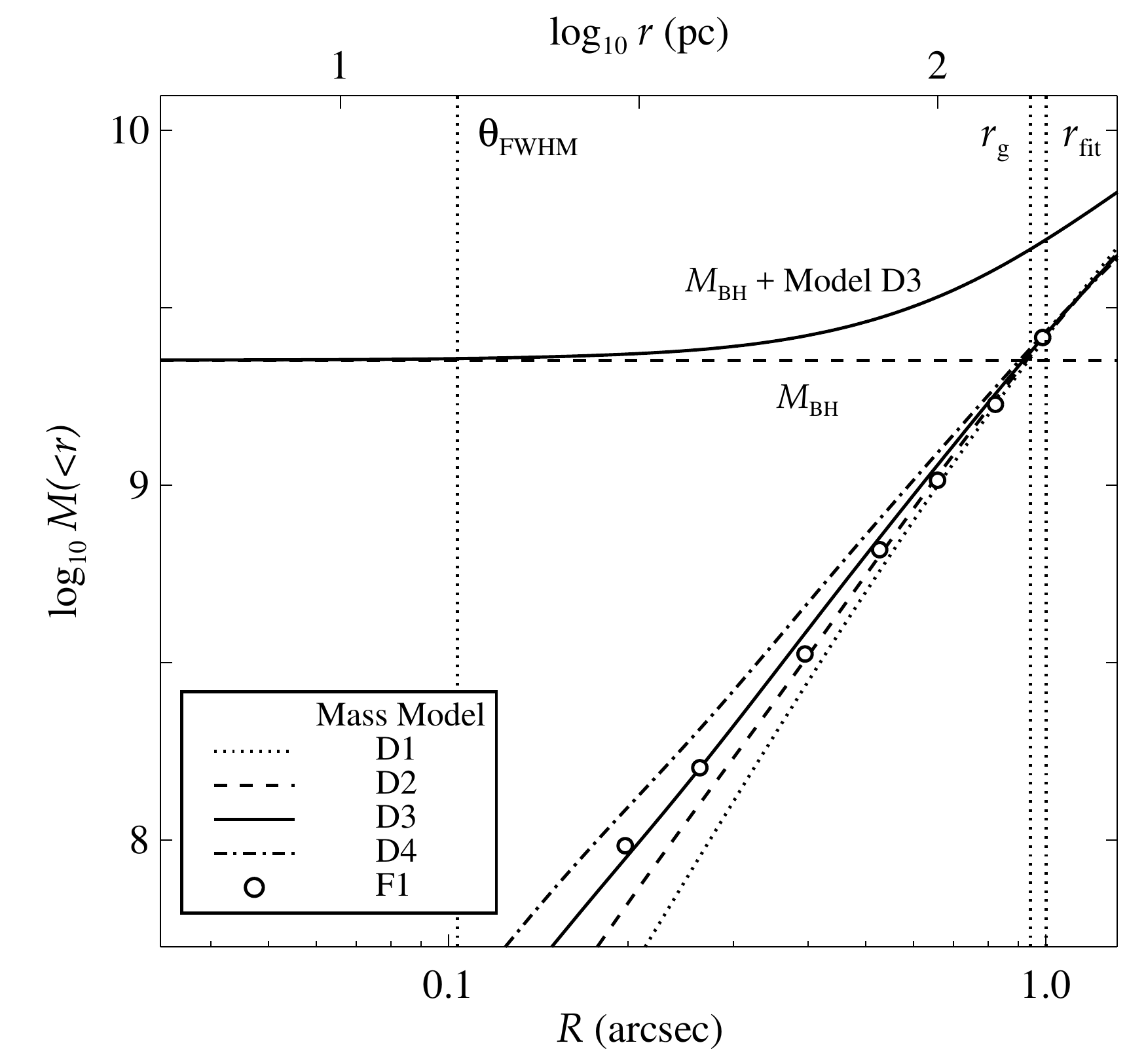}
\begin{singlespace}
  \caption{Enclosed mass $M(<r)=r v_{\rm c}^2/G$ in NGC 3258 as a function of physical radius for the various MGE galaxy mass distributions (with the corresponding \vcstar\ values scaled by best-fit $\sqrt{\Upsilon_H}$; see Tables~\ref{tbl:mods} and \ref{tbl:mod_res}) that are extrapolated beyond the edge of the CO rotation pattern. Dotted vertical lines indicate the ALMA Cycle 4 average beam size $\theta_{\rm FWHM}$ and fitting region radius \rfit, with $\rg=0\farcs94$ determined for a BH mass of $2.249\times10^9$ \msun.}
\end{singlespace}
\label{fig:luminosity}
\end{center}
\end{figure}

Very Long Baseline Interferometry (VLBI) observations of megamaser disks probe much closer (on sub-parsec scales) to their central BHs in absolute terms than do our ALMA observations. However, the BH mass and \rg\ for NGC 3258 are two and one orders of magnitude larger, respectively, than for many maser disk galaxies \citep[e.g.,][]{kuo11}. We follow \citet{rus13b} and compute the ratio $\xi=2\rg/\theta_\mathrm{FWHM}$ of the BH diameter of influence to the average beam size, which indicates the \textit{relative} resolution of \rg. Observations with larger values of $\xi$ are more amenable to producing a precise \mbh\ determination. While values of $\xi$ below $\sim$2 can still yield useful measurements of \mbh\ \citep[e.g.,][]{davis14}, such data will lead to larger \mbh\ uncertainties as the BH mass becomes increasingly susceptible to systematic biases from uncertainty in the stellar mass profile and other factors \citep{rus13a,kor13,bar16a,bar16b}.

For megamaser galaxies with well-measured values of $\sigma_\star$, VLBI observations typically achieve $\xi\sim10-100$ \citep{gre96,lod03,kon08,gre10,hur11,kuo11,yam12,gre16,gao16,gao17,zha18}, while for the prototypical megamaser disk in NGC 4258, $\xi\sim1000$ with high-velocity maser sources detected to within $\sim$0.02\rg\ of the active nucleus \citep{miy95,her05,hum13}. For comparison, published ALMA CO imaging of ETGs has typically reached relatively low $\xi$ values (e.g., $\xi\lesssim2$: \citealp{dav13a,oni17}; \citetalias{boi17}; \citealp{dav17b,dav18}) with one exception being the high resolution observations of NGC 1332 presented by \citet{bar16a}, which achieved $\xi\sim10$ along the disk's projected major axis. Our Cycle 4 imaging of NGC 3258 more fully resolves \rg\ than any previous ALMA observations, achieving $\xi\approx 17$ (see Figure~\ref{fig:luminosity}) with CO(2$-$1) emission detected down to $\sim0.14\rg$ from the disk center. This ALMA data set achieves greater relative resolution of \rg\ than about a third of all VLBI megamaser observations.

We note that this $\xi$ criterion ignores the adverse impact on BH mass measurement when the line surface brightness shows a central hole, or when beam smearing affects highly inclined disks. In \S\ref{sec:disc_co} we discuss central emission-line deficits in more detail. With regard to the latter case, \citet{bar16b} highlight problems that arise in model fitting of smooth disk emission when the kinematics are not sufficiently well resolved along the disk's projected \textit{minor} axis. In that situation, beam smearing of the disk's central kinematics spatially blends low-velocity emission with the high-velocity emission originating from along the disk's major axis, resulting in a broad ``fan'' of emission spanning a wide velocity range in the major-axis PVD. This situation may result in a degeneracy between rotation and dispersion in the disk's central region that can pose severe difficulties for model fitting.

For \mbh\ determination using ALMA data, \citet{bar16b} argue that observations should ideally resolve at least $\rg \cos i$ to fully mitigate these disk inclination effects. As a case in point, the high angular resolution ALMA observations of NGC 1332 achieve $\xi\sim 10$ but only $\xi\cos i\sim 1.3$ due to a high disk inclination \citep{bar16a}. As a result, minor-axis emission remains somewhat entangled with the rapidly rotating nuclear emission and is a factor that precludes very tight constraints on its BH mass. For NGC 3258, its more moderate disk inclination translates to $\xi\cos i \approx 12$, marking the first published case that a mm/sub-mm line tracer has fully resolved \rg\ over an entire circumnuclear disk as projected on the sky.

Even though most VLBI megamaser observations achieve large $\xi$, their few-percent \mbh\ uncertainties arise from maser source scatter about the disk midline and relative positional errors that complicate dynamical modeling of perhaps only 10$-$30 data points. The level of detail when modeling the disk structure and kinematics may further impact the final BH mass precision. In the best cases, gas-dynamical models can recover the parsec-scale disk structure of these nearly edge-on, moderately warped disks \citep[e.g.,][]{her05,hum13,gao16}. For megamasers with large source scatter or few data points, unconstrained disk warping will introduce additional systematic uncertainty to their final \mbh\ error budget.

Our ALMA Cycle 4 observations of NGC 3258 are not subject to these same limiting factors. The CO-bright disk area is covered by nearly 200 synthesized beams, resulting in an \mbh\ determination with very low statistical uncertainties that is also insensitive to locally irregular kinematics. As we describe in \S\ref{sec:disc_dust}, increasing the angular resolution much above $\xi \cos i \sim 2$ does not drastically affect the best-fit BH mass. However, highly resolving \rg\ enables detailed dynamical modeling to account for a more general disk structure and a flexible host galaxy mass profile. These additions eliminate the primary model systematics that would otherwise restrict the NGC 3258 BH mass precision to several percent (not including the distance uncertainty).

Another noteworthy feature of this measurement, in comparison with other CO-based BH mass measurements carried out to date, is that the molecular disk in NGC 3258 is almost entirely located within \rg. In other cases, the CO emission typically extends to scales far beyond \rg\ within the host galaxy, and the disk kinematics at $r>\rg$ are only minimally sensitive to \mbh. When models are fit to a spatial region dominated by pixels at $r>\rg$, the results will be more susceptible to systematic error in the determination of the spatially extended mass profile. NGC 3258 is the first ETG for which the combination of the disk structure and the high resolution of the ALMA observations allow for dynamical models to be constrained solely by fitting to kinematics within $r\lesssim\rg$, a situation that is optimal for carrying out a BH mass measurement that is both highly precise and minimally susceptible to systematic error.

\subsection{Dust extinction}
\label{sec:disc_dust}

Dust that accompanies the molecular gas disk in NGC 3258 suppresses the galaxy's central surface brightness and may result in substantial mischaracterization of the intrinsic circular velocity profile arising from its stellar mass distribution. From the dust modeling method detailed in \S\ref{sec:extinction}, we find strong evidence that the NGC 3258 disk is optically thick at visible wavelengths, with extinction reaching $A_V \sim 5$ mag near the disk center. However, we cannot confidently recover the intrinsic stellar luminosity profile from this dust model.

Our results imply that gas-dynamical models for dusty galaxies need to allow for a range of extinction levels (corresponding to different central stellar slopes) to capture the full uncertainty in the BH mass. To that end, we constructed and employed four extinction-corrected \vcst\ profiles to model the Cycle 2 and 4 data sets. The best-fit \mbh\ estimates derived from these \vcst\ models span $\sim$13\% and 10\% ranges in mass (see Table~\ref{tbl:mod_res}), respectively, indicating that the increase in angular resolution does not not significantly reduce the dust extinction uncertainties. As long as the host galaxy contribution to the total circular velocity profile remains dynamically important and is determined using optical/NIR imaging, a dusty galaxy nucleus will always introduce some irreducible systematic uncertainty to \mbh\ due to the uncertain dust correction, even when \rg\ is well resolved.

Radiative transfer modeling could produce a more detailed extinction map across the disk, but we anticipate that the recovered stellar surface brightness profile will retain some level of uncertainty on account of difficulties when attempting to fully account for complex dust geometries and multiple light sources. Without highly detailed extinction modeling, the \textit{only} way to eliminate the extinction uncertainty impact on \mbh\ is thus to obtain sufficiently high angular resolution observations to directly constrain $\vext (r)$ using the emission line kinematics, as we have demonstrated using model F1.

\subsection{CO emission in ETGs}
\label{sec:disc_co}

To date, CARMA and ALMA observing programs to measure BH masses have published maps of CO emission on $\sim$\rg\ scales for ten ETGs having high S/N detections of molecular line emission  (\citealp{dav13a,bar16a,bar16b}; \citetalias{boi17}; \citealp{oni17,dav17b,dav18,smi19}), and the sample continues to grow as further ALMA observations have been carried out in recent cycles. Additional ALMA observations have revealed disk-like gas rotation in a handful of other nearby ETGs \citep{oni15,zab18,san19,vil19}, but we do not consider these results in the current discussion due to much more coarse angular resolution or the use of a different molecular line species.

Based on these select targets, strong, high-velocity CO emission arising from deep within \rg\ appears to be uncommon for molecular gas disks in ETGs. Their central CO properties can be divided into three regimes: (1) those with no line emission from within \rg\ (i.e., due to large holes that may or may not be resolved); (2) those that show slight central upturns in emission-line velocities \citep[e.g., NGC 1332;][]{bar16a}, indicating the CO-bright gas does not populate very deep within \rg; and (3) those that exhibit very strong central velocity upturns, tracing quasi-Keplerian rotation.

For the set of ten ETGs with published CO maps at $\sim$\rg\ resolution, seven do not show clear evidence of a large central CO deficit. Only four targets from this set demonstrate either case (2) or (3) emission with at least some hint of rising central gas rotation speeds at small radii, as would be expected for gas disks extending down to small radii around large central BHs with $\mbh\sim10^8-10^9$ \msun. Unambiguous, case (3) detection of CO emission arising deep within \rg\ appears to be rare, with NGC 3258 being the only compelling case among the published targets to date. This paucity hints that central holes in CO emission with radii of order \rg\ are common for ETGs and are simply undetected due to beam smearing.

The ETGs observed by CARMA and ALMA for BH mass measurement were selected for high-resolution CO observations based on the known presence of gas disks either from prior CO observations or from the presence of well-defined circumnuclear dust disks in \hst\ imaging, and such disks are found to be present in only about 10\% of ETGs overall \citep[e.g.,][]{tra01,lau05}. Thus, the fact that strong high-velocity central rotation is not commonly observed for carefully selected targets suggests that case (3) emission will only be found in a very small percentage of the total ETG population.

The absence of CO emission in the inner regions of most ETG circumnuclear disks suggests that central molecular gas is either absent or poorly traced by low$-J$ lines. Several distinct processes may act to deplete the disk core of molecular gas, including photo-dissociation in an intense interstellar radiation field (perhaps due to central star formation), disk instabilities due to a non-axisymmetric potential, and episodic AGN activity that may dissociate and ionize the circumnuclear gas and perhaps drive it out in a wind. In addition, \citet{dav18} argue that the density of any remaining central molecular gas may be below the critical density (at least for the CO 2$-$1 and 3$-$2 transitions) due to strong BH tidal forces that prevent disk fragmentation into clouds \citep[see also][]{mar13}. Alternately, the molecular gas may become increasingly dense towards the galaxy center and be better traced by lines with larger critical densities. For NGC 3258, ALMA imaging of different CO lines at similar resolution as our Cycle 4 CO(2$-$1) observations, and optical spectroscopy to search for coincident ionized gas tracers, will provide further clues to the nature of the central hole in the CO(2$-$1) distribution.

As we argued in \citetalias{boi17}, imaging at a spatial resolution of $\sim$\rg\ is crucial to confidently identify rapid central gas rotation. Careful target selection may increase the probability of finding case (2) or (3) disks in future ETG surveys. Assuming CO-bright gas follows the optically thick dust, inspection of broadband imaging and color maps may help determine if the gas is likely to extend within \rg\ (with the caveat that observed color does not always track very optically thick regions). Moreover, surveys that select targets based on central stellar surface brightness may obtain a greater number of case (3) ETGs; for NGC 3258, its cored stellar surface brightness profile results in lower circular velocity contributions from stars (relative to the BH) and therefore a more distinct central rise in emission line velocities. We also note that focusing on disks with intermediate (between face-on and edge-on) inclination angles will facilitate more robust BH mass measurements. Regardless of the selection criteria, targeted BH surveys should first obtain initial line imaging at $\sim$\rg\ spatial resolution to increase the efficiency of case (2) and (3) detections, and higher-resolution observations can then be carried out for the most promising targets.

\section{Conclusions}
\label{sec:conclusions}

This paper presents the most precise BH mass measurement to date for an elliptical galaxy, using $\sim$0\farcs10$-$resolution ALMA Cycle 4 CO(2$-$1) imaging of NGC 3258's arcsecond-scale molecular gas disk. These new ALMA observations reaffirm our previous Cycle 2 findings of a dynamically cold disk with CO emission extending well within \rg\ and nearly to the galaxy center. At high spatial resolution, the disk appears to be mildly warped with a kinematic twist of $\sim$20\degr. Near the disk center, the line emission reaches the same $\sim$500 \kms\ rotation speed also detected in the Cycle 2 data set. In the Cycle 4 PVD, this rapid rotation is now resolved into a tight locus of emission tracing quasi-Keplerian rotation that extends inward to within $\sim$20 pc of the nucleus and terminates in a central hole in the CO(2$-$1) emission.

While these ALMA observations highly resolve \rg\ for the first time using mm/sub-mm gas tracers, we cannot neglect the host galaxy gravitational potential during gas-dynamical modeling. Using an inclined dust disk model to predict optical/NIR colors, we demonstrate that the extinction increases towards the disk center, reaching $A_V \sim 5$ mag at $R \sim 0\farcs5$. Incorporating extinction-corrected stellar mass profiles into our forward dynamical modeling procedure yields \mbh\ values that span a $\sim$10\% range in mass, which greatly exceeds the statistical uncertainty for any an individual mass model. As our Cycle 4 observations highly resolve the regular disk kinematics, we eliminate dust extinction systematic uncertainties by directly constraining the host galaxy mass profile in our final dynamical model using the observed CO(2$-$1) kinematics.

These results also demonstrate that, for mildly warped disks, fitting data with a flat disk model is not likely to lead to large systematic error in the BH mass. Nevertheless, our detailed gas-dynamical models directly constrain the warped disk structure when optimizing the tilted-ring model to the full NGC 3258 CO(2$-$1) data cube. The $\sim$3\% difference between flat and warped disk model BH mass measurements is large relative to the other sub-percent level modeling systematics. In more typical instances of gas-dynamical modeling, the difference in \mbh\ when measured using either flat or warped disk geometries should be well within their error budgets \citep[typically 10-20\% or larger;][]{kor13}. 

In our final gas-dynamical model, we determine the best-fit NGC 3258 BH mass to be $2.249\times 10^9$ \msun\ with sub-percent level modeling systematics that are equivalent to its statistical uncertainty. For an assumed distance, the high accuracy and precision of this BH mass measurement is commensurate with that obtained for the best-case megamaser disk in NGC 4258. Even after accounting for uncertainties in the galaxy distance, which introduces an additional 12\% contribution to the full \mbh\ error budget, this is the most precisely measured BH mass for any elliptical galaxy.

The current group of ETGs with published CO maps at high resolution suggests that high-velocity central rotation (extending to speeds well in excess of those due to the stellar mass distribution alone) is a feature only rarely present, and may be found in perhaps only a very small percentage of all luminous ETGs. Finding even a few more targets will therefore require ongoing lower-resolution ALMA imaging surveys to identify rapidly rotating gas well within \rg. For these targets, follow-up imaging at higher resolution will facilitate detailed gas-dynamical modeling that can determine BH masses to high precision.

ALMA-based BH mass measurements have already begun to provide direct comparisons with other techniques. For NGC 1332, our measurement of \mbh\ from high-resolution ALMA CO(2$-$1) data indicated a mass of $\mbh = 6.6\times10^8$ \msun\ with 10\% model-fitting uncertainty \citep{bar16a}, more than a factor of two smaller than the value of \mbh\ derived from stellar-dynamical modeling \citep{rus11}. The CO-based measurement was consistent, however, with an earlier determination of \mbh\ based on the hydrostatic equilibrium of the X-ray emitting gas in NGC 1332 \citep{hum09}.  For NGC 4697, on the other hand, BH mass measurements from ALMA CO disk dynamics \citep{dav17b} and from stellar dynamics \citep{geb03,schu11} are in good agreement. Carrying out additional direct comparisons between stellar dynamics and molecular disk dynamics remains a high priority, and the precision of ALMA BH mass measurements makes this the best available cross-check on stellar-dynamical BH mass measurements, which make up the majority of the locally measured BH census.



In the case of NGC 3258, future optical/NIR observations of this galaxy could enable direct comparison of our result with \mbh\ values measured via complementary techniques, independent of the systematic uncertainty in distance. Unfortunately, an available optical spectrum of NGC 3258 from the 6dF Galaxy Survey \citep{jon09} does not show evidence for significant H$\alpha$ or other optical emission lines, so NGC 3258 is probably not a good candidate for ionized gas kinematics observations with \hst. NGC 3258 has not previously been a target for stellar-dynamical BH mass measurement, but observations with laser guide-star AO may be feasible (using an $R\sim13$ mag star at 51\arcsec\ separation from the galaxy nucleus as a tip-tilt reference) and would allow for rigorous tests of stellar-dynamical modeling to understand the impacts of bulge triaxality, orbital anisotropy, stellar $M/L$ gradients, and dark matter on accurate BH mass measurements.

Highly precise BH mass measurements are also crucial to establish local BH demographics for ETGs. Of the small but growing sample of very massive ($\gtrsim$10$^9$ \msun) BH measurements, many are accompanied by substantial uncertainties, which may underrepresent the full error budgets due to potentially serious systematics. These factors inhibit any secure interpretation of the slope and scatter of the high-mass end of \mbh-host galaxy relationships. ALMA imaging of dynamically cold disk rotation is the most promising avenue to obtain precision \mbh\ values for luminous ETGs. A larger sample of such precise \mbh\ measured using CO kinematics will anchor these relationships at the highest BH masses. In addition, precision \mbh\ values across many ETGs will facilitate better constraints on the evolutionary processes \citep[e.g., by exploring the core vs.\ coreless elliptical dichotomy;][]{kor13} of these massive galaxies.

\acknowledgements
Research by B.D.B.\ and A.J.B.\ at UC Irvine was supported by NSF grant AST-1614212. J.L.W.\ was supported in part by NSF
grant AST-1814799. L.C.H.\ was supported by the National Key R\&D Program of China (2016YFA0400702) and the National Science Foundation of China (11721303). This paper makes use of the following ALMA data: ADS/JAO.ALMA\#2013.1.00229.S and ADS/JAO.ALMA\#2016.1.00854.S. ALMA is a partnership of ESO (representing its member states), NSF (USA) and NINS (Japan), together with NRC (Canada), MOST and ASIAA (Taiwan), and KASI (Republic of Korea), in cooperation with the Republic of Chile. The Joint ALMA Observatory is operated by ESO, AUI/NRAO and NAOJ. The National Radio Astronomy Observatory is a facility of the National Science Foundation operated under cooperative agreement by Associated Universities, Inc.
Support for \hst\ program \#14920 was provided by NASA through a grant from the Space Telescope Science Institute, which is operated by the Association of Universities for Research in Astronomy, Inc., under NASA contract NAS 5-26555.

\clearpage

\bibliographystyle{apj}
\bibliography{ms.bib}

\end{document}